\documentclass[english]{aa}
\usepackage{mathptmx}
\usepackage[T1]{fontenc}
\usepackage[latin9]{inputenc}
\usepackage{graphicx}
\usepackage{amsmath}
\usepackage[]{natbib}
\usepackage[]{longtable}
\bibpunct{(}{)}{;}{a}{}{,}

\makeatletter

\providecommand{\LyX}{L\kern-.1667em\lower.25em\hbox{Y}\kern-.125emX\@}

\DeclareRobustCommand*{\lyxarrow}{%
\@ifstar
{\leavevmode\,$\triangleleft$\,\allowbreak}
{\leavevmode\,$\triangleright$\,\allowbreak}}

\makeatother

\begin{document}
\title{Third generation stellar models\\ for asteroseismology of hot B
subdwarf stars}
\subtitle{A test of accuracy with the pulsating eclipsing
binary PG 1336--018}
\author{
V. Van Grootel\inst{1,2}
\and S. Charpinet \inst{3,4}
\and P. Brassard\inst{5}
\and G. Fontaine\inst{5}
\and E.M. Green \inst{6}
}

\institute{Institut d'Astrophysique et de G\' eophysique, Universit\' e
  de Li\` ege, 17 All\'ee du 6 Ao\^ ut, B-4000 Li\` ege, Belgium \\
\email{valerie.vangrootel@ulg.ac.be}
\and Charg\' e de recherches, Fonds de la Recherche Scientifique, FNRS,
  5 rue d'Egmont, B-1000 Bruxelles, Belgium 
\and Universit\'e de Toulouse, UPS-OMP, IRAP, Toulouse, France
\and CNRS, IRAP, 14 Av. E. Belin, 31400 Toulouse, France
\and D\'epartement de Physique, Universit\'e de Montr\'eal, CP 6128,
  Succursale Centre-Ville, Montr\'eal, QC H3C 3J7, Canada
\and Steward Observatory, University of Arizona, 933 North Cherry Avenue, Tucson, AZ, 85721, USA
}
\offprints{V. Van Grootel}

\date{Received 12 December 2012; Accepted 12 March 2013}

\newpage

\abstract{Asteroseismic determinations of structural parameters of hot B subdwarfs (sdB) have been carried out for more than a decade now. These analyses rely on stellar models whose reliability for the
required task needs to be evaluated critically.}
{We present new models of the so-called third generation (3G) dedicated to the asteroseismology of sdB stars, in particular to long-period pulsators observed from space. These parameterized models are complete static structures suitable for analyzing both $p$- and $g$-mode pulsators, contrary to the former second generation (2G) models that were limited to $p$-modes. While the reliability of the 2G models has been successfully verified in the past, this important test still has to be conducted on the 3G structures.}
{The close eclipsing binary PG 1336$-$018 provides a unique opportunity to test the reliability of hot B subdwarf models. We compared the structural parameters of the sdB component in PG 1336$-$018 obtained from asteroseismology based on the 3G models, with those derived independently from the modeling of the reflection/irradiation effect and the eclipses observed in the light curve.} 
{The stellar parameters inferred from asteroseismology using the 3G models are found to be remarkably consistent with both the preferred orbital solution obtained from the binary light curve modeling and the updated spectroscopic estimates for the surface gravity of the star. The seismology gives $M_*=0.471\pm0.006$ $M_\odot$, $R_*=0.1474\pm0.0009$ $R_\odot$, and $\log g=5.775\pm0.007$, while
orbital modeling leads to $M_*=0.466\pm0.006$ $M_\odot$, $R_*=0.15\pm0.01$ $R_\odot$, $\log g=5.77\pm0.06$, and spectroscopy yields $\log g=5.771\pm0.015$. In comparison, seismology from a former analysis based on the 2G models gave very similar results with $M_*=0.459\pm0.005$ $M_\odot$, $R_*=0.151\pm0.001$ $R_\odot$, and $\log g=5.739\pm0.002$. We also show that the uncertainties on the input physics included in stellar models have no noticeable impact, at the current level of accuracy, on the structural parameters derived by asteroseismology.} 
{The stellar models (both of second and third generation) presently used to carry out quantitative seismic analyses of sdB stars are reliable for the task. The stellar parameters inferred by this technique, at least for those that could be tested ($M_*$, $R$, and $\log g$), appear to be both very precise and accurate, as no significant systematic effect has been found.}  

\titlerunning{Third generation stellar models for asteroseismology of hot B subdwarf stars}
\authorrunning{V. Van Grootel et al.}

\maketitle

\section{Introduction}

Hot B subdwarf (sdB) stars are hot and compact objects with effective temperatures $T_{\rm eff}$ between 20 000 $-$ 40 000 K and surface gravities log $g$ in the range 5.0 $-$ 6.2 \citep[see, e.g.,][]{2008ASPC..392...75G}. They occupy the so-called Extreme Horizontal Branch (EHB), burning helium in the core and having a very thin residual hydrogen-rich envelope. The sdB stars dominate the population of faint blue objects down to $V \sim$ 16 and are mainly found in the thick galactic disk, but also in the halo, as globular clusters members and in the galactic bulge \citep{2004A&A...414..181A,2005ApJ...633L..29B}. Their presence is indirectly revealed in distant objects through their copious UV emission that is responsible for the UV-upturn phenomenon (an excess of UV emission) observed, for instance, in old elliptical galaxies \citep{1997ApJ...482..685}.

Major uncertainties remain about the details of the formation of the hot B subdwarfs. One important clue is the fact that the fraction of sdB stars in binaries with stellar companion is found to be relatively high, around $\sim$ 50\% \citep[see, e.g.,][]{1994AJ....107.1565A,2001MNRAS.326.1391M,2008ASPC..392...75G}. Plausible formation channels were modeled  in detail from binary population synthesis by \citet{2002MNRAS.336..449H,2003MNRAS.341..669H} and include binary evolution via a common envelope (CE), stable Roche lobe overflow (RLOF), and the merger of two helium white dwarfs, resulting in a single sdB star. The proposed scenario for forming single stars remains unsettled however, as they could also result from single star evolution through enhanced mass loss at the tip of the RGB at the moment of He-burning ignition \citep{1993ApJ...419..596D}, or through a delayed He-flash, at higher effective temperatures during the collapsing phase after having left the RGB \citep{1996ApJ...466..359D}. Another possibility is currently gaining observational support and involves the interaction of the star with close substellar companions (brown dwarfs and/or planets) in a common envelope evolution during the red giant phase. This scenario first proposed by \citet{1998AJ....116.1308S} is strongly suggested by the discovery of small, nearly Earth-sized planets around the isolated sdB star KPD 1943+4058 (KOI 55) that could be remnants of former giant planets that survived engulfment in the red giant envelope and contributed to expel the stellar envelope \citep{2011Natur.480..496C}. These distinct evolutionary scenarios leave a clear imprint on the binary fraction and distribution of sdB and their companion stars (CE evolution produces sdBs in very close binary systems preferably with white dwarfs companions, RLOF gives rise to longer period sdB + main-sequence star binaries), but also, most importantly, on the stellar mass distribution for sdB stars. After core-He exhaustion, the sdB stars evolve directly toward the white dwarf cooling sequence as {\it{AGB-Manqu\'e}} stars \citep{1993ApJ...419..596D}.
 
Asteroseismology should help clarify the question of the formation of sdB stars. The sdB stars indeed host two groups of nonradial pulsators. The first group, named the V361 Hya (originally EC 14026) stars \citep{1997MNRAS.285..640K}, exhibits short-period pulsations in the range 80 $-$ 600 s with amplitudes of a few milli-magnitudes. The pulsations correspond to low-order low-degree $p$-modes, which, in sdB stars, have significant amplitudes mainly in the outermost layers \citep{2000ApJS..131..223C}. The second group, discovered by \citet{2003ApJ...583L..31G} and referred to as V1093 Her stars, exhibits much longer periods from 45 min to a few hours. This corresponds to mid- and high-order low-degree $g$-modes, which can propagate in deep regions of the star, down to the convective He-burning core. Both $p$- and $g$-mode pulsations are driven by the same $\kappa$-mechanism powered by local envelope accumulations of heavy elements such as iron due to radiative levitation (\citealt{1996ApJ...471L.103C,1997ApJ...483L.123C}; \citealt{2003ApJ...597..518F}; see also \citealt{2011MNRAS.418..195H}). Quantitative asteroseismology of pulsating sdB stars has developed considerably during the past ten years, giving us new ways of revealing the properties of these objects at unprecedented level of accuracy. This includes, among many other structural parameters, tight constraints on their masses. The results obtained so far are compiled in Table 1 of \citet{2012A&A...539A..12F}, with the appropriate reference to each analysis having been carried out.

The method has long relied on stellar envelope models of the so-called second generation (2G) that incorporate the nonuniform abundance profiles of iron predicted by equilibrium between radiative levitation and gravitational settling (see \citealt{2002ApJS..139..487C} for details). These models are suitable for asteroseismology of pure $p$-mode pulsators whose oscillation modes are sensitive only to the details of the outermost regions of the star. A very important test of accuracy for the seismic results obtained on the basis of these models was made using the star PG 1336--018 \citep{2008A&A...489..377C}. This object is one of the only two known sdB + dwarf M close eclipsing binaries where the sdB component is a pulsating sdB star (the other one being 2M 1938+4603 recently discovered with {\it Kepler}; \citealt{2010MNRAS.408L..51O}). PG 1336--018 permitted to compare results obtained from the two independent techniques of asteroseismology and orbital light curve and eclipses modeling. It was found that asteroseismology is remarkably consistent with one of the orbital solutions uncovered by \citet{2007A&A...471..605V}. This close agreement showed that the seismic inference is both precise and accurate (i.e., free of significant systematics), at least for the stellar parameters that could be tested. This was of course an important finding attesting that the 2G models used for the quantitative asteroseismic studies of V361 Hya stars are sufficiently reliable for the task.  

Recently, the advent of the space missions CoRoT and {\it Kepler} opened up new opportunities to gather sufficiently high quality data for the very low-amplitude, long-period $g$-mode pulsations seen in sdB stars for asteroseismological purposes. Three $g$-mode sdB pulsators have indeed already been analyzed on this basis, bringing for the first time information on their He-burning cores -- including size and composition -- along with the determination of other structural parameters \citep{2010ApJ...718L..97V,2010A&A...524A..63V,2011A&A...530A...3C}. Because the 2G envelope models are no longer suitable for quantitative analyses of $g$-mode pulsators, a third generation (3G) of stellar models for accurate evaluations of the $g$-mode pulsation periods had to be developed for that purpose and were used in the studies cited above. These are complete stellar structures (as opposed to the 2G envelope models) that include a detailed description of the central regions, as well as, like their 2G counterparts, the nonuniform abundance profile of iron in the envelope predicted by microscopic diffusion. 

While the 2G models passed the test of reliability for asteroseismology of $p$-mode sdB pulsators with the case of PG 1336--018, the new 3G models still have to be verified in a similar way. In the following section (Section 2), we discuss the need for parameterized models in quantitative asteroseismology, and we describe the new 3G models that we use for analyzing pulsation data obtained on hot B subdwarfs. In Section 3, we reanalyze the available data on PG 1336--018 with these 3G models in order to confront the seismic inferences obtained on the fundamental parameters of this star with those resulting from 1) the former seismic analysis based on 2G models, and 2) the independent orbital light curve analysis of \citet{2007A&A...471..605V}. This test is important for linking older asteroseismic results obtained with 2G models with newer results based on 3G models in order to estimate potential systematics. We also carry out experiments in order to assess the importance of known model uncertainties on the stellar parameters determined by asteroseismology. A summary and conclusion are finally provided in Section 4.

\section{Third generation of sdB models for seismology}

\subsection{Background}

It may be worthwhile here to give a brief historical account of how our group came to the development of successive generations of models for the specific purpose of applying asteroseismological techniques to sdB stars. We refer to as first generation (1G) models those stellar equilibrium structures for sdB stars with a uniform metallicity. They include full evolutionary models such as those provided by \citet{1993ApJ...419..596D} at a time when no pulsating sdB stars were known, and when one of us (S.C.) had started a Ph.D. thesis to investigate the asteroseismological potential of these stars \citep{1999PhDT........26C}. They include as well static structures made of an envelope (calibrated in part on the basis of the evolutionary models of Dorman) sitting on top of an inert ball. Those envelope models allowed quite a bit of flexibility, in particular for investigating the effects of changing the metallicity on the overall stability of pulsation modes. It was thus found that uniform metallicity models (both evolutionary and static ones) are generally stable against pulsations, unless the metallicity is increased to some unrealistically high values \citep{1996ApJ...471L.103C}.

These investigations with 1G models identified a region of local instability associated with an opacity feature (the so-called Z-bump) even though the modes were globally stable. Given that sdB stars are
very hot objects, and given that they are known to be chemically peculiar, radiative levitation was the obvious new ingredient to incorporate into a next generation of equilibrium structures. The idea
was to verify if enough iron could accumulate in the driving region through radiative levitation and push some pulsation modes toward instability. Hence, we developed our second generation (2G) models,
which are again static envelope structures, but which incorporate a nonuniform distribution of iron (a representative Z-bump element) as determined by microscopic diffusion. In detail, the iron profiles are obtained through detailed calculations of radiative forces under the assumption of an equilibrium between radiative levitation and gravitational settling (resulting from the combined action of the pressure gradient and electric fields). This is called hereafter "diffusive equilibrium", disregarding, at this stage, other competing processes (see Section \ref{exp1}). The computations of the radiative accelerations themselves and of the equilibrium abundances of Fe are done following the procedure described in \citet{1995ApJS...99..189C}. These 2G models have been extremely successful at explaining the location of the $p$-mode sdB pulsators in the log $g$-$T_{\rm eff}$ plane as well as their nonadiabatic properties, and they have remarkably well resisted the test of time (\citealt{1997ApJ...483L.123C,2001PASP..113..775C,2006A&A...459..565C}; \citealt{2012ASPC..452..241R}). The building assumption that the local abundance of Fe in a sdB star could be estimated by equating the effects of radiative levitation with those of gravitational settling was ultimately confirmed through detailed time-dependent calculations \citep[see, e.g.,][]{2006MmSAI..77...49F,2008CoAst.157..168C}. 

Our driving motivation in the early years of sdB seismology was to explain the very existence of pulsators and account for the observed ranges of excited periods in these stars. Hence, ``simple'' envelope
structures (such as the 2G models) incorporating the appropriate details of the driving/damping zone proved to be perfectly suitable for investigating the nonadiabatic properties of pulsating sdB stars. At the same time, \citet{1999PhDT........26C} carried out a very detailed analysis with the evolutionary models of \citet{1993ApJ...419..596D}, and came to the conclusion that simple envelope models (obtained by cutting off the central zone of Dorman's evolutionary models or by building them with the same constitutive physics) would also be quite useful for quantitative adiabatic asteroseismology as long as only $p$-modes are involved \citep{2002ApJS..139..487C}. In contrast, systematic effects (the periods are increasingly overestimated with increasing radial order, reaching up to differences of 10$-$20\%) appear when dealing with $g$-modes, which is not surprising given that the central region of a sdB model carries a significant weight in terms of period formation for those modes contrary to $p$-modes. The use of the 2G models thus allowed us to combine both adiabatic and nonadiabatic seismology for quantitatively investigating the properties of the $p$-mode pulsators. Our initial effort in that direction was the study of the star PG 0014+067 which led to the first successful seismic inferences on the global structural parameters of a sdB pulsator, including its total mass \citep{2001ApJ...563.1013B}. This was before the discovery of the cooler and lower gravity long-period $g$-mode pulsators in 2003. The need for complete stellar models (incorporating the important central region for $g$-modes) became apparent when sufficiently high quality data  on $g$-mode pulsators became available from space missions. 

\subsection{Description}

The 2G models are envelope structures built under the assumption that the luminosity is constant throughout the entire envelope. This, it turns out, is an excellent approximation for He-core burning stars on the EHB, as can be seen in Fig. 2 of \citet{2000ApJS..131..223C}, for example. As a consequence, only three equations remain to define the structural properties of such an envelope, and the integration is carried out from the surface inward down, typically, to a fractional mass depth $\log q \equiv \log (1-M(r)/M_*) = -0.1$. There are four primary parameters to specify a 2G model: the effective temperature $T_{\rm eff}$, the surface gravity log $g$, the total mass of the star $M_*$, and the mass contained in the H-rich envelope, usually expressed as a logarithmic fractional mass, $\log (M_{\rm env}/M_*$). The first two ($T_{\rm eff}$ and log $g$) can be directly related to observable quantities obtained by spectroscopy.

\begin{figure}[!ht]
\begin{center}
\begin{tabular}{lll}
\includegraphics[scale=0.45,angle=0]{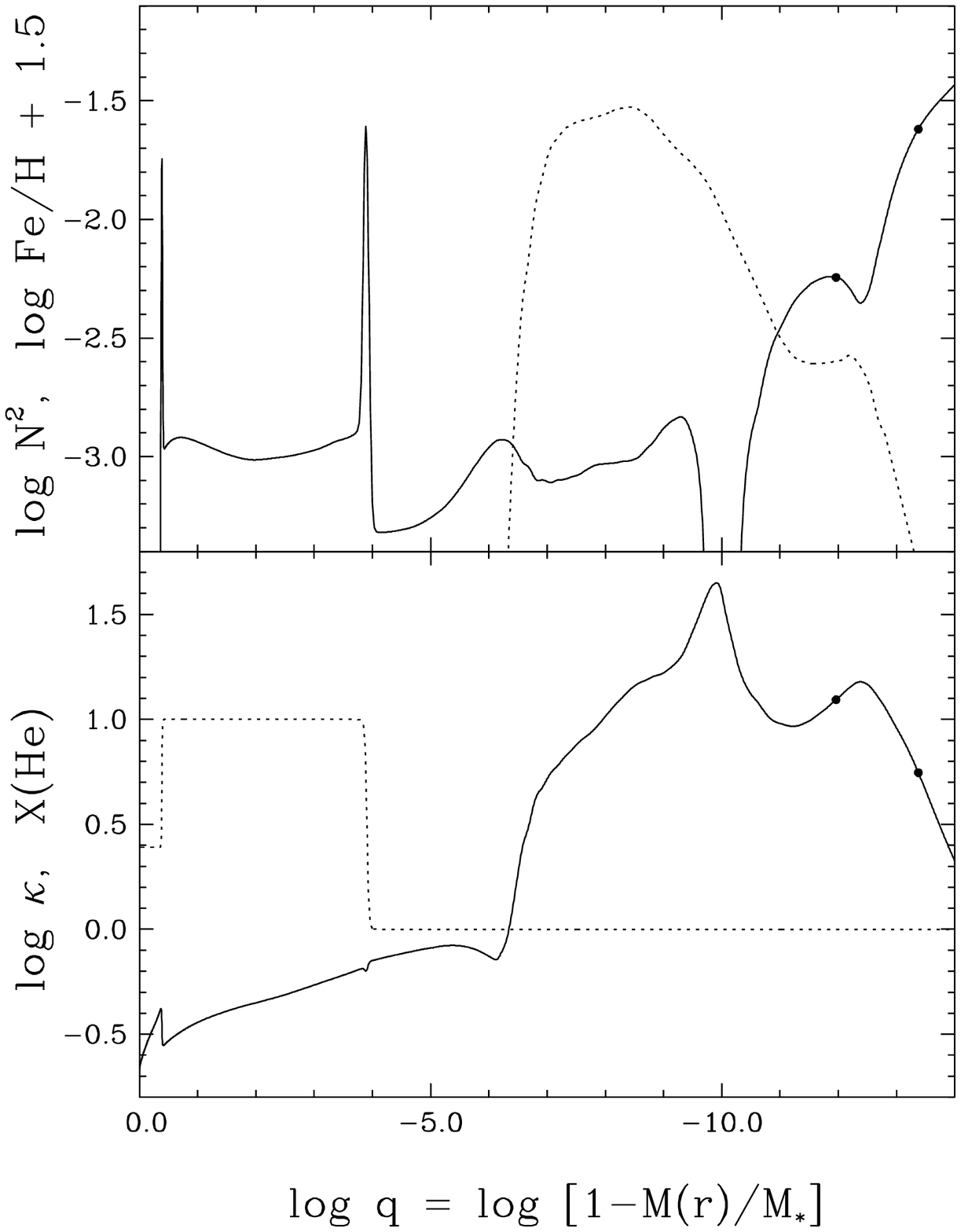}
\end{tabular}
\end{center}
\caption{\label{f0}Some structural properties for a representative 3G model. {\it Upper panel:} run of the logarithm of the square of the Brunt-V\"ais\"al\"a frequency as a function of fractional mass depth (solid curve) and equilibrium abundance profile of Fe (dotted curve, shifted upward by 1.5 dex). The solar ratio is log Fe/H = $-$4.5. ($-$3.0 on the shifted scale). {\it Lower panel:} run of the logarithm of the opacity as a function of depth (solid curve), and profile of the He distribution (dotted curve). In both panels the locations of the atmospheric layers, defined as those with an optical depth between log $\tau$ = 1.0 and log $\tau$ = $-$1.0 are indicated by dots.}  
\end{figure}

In the case of the 3G models, these are now complete stellar structures assumed to be in strict thermal and mechanical equilibrium. Thus, the luminosity of such a static model is exactly balanced by core He burning (and the very small contribution of H burning at the base of the H-rich envelope). The four usual equations of 1D stellar structure (minus the time-dependent term) are simultaneously solved with the help of a relaxation scheme. The physical conditions at the outer boundary are handled through a gray atmosphere scheme. A thin convection zone is usually created around the opacity peak in the envelope of a sdB model, but the flux carried by convection across that thin zone is quite small, and well below the base of the atmosphere (Fig. \ref{f0}). When present, convection is handled through one or another of the various versions of the mixing-length theory. In the present application, convection is modeled in terms of the standard ML1 version of \citet{1958ZA.....46..108B}.

The constitutive physics incorporates the nuclear reaction rates of \citet{1988ADNDT..40..283C} for H and, particularly, He burning. Electron screening is handled through the prescription of \citet{1973ApJ...181..457G}. Since it was developed to model white dwarf stars as well, it also incorporates neutrino production rates as described by \citet{1996ApJS..102..411I}, and references therein, and conductive opacities, also made available by \citet{2008ApJ...677..495I}, and references therein. The radiative opacities are OPAL data of 1995, including additional tables for iron in a hydrogen background especially computed for our needs by Forrest Rogers in 1996 and briefly described in \cite{1997ApJ...483L.123C}. These special tables are essential for properly mapping the opacity profile of Fe associated with radiative levitation. As indicated above, the nonuniform distribution of Fe with depth is obtained under the assumption of a perfect equilibrium between radiative levitation and gravitational settling following the procedure described in \citet{1995ApJS...99..189C}. The equation of state (EOS) is an in-house construction covering a large domain of the temperature-density plane, sufficiently large to cover all types of stars, including white dwarfs. Three regimes are involved. For the low-density regions, a network of Saha equations is solved for a mixture of radiation and an almost ideal (including Coulomb corrections), nondegenerate, partially ionized gas composed of a mixture of H, He, C, and O in various proportions as needed. In the region of partial ionization where nonideal and degeneracy effects are important, we use the EOS of \citet{1995ApJS...99..713S} for H and He, an improved version of the EOS of \citet{1977ApJS...35..293F} for C, and an unpublished EOS
for O developed at Universit\'e de Montr\'eal in 1998. The third regime covers the high-density domain corresponding to the fully ionized plasma in the liquid and, ultimately, in the solid phase according to the basic physics described by \citet{1974PhDT........56L} and improved by \citet{2005ASPC..334...65K}. Efforts were made to smoothly bridge the different EOS surfaces across the boundaries of the different regimes. This is of particular importance for pulsation studies. Let us note, however, that the third regime covering the high-density domain is not relevant here for sdB stars. Interpolation in composition is handled following the additive volume prescription of \citet{1977ApJS...35..293F}. 

The primary parameters to define a 3G model are: the total mass of the star $M_*$, the mass contained in the outer H-rich envelope $\log (M_{\rm env}/M_*$), the mass contained in the (usually convective) core $\log (M_{\rm core}/M_*$), and the chemical composition in the core (with the constraint $X$(He) + $X$(C) + $X$(O) + $Z =$ 1.0). The stratified overall structure consists of a core, surrounded by a He-rich mantle, itself surrounded by a H-rich envelope in which Fe radiatively levitates under the assumption of diffusive equilibrium. The composition transition layers at the interface between the H-rich envelope and the He mantle, and those at the interface between the He-rich mantle and the core are modeled in terms of two additional parameters which become of relevance in presence of diffusion. In the present application, the composition transition zones have been calibrated on the basis of full evolutionary models such as those of \citet{1993ApJ...419..596D}, which do not take diffusion or possible sources of turbulence into account. These extra parameters are therefore held fixed in the present experiment (see, however, Section \ref{exp2}). Likewise, the value of $Z$ is fixed at 0. 

As an illustrative example, Fig. \ref{f0} depicts some of the structural properties of the 3G model identified below as the model that best accounts for the available seismic data on the pulsating sdB component of the PG 1336$-$018 system. The relatively sharp peaks which characterize the profile of the Brunt-V\"ais\"al\"a frequency near log $q$ $\simeq$ $-$0.35 and log $q$ $\simeq$ $-$3.8 are clearly associated with the abrupt change of chemical composition in the composition transition zones. The signature of these transition zones is also evident on the opacity profile, although the magnitude of that signature is much less important in that case. The model features a convective core and a much less significant convection zone in the outer envelope, in the sense that the latter carries at the most only $\sim$0.6\% of the flux. This very weak convection zone is clearly associated with the maximum of the opacity profile, but does not change in any significant way the basic character of the driving mechanism, which remains essentially a pure $\kappa$-mechanism. It is the location and shape of the Fe opacity peak that provide the conditions for exciting $p$-mode pulsations in the model. Again, radiative levitation is the key element for boosting the abundance of Fe in the driving region, which is necessary for destabilizing pulsation modes. The opacity profile in the envelope not only affects the driving/damping processes, but also the adiabatic properties (particularly the pulsation periods for $p$-modes) through its direct effects on the mechanical structure of the model. Finally, this 3G model is a complete model and, therefore, provides a reliable equilibrium structure for deriving not only $p$-mode periods, but also $g$-mode periods. 

\subsection{The need for parameterized models}
When we started investigating the application of the forward asteroseismological method to pulsating hot subdwarf and white dwarf stars a decade ago (see, e.g., \citealt{2001ApJ...563.1013B} or
\citealt{2002ApJ...581L..33F}), we realized at the outset that the use of full evolutionary models would not be practical. This is because it is nearly impossible, even with large computer clusters, to cover 
finely all of the relevant domains of parameter space with evolutionary sequences; it would take too much computing time. The practical consequence is that evolutionary sequences, necessarily limited in
number, may actually miss the correct region of parameter space where resides the best seismic model.

As a fallback position, we turned to less realistic, static models of the kind described here. Those allow maximum flexibility for very exhaustive searches in parameter space and the eventual identification of an optimal seismic model as the case may be. Of course, such a seismic model must ultimately be validated by detailed evolutionary calculations. The parameterized model approach can also be useful for
identifying readily the dependences of the pulsation periods on specific model parameters. They can also help identifying rather directly the parts of the input constitutive physics that may need improvement.

The parameterized static 3G models that we discuss in this paper are still crude models of hot subdwarf B stars. We believe, however, that they are defined in terms of the most sensitive parameters from a
seismic point of view. It is important that they be tested as thoroughly as possible. \citet{2008ASPC..392..261B} presented the results of a first test connecting the seismic model of PG 0014+067 derived from 2G structures as described by \citet{2001ApJ...563.1013B} with the new seismic solution based on the first 3G models that were being developed for that star. Interestingly and encouragingly, both solutions turned out to be essentially the same. A still more stringent test is described in the next section.  

\section{Seismic test of accuracy for the new models}

\subsection{PG 1336--018: A Rosetta stone for asteroseismology}

The system PG 1336$-$018 (NY Virginis) is one of the very few known sdB + dM close eclipsing binaries, and one of the only two known objects of this type where the sdB component is a pulsating star
\citep{1998MNRAS.296..329K}. This makes PG 1336$-$018 the equivalent of a Rosetta stone permitting independent techniques used to derive some of the stellar parameters (namely, orbital modeling, asteroseismology, and spectroscopy) to be directly confronted. One can therefore check the reliability of the methods and models, as well as the presence (or lack thereof) of biases in the derived parameters. This rare opportunity provides extremely important information on the accuracy actually achieved, as generally only the internal precision of the measurements given by the error estimates can be obtained.  

Parameters for the two components of PG 1336--018 have been determined by combining multicolor ULTRACAM/VLT light curves and the radial velocity (RV) curve obtained from UVES/VLT spectra \citep{2007A&A...471..605V}. These authors modeled the reflection/irradiation effect and eclipses seen in the light curves and found three solutions -- each deriving among other parameters a mass, a radius, and a surface gravity for the sdB star -- of equal statistical significance, due to the large parameter space and correlations between some parameters (see Table~\ref{tabcomp} for a summary of the relevant values). Independently, the first asteroseismic analysis of the sdB component carried out by \citet{2008A&A...489..377C} was based on the pulsation spectrum provided by \citet{2003MNRAS.345..834K}. These authors identified 28 periodicities in the 96 $-$ 205 s range by analyzing the multisite Whole Earth Telescope (WET) campaign held in April 1999, which constitutes the most extensive photometric monitoring available to date for this star. The seismic analysis was based on the 2G envelope models and yielded a solution with parameters falling very close to one of the solution of \citet{2007A&A...471..605V}. The latter derives for the sdB component a total stellar mass of $M_* =$ 0.466 $\pm$ 0.006 $M_{\odot}$ and a radius of $R_* =$ 0.15 $\pm$ 0.01 $R_{\odot}$, while the seismic analysis of the observed $p$-modes led to a total stellar mass of $M_* =$ 0.459 $\pm$ 0.005 $M_{\odot}$ and a radius of $R_* =$ 0.151 $\pm$ 0.001 $R_{\odot}$. This close agreement is within the $1\sigma$ error estimates of each technique.

In the following, we present a new asteroseismic analysis of the sdB component of PG 1336--018 based on the same set of observed frequencies, but using, this time, the 3G complete models previously described. We also take this opportunity to update the spectroscopic estimates of the atmospheric parameters of PG 1336--018 reported in \citet{2008A&A...489..377C} on the basis of better spectra and of new NLTE model atmospheres including metals (subsection \ref{spectroscopy}; see also Table~\ref{tab}). Our goal is to provide a thorough comparison of the various results and test the validity of the 3G models for asteroseismology of sdB stars.

\subsection{Improved spectroscopy for PG 1336--018}
\label{spectroscopy}
\begin{table*}[!ht]
\caption{\label{tab} Summary of all available atmospheric parameters of PG 1336$-$018 derived from spectroscopy.}
\begin{center}\begin{tabular}{llcccc}
\hline\hline
Spectrum & Type of model & $T_{\rm eff}$ (K) & log $g$ & log $N$(He)/$N$(H) 
& Reference \tabularnewline
\hline
SAAO1 & LTE, H & 33,139 $\pm$ 1000 & 5.78 $\pm$ 0.10 & ... & Kilkenny et
al. (1998) \tabularnewline
SAAO2 & LTE, H & 32,895 $\pm$ 1000 & 5.67 $\pm$ 0.10 & ... & Kilkenny et
al. (1998) \tabularnewline
VLT/UVES & LTE, H, He & 31,300 $\pm$ 250 & 5.60 $\pm$ 0.05 & $-$2.93
$\pm$ 0.05 & Vuckovi\'c et al. (2007) \tabularnewline
PB6 & NLTE, H, He & 33,220 $\pm$ 170 & 5.75 $\pm$ 0.04 & $-$3.11 $\pm$ 0.25 &
Charpinet et al. (2008) \tabularnewline
BG9 & NLTE, H, He & 32,380 $\pm$ 150 & 5.77 $\pm$ 0.03 & $-$2.87 $\pm$ 0.17 &
Charpinet et al. (2008) \tabularnewline
Bok6 & NLTE, H, He & 32,874 $\pm$ 148 & 5.705 $\pm$ 0.028 & $-$3.066
$\pm$ 0.250  & This work \tabularnewline
Bok6 & NLTE, H, He, metals & 33,285 $\pm$ 169 & 5.707 $\pm$ 0.028 & $-$3.016
$\pm$ 0.216  & This work \tabularnewline
Bok9 & LTE, H, He & 31,959 $\pm$ 71 & 5.823 $\pm$ 0.018 & $-$2.847
$\pm$ 0.101  & This work \tabularnewline
Bok9 & NLTE, H, He & 32,359 $\pm$ 84 & 5.793 $\pm$ 0.017 & $-$2.879
$\pm$ 0.095  & This work \tabularnewline
Bok9 & NLTE, H, He, metals & 32,649 $\pm$ 97 & 5.794 $\pm$ 0.017 & $-$2.892
$\pm$ 0.098  & This work \tabularnewline
Bok9H & NLTE, H, He, metals & 32,608 $\pm$ 104 & 5.798 $\pm$ 0.018 & $-$2.839
$\pm$ 0.098  & This work \tabularnewline
Bok9L & NLTE, H, He, metals & 32,748 $\pm$ 111 & 5.800 $\pm$ 0.020 & $-$2.893
$\pm$ 0.113  & This work \tabularnewline
\hline
&\\
Former adopted value & NLTE,H,He  & 32,780 $\pm$ 200 & 5.76 $\pm$ 0.03 & $-2.94$
$\pm$ 0.14  &  Charpinet et al. (2008) \\ 
New adopted value & NLTE,H,He,metals & 32,807 $\pm$ 82 & 5.771 $\pm$ 0.015 & $-$2.918
$\pm$ 0.089  & This work \\
&\\
\hline
\end{tabular}\end{center}
\end{table*}

There are two reasons why we were prompted to reinvestigate this issue. First, in view of the high importance of PG 1336$-$018 as a testbed for stellar modeling and asteroseismology, we kept obtaining 
spectra over the orbital phase and ended up with a combined zero-velocity optical spectrum reaching a very high value of S/N $\sim$ 491 around 4000 {\AA} (compared to $\sim 175$ in \citealt{2008A&A...489..377C}). We refer to that spectrum as Bok9 in what follows, reflecting the fact that it was gathered at the 2.3-m Bok Telescope of the Steward Observatory Kitt Peak Station, and that it has a resolution of 8.7 {\AA} at 4000 {\AA}\footnote{The spectrum named BG9 in \citet{2008A&A...489..377C} represents the sum of our first few spectra and is, therefore, part of Bok9.}. It is the sum of 14 individual spectra, 6 obtained when the reflection effect in the system was near minimum, and 8 taken at other orbital phases. We also combined the six former spectra into a single one named Bok9L (as in "low" reflection effect; S/N $\sim$ 344 in the blue), and the eight others to obtain Bok9H (S/N $\sim$ 350 in the blue) in order to investigate the effects, if any, of the irradiation of the secondary on the derived atmospheric parameters. 

The second reason is that we have now developed the capacity to build NLTE model atmospheres and synthetic spectra with arbitrary heavy element abundances for a wide range of H and He compositions within reasonable computation times. It is well known that hot B subdwarfs are all chemically peculiar stars and that their atmospheric compositions are nonstandard and vary from star to star. The effects of metals on the determination of the atmospheric parameters of sdB stars have been studied in the past \citep[e.g.,][]{2000A&A...363..198H}, but this has been done only within the LTE approximation. In the present case, we use two of our recent NLTE model grids to investigate potential effects in the context of PG 1336$-$018. 

The first grid is similar to the NLTE, H and He model spectra used in \citet{2008A&A...489..377C}, except that our new synthetic spectra extend out to the red to include the H$\alpha$ region (not previously considered). The second NLTE grid includes a representative heavy element mixture inspired from the work of \citet{2008ApJ...678.1329B}. These authors presented a NLTE analysis of {\it FUSE} spectra to derive the abundances of several astrophysically important elements in the atmospheres of five typical long-period pulsating sdB stars. The five stars analyzed by \citet{2008ApJ...678.1329B} show very similar abundance patterns, and it is from these results that we derived a representative composition using the most abundant heavy elements. Hence, we assumed atmospheres containing C (1/10 solar), N (solar), O (1/10 solar), Si (1/10 solar), S (solar), and Fe (solar). These two 3D grids ($T_{\rm eff}$, log $g$, and log $N$(He)/$N$(H)) were computed with the public codes TLUSTY and SYNSPEC 
\citep[][]{1995ApJ...439..875H,1995ApJ...439..905L,1997ApJ...485..843L}. These codes have been parallelized by one of us (P.B.) to efficiently run on a computer cluster. With these two grids of synthetic spectra, as well as a corresponding LTE grid for comparison purposes, we analyzed the spectra Bok9, Bok9L, and Bok9H. We also reanalyzed the blue spectrum PB6, renamed Bok6 from then on, as it has a higher resolution ($\sim 6$ {\AA}), but a somewhat lower S/N of $\sim 80$.

\begin{figure*}[!ht]
\begin{center}
\begin{tabular}{lll}
\includegraphics[scale=0.5,angle=270]{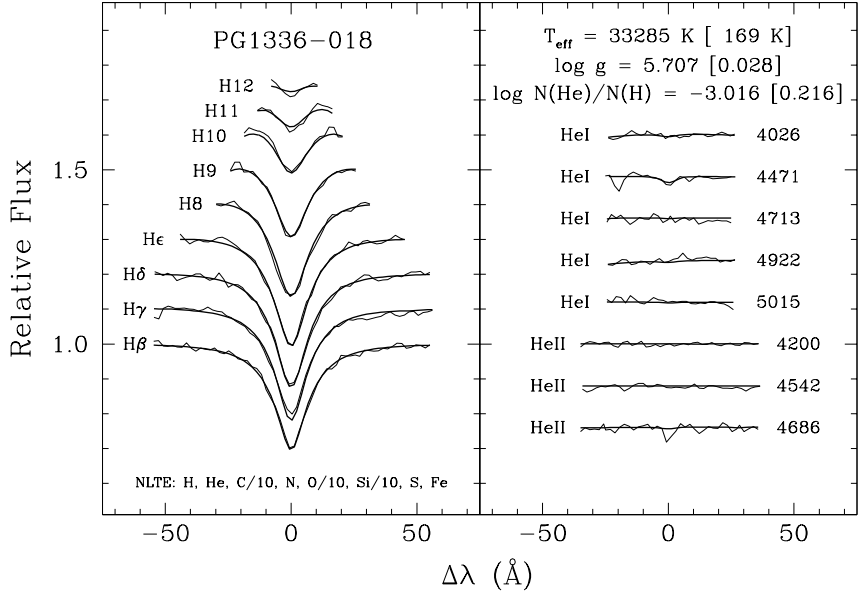}
\end{tabular}
\end{center}
\caption{\label{f1}Model fits (thick curves) to the hydrogen and helium lines (thin curves) available in our time-averaged, mid-resolution spectrum Bok6.} 
\end{figure*}

\begin{figure*}[!ht]
\begin{center}
\begin{tabular}{lll}
\includegraphics[scale=0.5,angle=270]{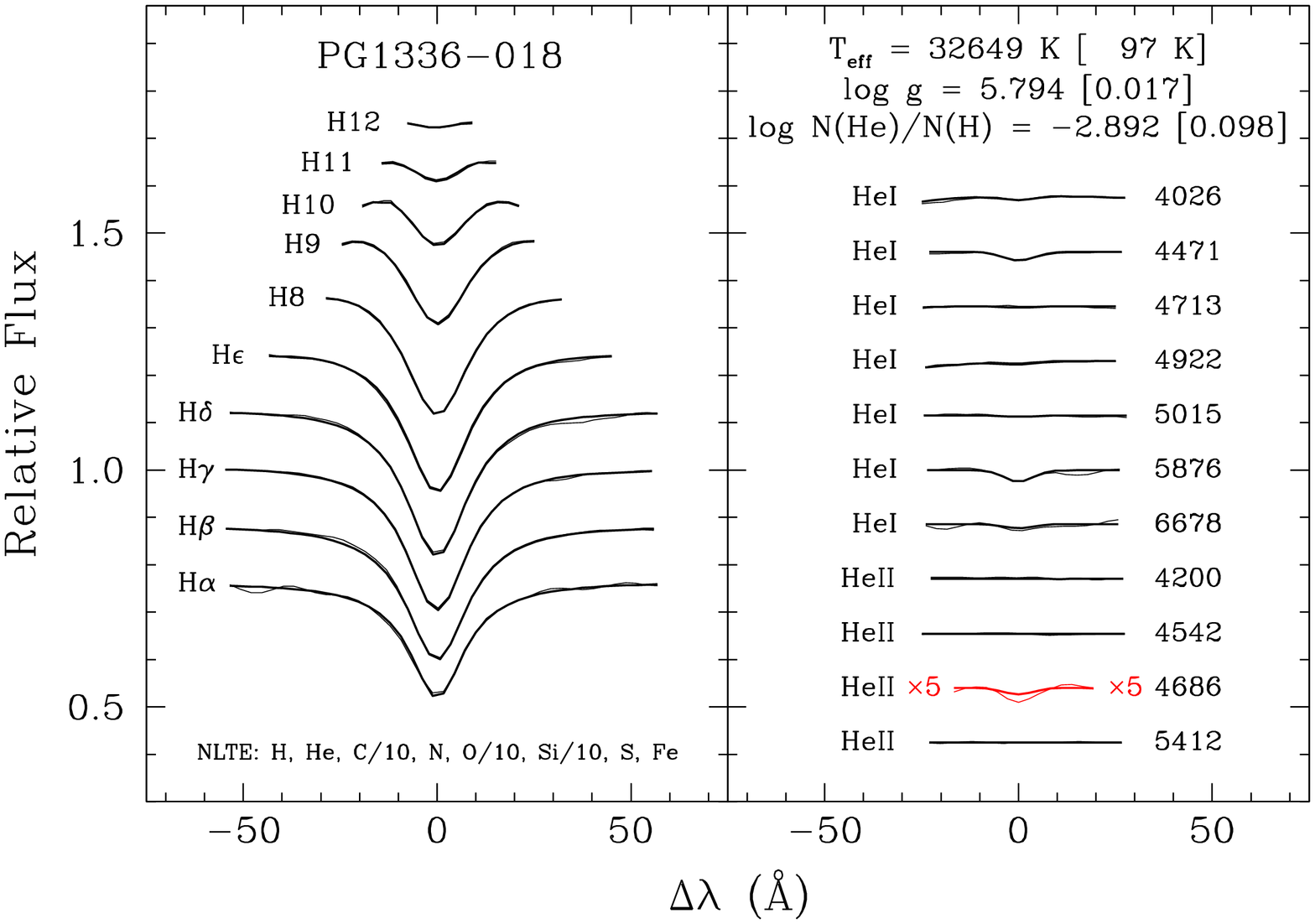}
\end{tabular}
\end{center}
\caption{\label{f2}Model fits (thick curves) to the hydrogen and helium lines (thin curves) available in our time-averaged, very high signal-to-noise, low-resolution spectrum Bok9.}  
\end{figure*}

We summarize, in Table \ref{tab}, the results of several analyses of the Bok6 and Bok9 spectra, including a test with a metal-free LTE grid for comparison purposes. For practical use (especially in the seismic investigation that follows), the solutions obtained with the NLTE grid including metals are to be preferred. Figures~\ref{f1} and \ref{f2} show the spectral fit obtained with that grid for the Bok6 and Bok9  spectra, respectively. Given the exceptional S/N ratio of that latter spectrum, Fig. \ref{f2} reveals an extraordinary good match, notwithstanding the He II 4686  {\AA} line (although we had to increase the vertical scale by a factor of 5 in that case or, otherwise, the differences would not have been seen in the plot). Except for the redder regions where the S/N has significantly decreased (specifically,
near the H$\alpha$ and He I 6678  {\AA} lines), the other weak features seen in the observed spectrum are real. For instance, the weak structures in the red wings of  H$\gamma$,  H$\delta$, and  H$\epsilon$ are due to metallic absorption of unknown origin. Likewise, the absorption structure seen in the red wing of He I 5876 {\AA} is also real and is due, in that case, to weak interstellar reddening associated with the well known Na I doublet. For our purposes, we will adopt the weighted means of the atmospheric parameters as inferred in Fig. \ref{f1} and \ref{f2}. Hence our updated estimates of these parameters for the sdB component of PG 1336$-$018 are $T_{\rm eff}$ = 32 807 $\pm$ 82 K, log $g$ = 5.771 $\pm$ 0.015 (in cgs units, as elsewhere in the text and in the tables), and log $N$(He)/$N$(H) = $-$2.918 $\pm$ 0.089. These estimates are very close to the former values adopted in \citet{2008A&A...489..377C}, but are more precise.

Finally, we note, from Table~\ref{tab}, that the solutions obtained with Bok9H and Bok9L are, well within the uncertainties, the same as those obtained with the full spectrum using the same grid of models. This shows that the inferred atmospheric parameters do not depend in any significant way on the orbital phase at which the spectra are taken, i.e., the results are not affected by the irradiation and reflection effects on the secondary. Moreover, the NLTE with metals vs NLTE without metals comparison for both the Bok6 and Bok9 spectra indicates that the presence of metals (at least at the level that we considered) does not strongly change the values of the inferred atmospheric parameters. We find only a small systematic effect in effective temperature (the inclusion of metals slightly increases it), and negligible effects for the two other quantities. This shows that our spectroscopic determination of the atmospheric parameters of PG 1336--018 is very robust.

\subsection{Parameters from asteroseismology using 3G models}
\label{analysis}
The procedure to derive the structural parameters of the pulsating sdB star in PG1336--018 is the same as the one described in \citet{2008A&A...489..377C}. It is a double optimization scheme aimed at isolating the model(s) that best fit the pulsation periods identified in this star. This is done by minimizing a $\chi^2$-type merit function of general form
\begin{equation}
\label{eqs2}
S^2=\sum_{i=1}^{N_{\rm obs}}\Big(\frac{P_{\rm obs}^{i}-P_{\rm
th}^{i}}{\sigma_{i}}\Big)^2
\end{equation}
where $N_{\rm obs}$ is the number of observed periods and $\sigma_{i}$ a weight that can be associated to each pair of observed/computed periods $\{P^i_{\rm obs},P^i_{\rm th}\}$ for a given model. The minimization is done both at the level of matching the periods and in model parameter space. We refer the reader to \citet{2008A&A...489..377C} for further details. The main difference resides in the fact that the stellar structures are now calculated using 3G models while in the previous study 2G envelope structures have been used for the analysis. This has two consequences. First, the natural parameters needed to specify a 3G model (and therefore the parameter space to explore) are different. As indicated above, these parameters are the total stellar mass $M_*$, the fractional mass of the outer hydrogen-rich envelope $\log (M_{\rm env}/M_*)$ also referred to as $\log q(H)$, the fractional mass of the convective core $\log (M_{\rm core}/M_*)$ or $\log q({\rm core})$, and the chemical composition in the core (under the constraint $X({\rm He})+X({\rm C})+X({\rm O}) =1$). Second, the external constraints provided by spectroscopy ($T_{\rm eff}$ and $\log g$) can no longer be used to define a priori the relevant range for the explored parameter space. This is because $T_{\rm eff}$ and $\log g$ are no longer natural parameters of the 3G stellar models, contrary to 2G envelope structures. Instead, these quantities now depend on the 4 parameters mentioned above and their values are known only a posteriori, after a model is converged. Since spectroscopic constraints are essential to guide the search for a meaningful asteroseismic solution in the vast parameter space and avoid the multiplication of inconsistent solutions, the adopted solution is to incorporate these constraints within the optimization procedure itself, by eliminating de facto during the search the models in parameter space that differ too much in $T_{\rm eff}$ and/or $\log g$ from the spectroscopic values (by applying a correction factor to the $S^2$ value). In what follows we used the updated atmospheric values adopted in the previous section with a tolerance of $3\sigma$ (see Table~\ref{tab}). This approach ensures by construction a fair consistency with spectroscopy but of course there is no guarantee, a priori, that a good period fit exists within these constraints.

\begin{figure*}[!ht]
\begin{center}
\includegraphics[scale=0.5,angle=0]{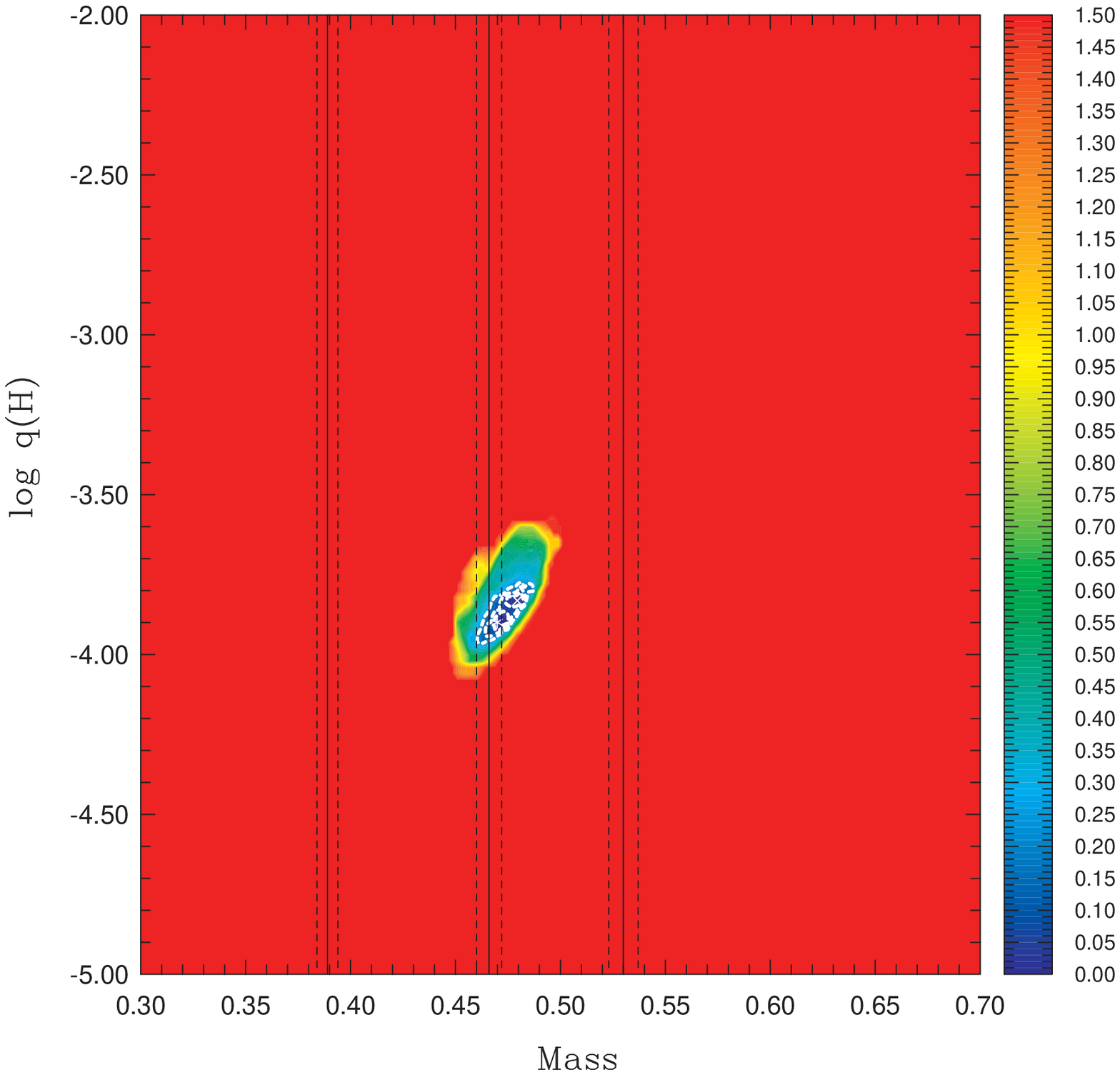}
$\;$
\includegraphics[scale=0.5,angle=0]{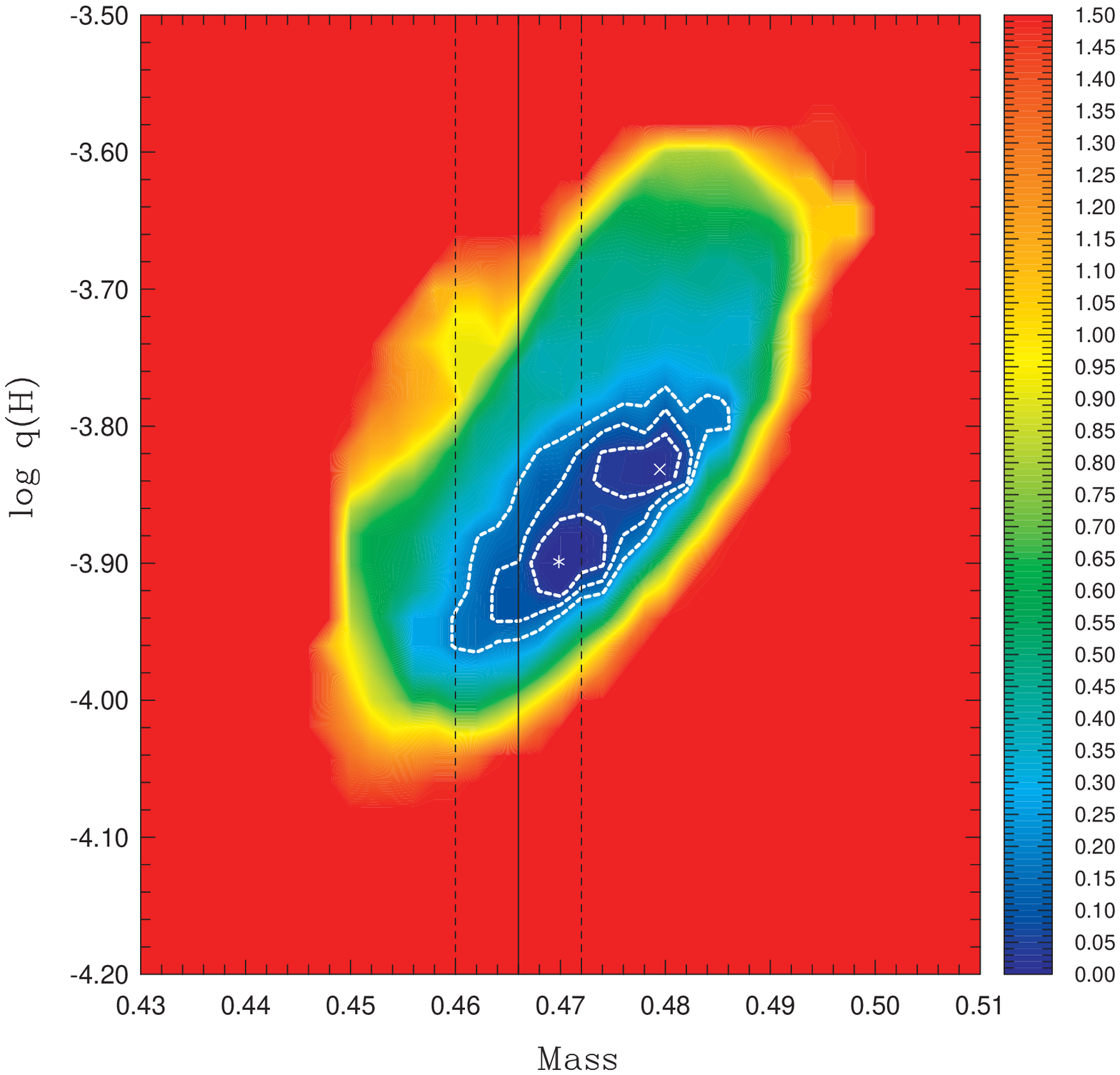}
\end{center}
\caption{\label{f5}\textit{Left panel:} map of the projected merit function (on a log scale) along the $M_*$- log $q$(H) plane. \textit{Right panel:} zoom of the left panel in the region of interest. At each $M_*$, log $q$(H) position, the value given is the projected merit function, i.e., the minimum of log $S^2$ that can be found among the values obtained for all log $q$(core) and $X_{\rm core}$(C+O). The best-fit model is indicated by an asterisk on the figures, while the model corresponding to the "second" best-fit model is indicated by a cross. White contours show regions where the frequency fits have $S^2$ values within, respectively, the 1$\sigma$, 2$\sigma$, and 3$\sigma$ confidence levels relative to the best-fit solution.}  
\end{figure*}

The search for best-fit solutions was launched in the largest possible parameter space relevant for sdB stars:  0.30 $\leq$ $M_*/M_{\odot}$ $\leq$ 0.70, $-$5.0 $\leq$ $\log q$(H) $\equiv$ $\log (M_{\rm env}/M_*)$ $\leq$ $-$2.0, $-$0.40 $\leq$ $\log q$(core) $\equiv$ $\log (1-M_{\rm core}/M_*)$ $\leq$ $-$0.15, and 0 $\leq$ $X_{\rm core}$(C+O)  $\leq$ 0.99, where $X_{\rm core}$(C+O) is the fractional part of carbon and oxygen in the convective core\footnote{We have found that theoretical periods are not much sensitive to the exact core composition of C and O. Grouping (C+O) in one parameter facilitates and speeds up the optimization procedure. This is equivalent of choosing the mass fraction of helium left in the core as the free parameter, since $X_{\rm core}({\rm He}) = 1-X_{\rm core}({\rm C+O})$.}. The constraints on log $q$(H) and $M_{*}$ rely on expectations from stellar modeling and various formation scenarios for hot B subdwarfs \citep[see][]{2002MNRAS.336..449H,2003MNRAS.341..669H}, whereas the range for the core size is loosely inspired by horizontal branch stellar evolution calculations \citep{1993ApJ...419..596D}. 

We assume that the PG 1336$-$018 system has reached full spin-orbit synchronism, such that the sdB component rotates as a solid body with $P_{\rm rot} = P_{\rm orb} =$ 2.42438 h \citep{2000Obs...120...48K,2011MNRAS.412..487K}. This hypothesis relies on very firm grounds, as demonstrated by \citet{2008A&A...489..377C}. The rotationally split pulsation modes are calculated within the first-order perturbative approach. The importance of higher-order perturbation effects and tidal deformation due to the M dwarf companion have been also discussed in \citet{2008A&A...489..377C}, showing that they do not alter in a significant way the asteroseismic analysis, at the precision of the current modeling. 

As in \citet{2008A&A...489..377C}, we use 25 out of the 28 periods listed in Table 4 of \citet{2003MNRAS.345..834K}, leaving aside 3 periods ($f_{10}$, $f_{13}$, and $f_{8}$) that may be spurious. All theoretical modes of degree $\ell =$ 0, 1, 2 and 4 are considered in the 90$-$230 s period range. Those with $\ell =$ 3 are explicitly excluded, having an extremely low visibility in the optical domain due
to cancellation effects because of their specific surface geometry (\citealt{2005ApJS..161..456R}). As an additional constraint, we also limited the association of the dominant modes $f_1$ to $f_5$ to
$\ell \leq 2$, while keeping it open for the lower amplitude modes. This assumption simply filters out eventual, unconvincing solutions where some observed periods of highest amplitude would be associated to $\ell=4$ modes. This is fully justified in view of independent constraints on mode identification obtained from multicolor photometry for some bright sdB stars, from which all the dominant modes so far have been identified to $\ell=0$, 1 and/or 2 modes (see, e.g., \citealt{2006ApJS..165..551T,2008A&A...488..685V}). 

\begin{figure}[!ht]
\begin{center}
\includegraphics[scale=0.48,angle=0]{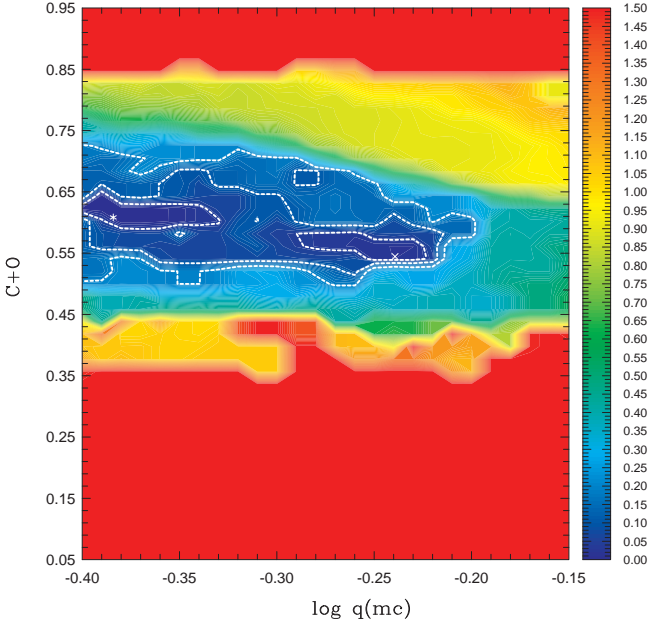}
\end{center}
\caption{\label{f6}Map of the projected merit function (in a log scale) along the log $q$(core)- $X_{\rm core}$(C+O) plane. At each log $q$(core), $X_{\rm core}$(C+O) position, the value given is the minimum of log $S^2$ that can be found among the values obtained for all $M_*$ and log $q$(H). The best-fit model is indicated by an asterisk on the figures, while the model corresponding to the second minimum of the merit function is indicated by a cross. White contours show regions where the frequency fits have $S^2$ values within, respectively, the 1$\sigma$, 2$\sigma$, and 3$\sigma$ confidence levels relative to the best-fit solution.}  
\end{figure}

Within the search domain specified, the multimodal optimization code (see \citealt{2008A&A...489..377C} for details) spotted a region of parameter space where, strictly speaking, two close but distinct
families of models corresponding to minima of the merit function exist. The absolute minimum ($S^2 \sim 4.81$) has $M_*= 0.4699$ $M_{\odot}$, $\log q({\rm H}) =-3.899$, $\log q({\rm core})
=-0.384$ and $X_{\rm core}({\rm C+O}) = 0.609$. The second slightly shallower minimum ($S^2 \sim 5.04$) has quite similar stellar parameters, with $M_*= 0.4795$ $M_{\odot}$, $\log q({\rm H}) =-3.831$,
$\log q({\rm core}) = -0.239$ and $X_{\rm core}({\rm C+O}) = 0.544$. With effective temperatures of 33 038 K (best-fit model) and 33 053 K (second-minimum model), and surface gravities of 5.7713 and 5.7772, respectively. Despite this distinction, Fig. \ref{f5} and Fig. \ref{f6} show however that the two above mentioned families of solutions cannot be statistically disconnected. These maps are projections of the merit function (renormalized to one at its minimum and shown on logarithmic scale) on the 2-dimensional plane, i.e., at a given point of a map, the $S^2$ value is the minimum among all the values when the two other parameters are varied independently. The regions of the best-fit model and the second-minimum model are, indeed, connected at the 2$\sigma$ level (white dotted contours). It is therefore appropriate to treat them as a unique family, adopting the absolute minimum as the model of reference that best match the periods observed in PG 1336--018. More information can be given on this best-fit model, including the details of the period match and mode identification, but since this is not essential for our purposes, we relegate this information to Appendix A. The only relevant point for the discussion is that the relative average dispersion achieved for this best fit model is $\overline{\Delta X/X} = 0.18 \%$ (where $X$ is either the period $P$ or the frequency $\nu$), corresponding to $\overline{\Delta P} = 0.30 $ s and $\overline{\Delta \nu} = 11.4 $ $\mu$Hz. The average precision in simultaneously reproducing the observed periods of PG 1336--018 with the 3G models is very close to that achieved with the 2G models ($\overline{\Delta X/X} = 0.17 \%$, $\overline{\Delta P} = 0.27 $ s and $\overline{\Delta \nu} = 10.5 $ $\mu$Hz; \citealt{2008A&A...489..377C}). Hence the optimal seismic 3G model is essentially as good as its 2G counterpart to that respect.

In Fig. \ref{f5}, the solid and dotted vertical lines correspond to the three orbital solutions and their associated $1\sigma$ uncertainties for the mass of the sdB component derived by \citet{2007A&A...471..605V}: 0.389 $\pm$ 0.005 $M_{\odot}$, 0.466 $\pm$ 0.006 $M_{\odot}$, or  0.530 $\pm$ 0.007 $M_{\odot}$. The right panel of Fig.~\ref{f5} is a close-up view of the $M_*-\log q({\rm H})$ plane in the region of interest. The valley of low $S^2$-values clearly points toward the second orbital solution of \citet{2007A&A...471..605V}, as it was already found with the 2G envelope models \citep{2008A&A...489..377C}. Fig. \ref{f6} shows an elongated valley along the $\log q$(core) parameter between $-0.40$ and $-0.20$, which reflects the fact that the $p$-modes observed in PG 1336$-$018 are not sensitive to the stellar innermost layers, and therefore the size of the core cannot be constrained. The core composition is however indirectly constrained through its impact on the overall hydrostatic equilibrium of the star which affects the pulsation periods and the atmospheric parameters ($T_{\rm eff}$ and $\log g$) for a given stellar mass and envelope mass (see evolutionary tracks of sdB stars from, e.g., \citealt{1993ApJ...419..596D}).

More quantitative statements can be made by statistically estimating the value for each parameter of PG 1336-018 and its associated error from the asteroseismic fit. We adopt here a new procedure
differing from our former methods to estimate errors based on contour maps and $\chi^2$ significance levels (see, e.g., \citealt{2008A&A...489..377C}). We however note that the two approaches are essentially equivalent and provide quantitatively comparable error estimates (see below). We calculate the likelihood function 
\begin{equation}
\mathcal{L}(a_1, a_2, a_3, a_4)\propto e^{-\frac{1}{2}S^2}
\end{equation}
from the $\chi^2$-type merit function, $S^2(a_1,a_2,a_3,a_4)$, that has been sampled by the optimization code during the search for the best-fit models along with additional grid calculations covering the regions of interest in parameter space. For each parameter of interest, say $a_1$, this function is integrated over the full parameter range covered by the other free parameters, thus defining a density of probability function for the chosen parameter:
\begin{equation}
\mathcal{P}(a_1)\mathrm da_1\propto
\mathrm da_1 \iiint \mathcal{L}(a_1,a_2,a_3,a_4)
\mathrm da_2 \mathrm da_3 \mathrm da_4 \qquad .
\end{equation}
This density of probability function is then normalized assuming that the probability is equal to 1 that the value of $a_1$ is in the range specified for the search of a solution. In other words, the normalization factor is such that
\begin{equation}
\int \mathcal{P}(a_1)\mathrm da_1 = 1
\end{equation}
over the allowed parameter range. This method permits the construction of histograms for the probability distribution of each primary model parameter. Secondary model parameters, like $T_{\rm eff}$, $\log g$, the radius $R$, and the luminosity $L$ can also be evaluated in a similar way. Let $b_1$ be one of these secondary parameters. The density probability function is then 
\begin{equation}
\mathcal{P}(b_1)\mathrm db_1 \propto \\
\mathrm db_1 \!\!\!\!\!\!\!
\iiiint\limits_{b\in[b_1,b_1+\mathrm db_1]} 
\!\!\!\!\!\!\!
\mathcal{L}(a_1,a_2,a_3,a_4;b) \mathrm da_1 \mathrm da_2 \mathrm da_3
\mathrm da_4
\end{equation}
where the integration is done with the additional constraint that for a given set of primary parameters, the corresponding value $b$ of the secondary parameter considered must be within $b_1$ and
$b_1+\mathrm db_1$. Again, the density of probability for the secondary parameters is normalized such that 
\begin{equation}
\int \mathcal{P}(b_1)\mathrm db_1 = 1 \qquad.
\end{equation}

\begin{figure}[!ht]
\begin{center}
\includegraphics[scale=0.45,angle=0]{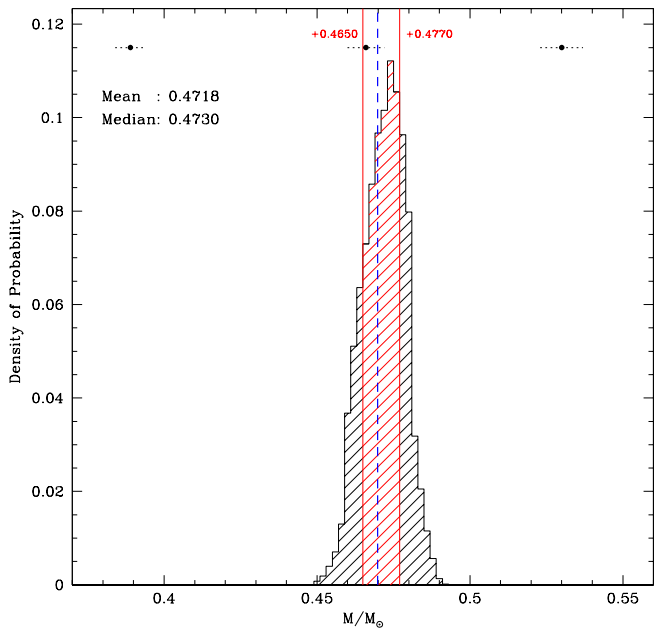}
\end{center}
\caption{\label{f7}Probability density function for the stellar mass from asteroseismology. The filled circles with the dotted lines are the 3 orbital solutions for the mass of the sdB component proposed by \citet{2007A&A...471..605V} with their $1\sigma$ uncertainties. The red hatched part between the two vertical solid red lines defines the 1$\sigma$ range, containing 68.3\% of the mass distribution. The blue vertical dashed line indicates the mass of the optimal model solution of lowest $S^2$-value.} 
\end{figure}

\begin{figure}[!ht]
\begin{center}
\includegraphics[scale=0.45,angle=0]{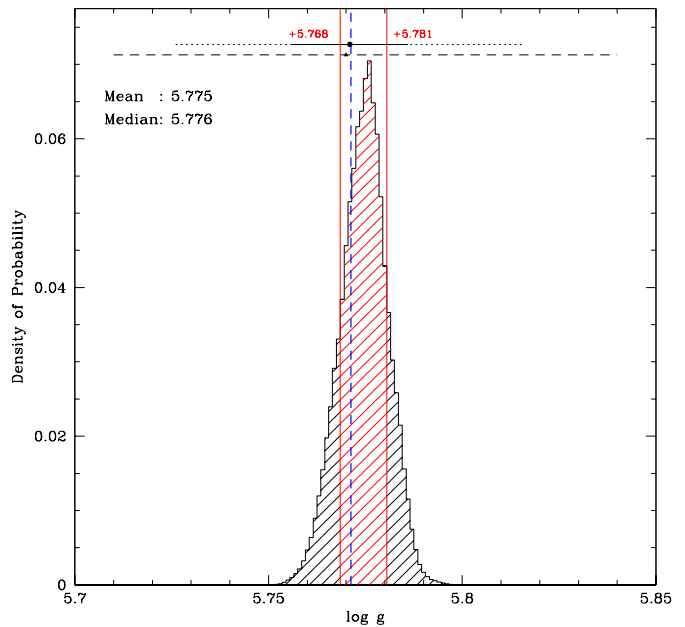}
\end{center}
\caption{\label{f8}Probability density function for the surface gravity $\log g$ from asteroseismology. The filled circle with the solid line and dotted line indicate the value derived from spectroscopy with its
$1\sigma$ and $3\sigma$ uncertainties, respectively. The red hatched region between the two vertical solid red lines is the 1$\sigma$ range, containing 68.3\% of the $\log g$ distribution. The blue vertical
dashed line corresponds to the $\log g$ value of the optimal model solution of lowest $S^2$-value.} 
\end{figure}

\begin{figure}[!ht]
\begin{center}
\includegraphics[scale=0.45,angle=0]{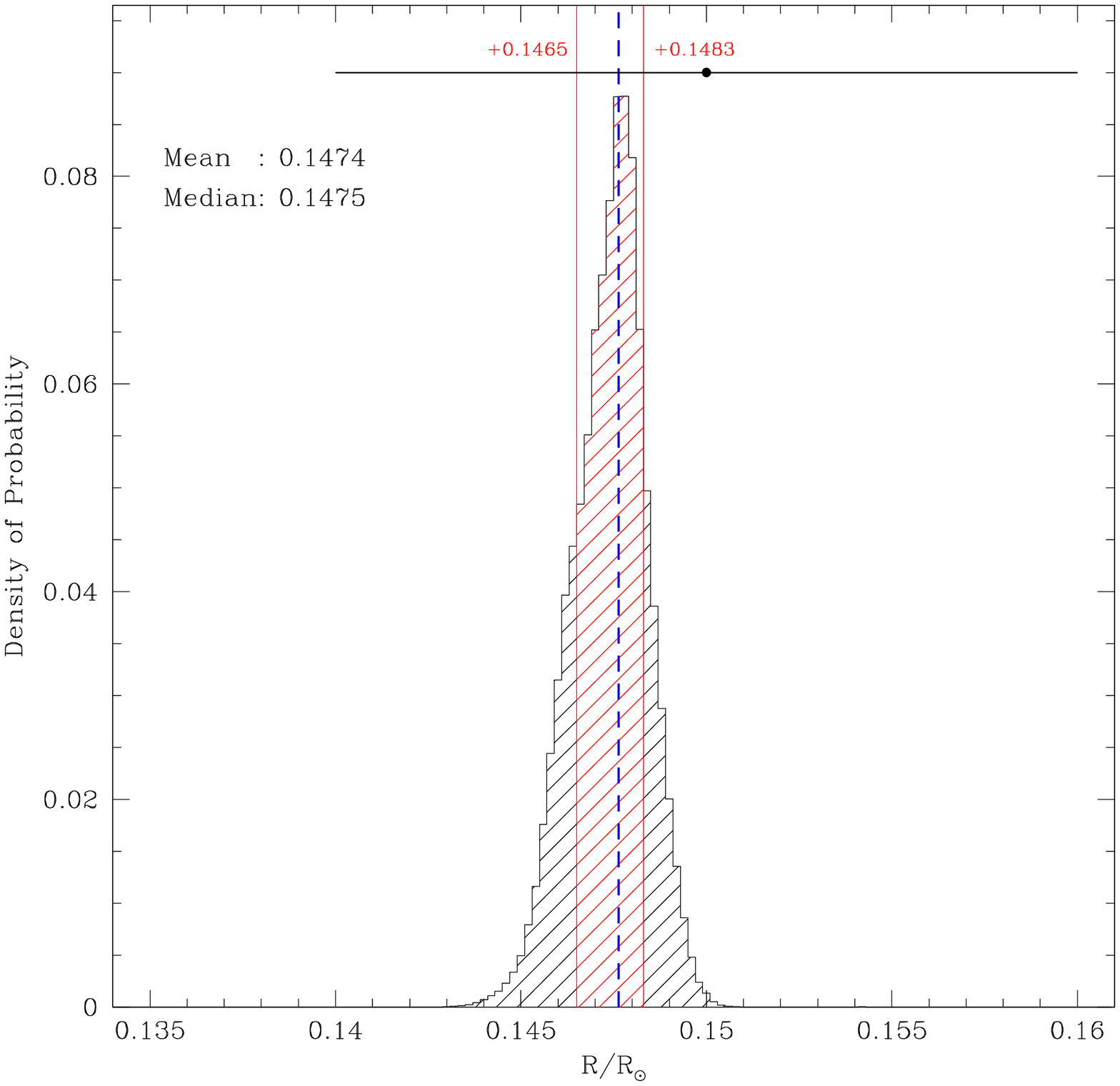}
\end{center}
\caption{\label{f9}Probability density function for the stellar radius from asteroseismology. The filled circle with the solid horizontal line indicates the $1\sigma$ error associated to the relevant orbital
solution from \citet{2007A&A...471..605V}. The red hatched region between the two vertical solid red lines is the 1$\sigma$ range containing 68.3\% of the radius distribution. The blue vertical dashed line corresponds to the radius of the optimal model solution of lowest $S^2$-value.} 
\end{figure}

Figure \ref{f7} shows the histogram obtained for the probability distribution of the total mass of PG 1336--018 inferred from asteroseismology following the procedure described above. The range shown in this diagram only covers the region of interest, where a peak in the distribution is found (there is nothing below 0.37 $M_\odot$ and above 0.56 $M_\odot$) and where the 3 orbital solutions for the mass of
the sdB component derived by \citet{2007A&A...471..605V} can be represented (filled circles with a horizontal dotted line indicating the $1\sigma$ error). Clearly, we find a very striking correspondence between the mass value derived from asteroseismology based on the new 3G models and the second orbital solution from \citet{2007A&A...471..605V}. The two other orbital models giving respectively a much lower mass or a much higher mass for the sdB component are totally inconsistent with the seismic models and can be clearly excluded. From the histogram of the probability distribution, we can
draw a statistical inference for the mass of the sdB star. The region between the two vertical lines (in red) contains 68.3\% of the probability distribution and therefore defines, by convention, the $1\sigma$ range for the value of the represented parameter. This range covers from 0.465 $M_\odot$ to 0.477 $M_\odot$, or equivalently we infer that, statistically, $M_*=0.471\pm0.006$ $M_\odot$ for the sdB star in PG 1336--018, taking the central value of the $1\sigma$ range as the reference for simplicity. We recall that the relevant orbital solution of \citet{2007A&A...471..605V} yielded $M_*=0.466\pm0.006$ and that the seismic solution obtained by \citet{2008A&A...489..377C} based on 2G models led to $M_*=0.459\pm0.005$. Interestingly, the mean and median values of the mass distribution are equal to 0.472 $M_{\odot}$ and 0.473 $M_{\odot}$, respectively. The small differences in these values indicate that the distribution is close to, but not exactly, a normal (gaussian) distribution, as can be easily seen in Fig. \ref{f7}. The small visible "distortion" is to be related to the shape of the $S^2$ function illustrated in the right panel of Fig.~\ref{f5} for instance. Finally, we note that the best-fit seismic model (of minimum $S^2$-value) shown as a vertical blue dashed line lies within the $1\sigma$ range of the derived statistical mass distribution, meaning that this optimal model is not an outlier of the distribution
and is representative of the star, as far as this parameter is concerned.

In the spirit of confronting the results of asteroseismology with other techniques, two additional parameters are of specific interest. First, the probability distribution obtained for the surface gravity
$\log g$ (Fig.~\ref{f8}) can be compared with both the values derived from orbital modeling and spectroscopy. This distribution shows a very narrow peak with a $1\sigma$ interval (containing 68.3\% of the distribution) between 5.768 and 5.781, or a statistically inferred value of $\log g=5.775\pm0.007$ (taking the centre of the 68.3\% significance interval as the reference value). The comparison with other determinations shows a perfect agreement, within the $1\sigma$ errors of each technique, although the value determined from asteroseismology is much more precise. In particular we recall that the relevant orbital solution gives $\log g=5.77\pm0.06$ (filled triangle and horizontal dashed line in Fig.~\ref{f8}) and the spectroscopic analysis (Section 2) leads to $\log g=5.771\pm0.015$ (filled circle and horizontal solid line in Fig.~\ref{f8}). The former asteroseismic solution obtained with 2G models gave $\log g= 5.739\pm0.002$ \citep{2008A&A...489..377C}, also consistent with the other techniques (at the achieved precision), but somewhat shifted compared to the new estimate based on 3G models. Of course, the spectroscopic value was used to constrain the search of an asteroseismic solution. However, Fig.~\ref{f8} shows that the adopted $3\sigma$ tolerance for this constraint (dotted horizontal line) is much larger than the obtained probability distribution, indicating that the search was not overly constrained by spectroscopy. We also checked that relaxing further the spectroscopic constraints does not change at all the location and shape of this distribution. Finally, we note that the mean and median values of the distribution (5.775 and 5.776, respectively) are essentially equal and the derived asteroseismic values for $\log g$ are normally distributed to a very good approximation. Moreover, the $\log g$ value for the optimal model (the vertical dashed line in blue) is representative of the distribution, as it is found within the $1\sigma$ range. 

The other interesting parameter for comparison purposes is the radius $R$ of the star (Fig.~\ref{f9}) derived from both asteroseismology and the analysis of the orbital light curve of the system (in particular the eclipses). The distribution for $R$ also shows a narrow peak with a $1\sigma$ interval (containing 68.3\% of the distribution) between 0.1465 $R_\odot$ and 0.1483 $R_\odot$, or a statistically inferred value of $R=0.1474\pm0.0009$ (taking the centre of the 68.3\% significance interval as the reference value). This is a very precise estimate which is in total agreement with the value derived for the
relevant orbital solution of \citet{2007A&A...471..605V}. The latter is however an order of magnitude less precise with an estimated value of $R=0.15\pm0.01$ $R_\odot$. The previous asteroseismic solution obtained with 2G models led to $R = 0.151\pm0.001$ \citep{2008A&A...489..377C}, also consistent with the orbital value, but shifted notably (in view of to the high precision measurements) compared to the new estimate based on 3G models. Again, we note that the mean and median values of the distribution (0.1474 and 0.1475, respectively) are nearly equal and the statistically derived asteroseismic value for $R$ essentially follows a normal distribution. We also note that radius of the optimal model (the vertical dashed line in blue) is representative of this distribution (as it is within the $1\sigma$ range).

\begin{figure*}[!ht]
\begin{center}
\includegraphics[scale=0.37,angle=0]{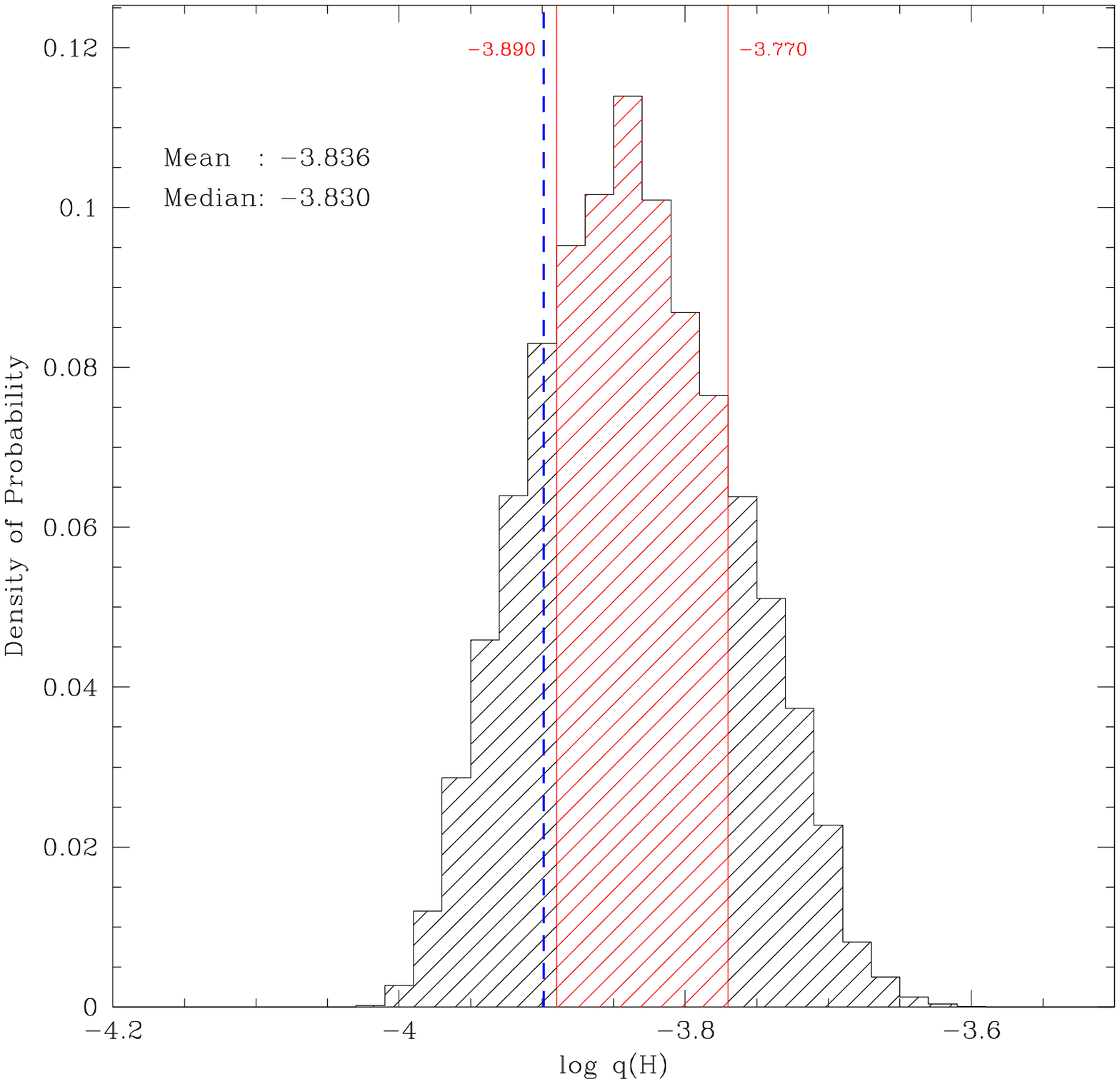}
\includegraphics[scale=0.37,angle=0]{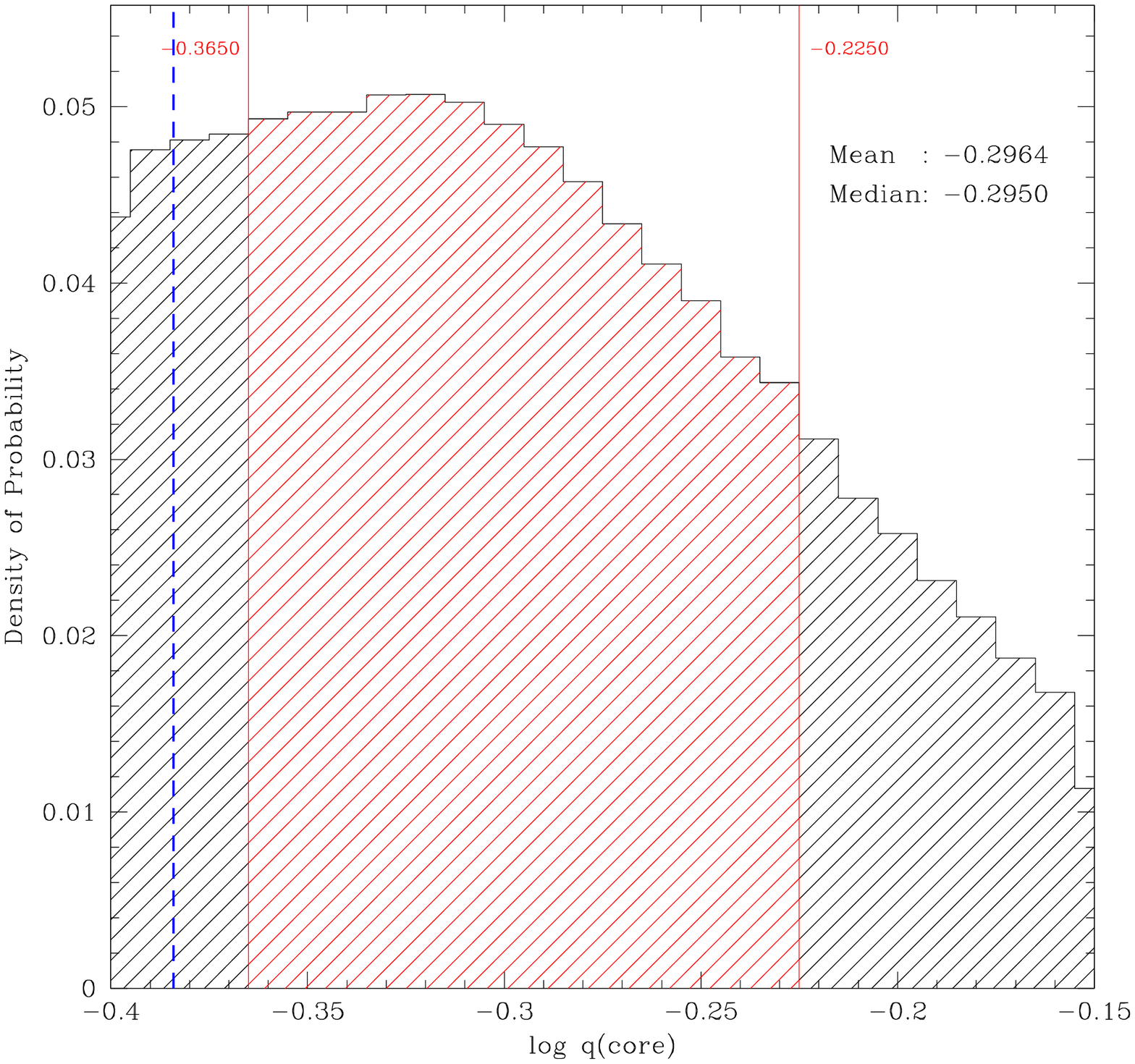}\\
\includegraphics[scale=0.37,angle=0]{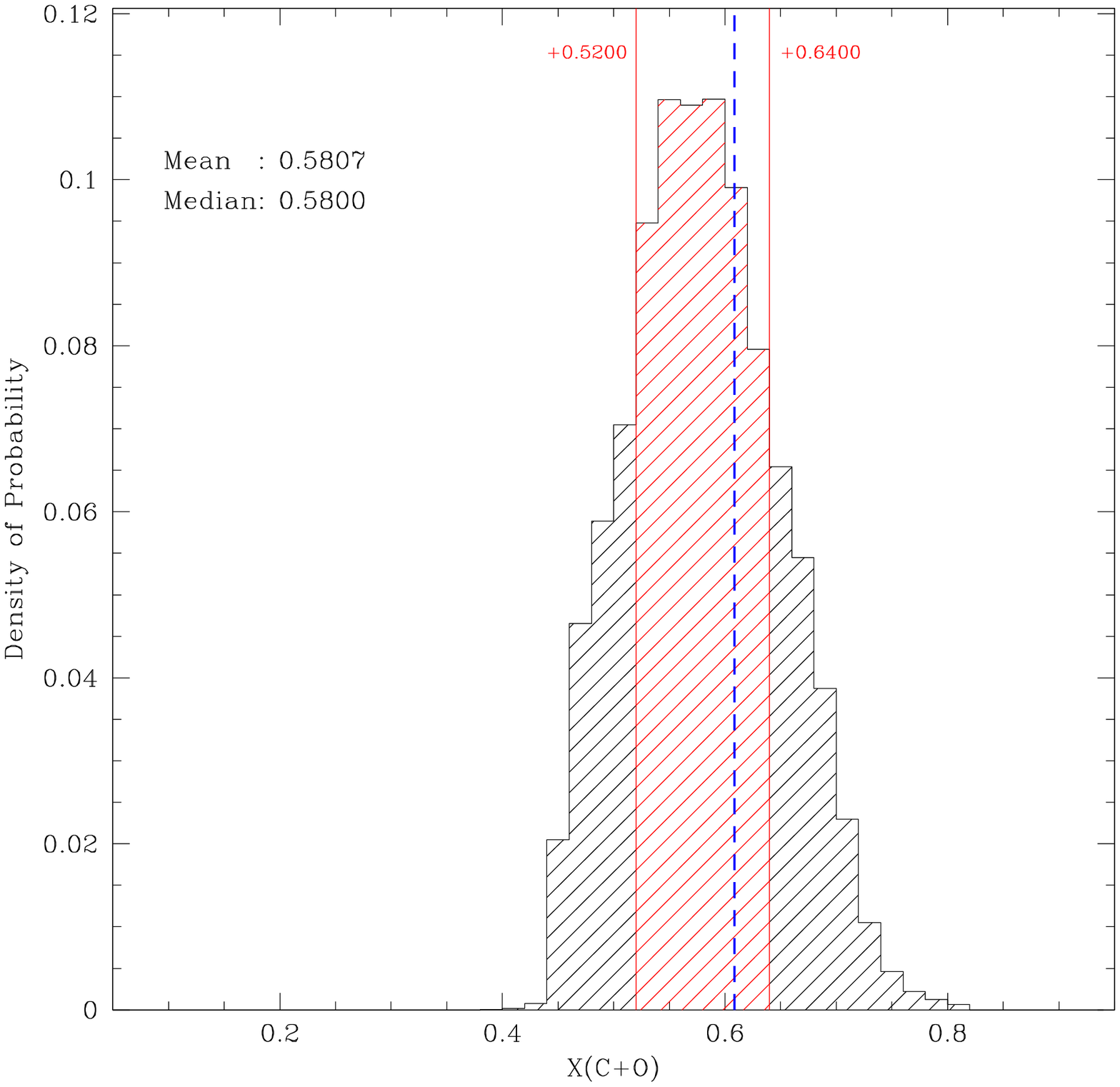}
\includegraphics[scale=0.37,angle=0]{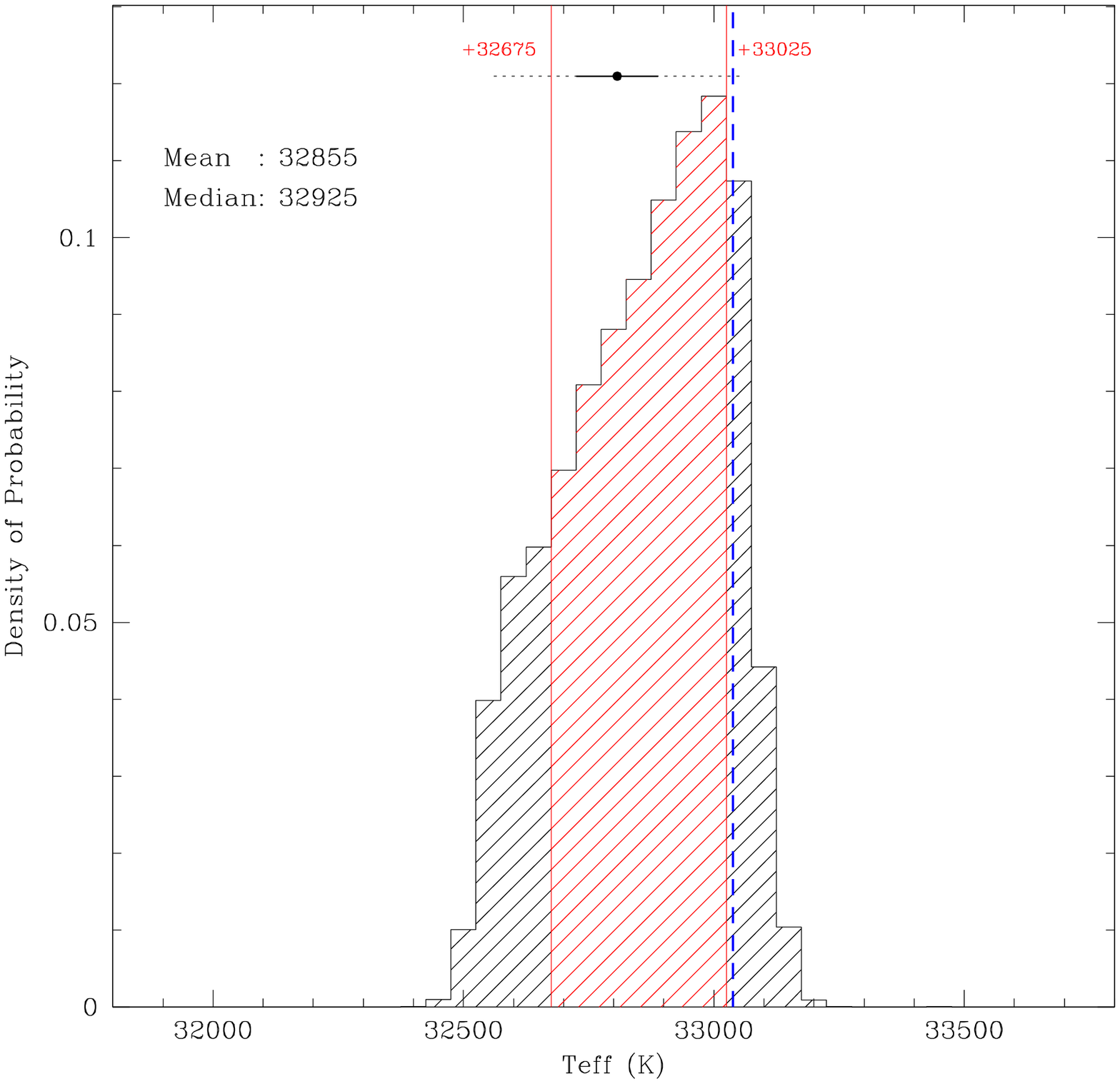}\\
\includegraphics[scale=0.37,angle=0]{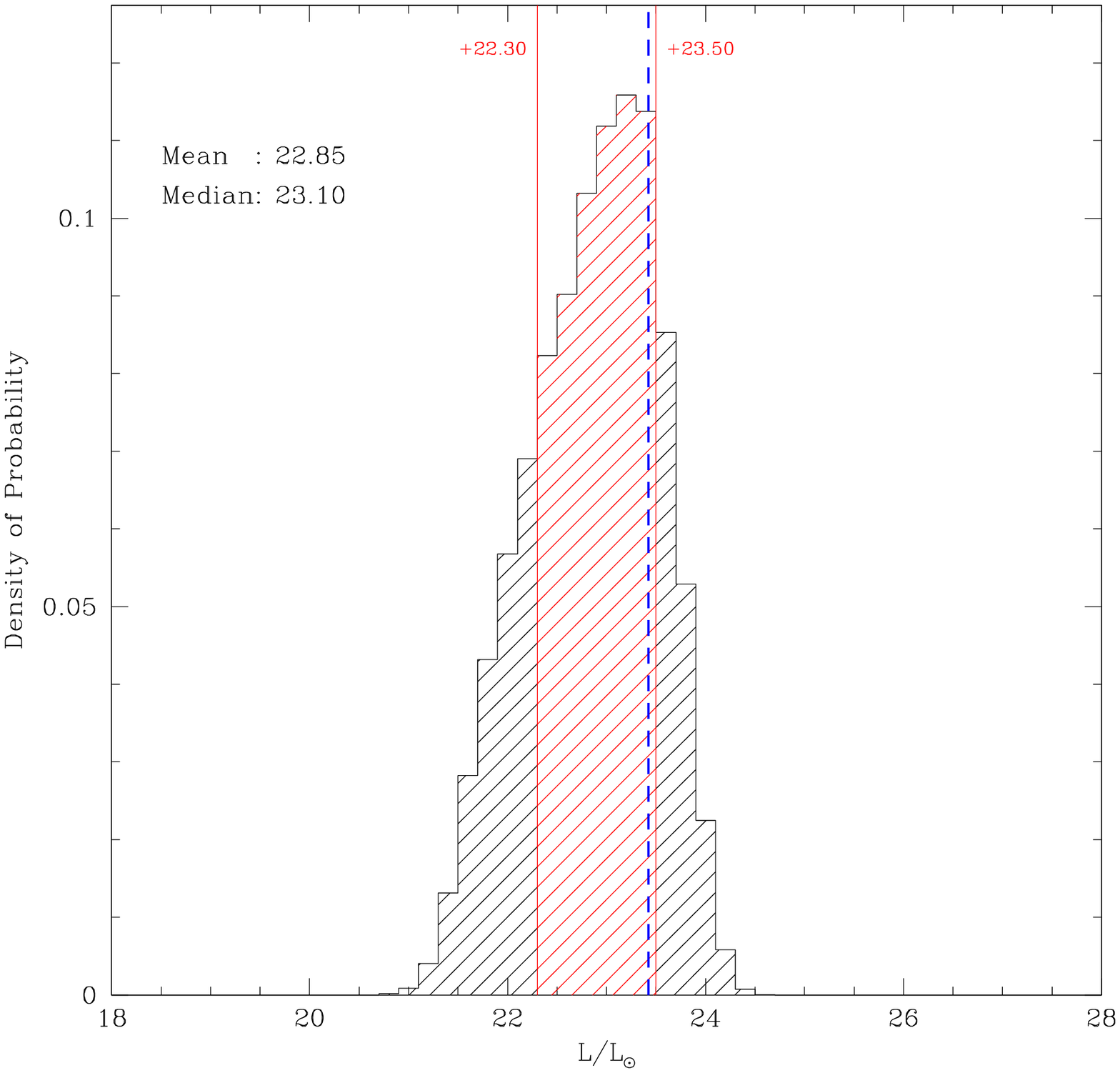}
\end{center}
\caption{\label{f10}Probability density functions derived from asteroseismology for the parameters loq $q$(H), log $q$(core), $X_{\rm core}$(C+O), $T_{\rm eff}$, and $L_*/L_{\odot}$. The red hatched regions between the two vertical solid red lines shows the 1$\sigma$ range containing 68.3\% of the distribution. The blue vertical dashed line corresponds to the value of the optimal model. In the panel showing the distribution for $T_{\rm eff}$, the filled circles is the spectroscopic value with its 1$\sigma$ (solid horizontal line) and 3$\sigma$ (dotted horizontal line) uncertainties.}  
\end{figure*}

\begin{table*}[!ht]
\caption{\label{tabcomp} Structural parameters of PG 1336$-$018 derived
from asteroseismology, spectroscopy, and orbital light curve analysis.}
\small
\begin{center}\begin{tabular}{lcccccc}
\hline\hline
Quantity & \multicolumn{2}{c}{Asteroseismology} & Spectroscopy &
\multicolumn{3}{c}{Orbital light curve modeling} \tabularnewline
\hline
& This study & C2008$^\dag$ & Misc.
&\multicolumn{3}{c}{\citet{2007A&A...471..605V}}  \tabularnewline
& (3G models) & (2G models) & (see Sect. \ref{spectroscopy})& Model I & Model
II & Model III\tabularnewline
\hline
&\\
$M_*/M_{\odot}$ & $0.471$ $\pm$ $0.006$ (1.3\%) & 0.459 $\pm$ 0.005 &
... & 0.389 $\pm$ 0.005 & 0.466 $\pm$ 0.006 & 0.530 $\pm$ 0.007 
\tabularnewline
$R/R_{\odot}$ & $0.1474 \pm 0.0009$ (0.6\%) & 0.151 $\pm$ 0.001 & ...
& 0.14 $\pm$ 0.01 & 0.15 $\pm$ 0.01 & 0.15 $\pm$ 0.01 \tabularnewline
$\log g$ & $5.775 \pm 0.007$ (0.1\%) & 5.739 $\pm$ 0.002 & 5.771 $\pm$
0.015 & 5.74 $\pm$ 0.05 & 5.77 $\pm$ 0.06 & 5.79 $\pm$ 0.07
\tabularnewline
& \\
 $T_{\rm eff}$ (K) & $32 850 \pm 175$ (0.5\%) & 32 740 $\pm$ 400 & 32
807 $\pm$ 82 & ... & ... & ... \tabularnewline
 $\log N({\rm He})/N({\rm H})$ & ... & ... & $-2.918\pm0.089 $
 & ... & ... & ... \tabularnewline

& \\
$\log (M_{\rm env}/M_*)$ & $-3.83 \pm 0.06$ (1.6\%) & $-4.54$ $\pm$
$0.07$ & ... & ... & ... & ... \tabularnewline
$\log (1-M_{\rm core}/M_*)$ & unconstrained & ... & ... & ... & ... &
... 
\tabularnewline
$X_{\rm core}$(C+O) & $0.58 \pm 0.06$ (10\%) & ... & ... & ... & ... &
... \tabularnewline
$\log (L/L_\odot)$ & $22.9 \pm 0.6$ (2.6\%) & $23.3 \pm 1.5$
& ... & ... & ... & ... \tabularnewline
&\\
$M_V(g,T_{\rm eff}, M_*)$ & $4.60$ $\pm$ $0.04$ (0.8\%) & $4.49\pm 0.04$ &
... & ... & ... & ... \tabularnewline
$d(V,M_V)$ (pc) & $571^{\star}$ $\pm$ $35$ (6.1\%) & $619 \pm 38$
& ... & ... & ... & ... \tabularnewline
$P_{\rm rot}$ (s)$^{\ddag}$ & $8727.7823 \pm 0.0002$
& ... & ... & ... & ... & ... \tabularnewline
$V_{\rm eq}(P_{\rm rot}, R)$ (km/s) & $73.9 \pm 0.5$ (0.6\%) & $75.9
\pm 0.6$ & ... & ... & ... & ... \tabularnewline
$i$ ($^{\circ}$) & ... & ... & ... & $80.67\pm0.06$ & $80.67\pm0.06$ &
$80.67\pm0.06$ \tabularnewline
$V\sin i$ (km/s) & $72.9 \pm 0.5$ (0.6\%) & $74.9 \pm 0.6$
& $<79^{\ast}$ & ... & ... & ... \tabularnewline

&&\\
\hline
\multicolumn{7}{l}{
{\footnotesize $^{\dag}$ From \citet{2008A&A...489..377C}.}
}\\
\multicolumn{7}{l}{
{\footnotesize $^{\star}$ Corrected for a reddening of $E(B-V) = 0.021
  \pm 0.009$. The 2008 distance estimate was not corrected for reddening.}
}\\
\multicolumn{7}{l}{
{\footnotesize $^{\ddag}$ Synchronized at the orbital period value (from
\citealt{2011MNRAS.412..487K}).}
}\\
\multicolumn{7}{l}{
{\footnotesize $^{\ast}$ From \citet{2010A&A...519A..25G}.}
}

\end{tabular}\end{center}
\normalsize
\end{table*}

Asteroseismology provides much more information on the star than the three parameters discussed above. For completeness, we show in Fig.~\ref{f10} the histograms obtained for the probability distributions of the other derived parameters. In a nutshell, Fig.~\ref{f10} shows that several parameters, namely the mass of the H-rich envelope $\log q(H)$, the mass fraction of C+O in the core $X_{\rm core}$(C+O) (or equivalently the mass fraction of helium left there), and the luminosity of the star are rather well measured by asteroseismology at a quite interesting precision (see Table~\ref{tabcomp} for the corresponding values and error estimates). It is important to recall that at least two of the above mentioned parameters cannot be measured by other means. For the three parameters discussed above, we also point out that the histograms shown in Fig.~\ref{f10} are all close to normal (gaussian) distributions. In contrast, the size of the mixed core $\log q({\rm core})$ remains unconstrained, as the corresponding panel in Fig.~\ref{f10} clearly shows. The distribution for this parameter is almost uniform and no useful information can be obtained. This is not surprising since the $p$-modes observed in PG 1336--018 are essentially envelope modes that are largely insensitive to the detailed structure of the deepest regions, including in particular the core. It is only with $g$-mode pulsators that this quantity can be effectively measured. The last parameter is the effective temperature $T_{\rm eff}$ which is well known to be poorly measured by the $p$-modes in sdB stars \citep{2005A&A...437..575C}. Indeed, in that case the corresponding panel in Fig.~\ref{f10} shows that it is mainly the spectroscopic constraint imposed during the search for an asteroseismic solution that limits the range of the distribution. Finally, we stress that all the parameter values characterizing the optimal model that provides the absolute best match to the pulsation periods of PG 1336--018 fall within (or very close to) the $1\sigma$ range of their corresponding statistical distributions (Fig. \ref{f7}, \ref{f8}, \ref{f9}, and \ref{f10}; blue vertical dashed line). It proves that this optimal model is not an outlier of the distribution of models that could
potentially fit reasonably well the pulsation periods, but instead a realization of the statistics which is representative of the hot B subdwarf in PG 1336--018 and can safely be considered as the seismic reference for this star. 

\subsection{Comparison between various estimations}

We summarize in Table~\ref{tabcomp} all the values inferred for the parameters of PG 1336--018 from the various techniques employed. This sets the stage for a comparative discussion of the various results and an assessment of eventual systematic effects. As already mentioned, the best tested parameters are $M$, $R$, and $\log g$ because they have been measured quite precisely from both
asteroseismology and orbital modeling, as well as from spectroscopy for the surface gravity. For these 3 parameters, all measurements indicate a particularly strong (within the $1\sigma$ errors) consistency. Notably, the stellar mass is precisely \textit{and accurately} determined from asteroseismology using the 3G models according to model II of \citet{2007A&A...471..605V}. This consistency was already proven for the 2G models \citep{2008A&A...489..377C}, indicating that no important bias exists in the determination of this parameter from asteroseismology. This constitutes a very strong test demonstrating the reliability of the models (both 2G and 3G) for this type of seismic analyses. This is also a crucial point for the project of building an empirical mass distribution of sdB stars from asteroseismology \citep{2012A&A...539A..12F}. Such empirical mass distributions are essential in order to clarify the question of the formation of sdB stars, and unbiased measurements are of course a key factor.

A closer comparison of the two sets of seismic parameters derived from 2G and 3G models, respectively, shows that in practical terms no important difference exists between the two evaluations. The intrinsic precision of these measurements is very high by usual standards, but in general the differences between the two analyses remain below the $3\sigma$ errors (although several parameters can differ by more than $1\sigma$). There is one exception, $\log (M_{\rm env}/M_*)$, showing a significant systematic effect, as this parameter is estimated to be somewhat larger with the 3G models. This is related to the presence of a realistic stellar core in 3G models, which has a small but noticeable impact on the frequencies of $p$-modes, especially for degrees $l=0$ \citep{2002ApJS..139..487C}. These small frequency shifts slightly modify mode trapping, which depends on the positions of the eigenfunctions nodes relative to the envelope He/H transition \citep{2000ApJS..131..223C}. Mode trapping therefore primarily determines the value of log q(H), which thus slightly but significantly changes from 2G to 3G models. Overall, if the improved 3G complete stellar models of sdB stars are now to be preferred for asteroseismology purposes, this comparison underlines that the former 2G envelope models already provided quite robust estimations of the structural parameters of PG1336--018. By extension, this considerably increases our confidence in the seismic parameters derived for all the other $p$-mode sdB pulsators based on these 2G models.

Finally, we point out that this reliability check for the 3G models using PG1336--018 as a privileged testbed remains partial in the sense that only $p$-modes are observed in this pulsating sdB star. An
extension of this test to a $g$-mode pulsator in a close eclipsing binary may become possible with the sdB+dM system 2M~1938+4603 recently discovered by the {\sl Kepler} space mission \citep{2010MNRAS.408L..51O}. \citet{2012ApJ...753..101B} proposed an orbital solution from \textit{Kepler} eclipse timings, assuming a perfect circular orbit. The mass ratio, and therefore the masses of the two components, are however very sensitive to the eccentricity. The light curve modeling of 2M 1938+4603 still has to be carried out in a fully consistent way, i.e., by using additional multicolor photometry like in \citet{2007A&A...471..605V}. On the asteroseismology front, the sdB component of 2M 1938+4603 exhibits an extremely rich pulsation spectrum of both $p$- and $g$-modes. Moreover, the star is presumably a fairly fast rotator if synchronization is achieved (which is likely considering the very short orbital period of 3.024 hours). For $g$-modes that have periods comparable to the expected rotation period of the sdB star (if synchronized), the simple perturbative approach to compute the pulsation spectrum can no longer be applied. One must then investigate the problem with more sophisticated methods based either on the traditional approximation or on nonperturbative treatments to compute the effects of rotation on stellar pulsations (\citealt{2009A&A...506..189R}; \citealt{2012A&A...547A..75O}). This star therefore currently poses a difficult challenge for the seismic modeling, but could become exploitable in the future, as work being done in that sense progresses.

\subsection{Impact of known model uncertainties on the seismic inference of stellar parameters}
\label{uncer}
The static 2G and 3G models developed for detailed asteroseismic studies of sdB stars include a simplified treatment of the effects of diffusion. Diffusion, along with other potentially competing mixing mechanisms, is a process that can modify the chemical stratification inside the star over time. It is therefore more difficult to implement in a strategy based on static stellar structures and approximate ways of dealing with it must be used. Diffusion will most notably have two main effects. First, it will control the heavy metal abundance profiles in the envelope of a sdB star under the pressure of selective radiative forces (radiative levitation). Second, it will also modify (smooth), as time passes, the profile of the chemical transition at the interface between the He mantle and the H-rich envelope. We investigate below the impact of uncertainties associated with these two effects on the seismic determination of stellar parameters. We also investigate the impact of uncertainties on He-burning rates, namely, the triple-$\alpha$ and $^{12}$C($\alpha$,$\gamma$)$^{16}$O nuclear reactions.

\subsubsection{Impact of the iron composition profile}
\label{exp1}
In the simplified approach used in our models, only iron is considered (as the main contributor to the Rosseland mean opacity) and the stratification adopted is assumed to be the profile computed at diffusive equilibrium, i.e., equilibrium between radiative levitation and gravitational settling. This approximation is both practical (in a hydrostatic modeling context) and physically justified by the fact that diffusion timescales are short in the relevant region of the envelope, thus presumably leading almost "instantly" (relative to the secular evolution timescale) to an equilibrium state \citep{2006MmSAI..77...49F,2008CoAst.157..168C}. Of course, in reality, diffusive equilibrium is never completely reached and other competing phenomena can potentially affect the metal distribution such as stellar winds, turbulent mixing (induced, e.g., by differential rotation), and/or double diffusive (thermohaline) convection due to $\mu$-gradient inversions. These effects add to the difficulty of incorporating diffusion in static models and are quite uncertain in their details in evolutionary calculations as well. Therefore, strong uncertainties remain on the exact abundance distributions of chemical species inside the envelope of sdB stars. Because heavy elements (in particular iron) are the dominant source for the gas opacity, their distribution will affect the thermal structure of the star and consequently modify the periods of oscillation modes. However, this assessment has to be made more quantitative in order to truly appreciate the impact in the context of deriving structural parameters by asteroseismology. We demonstrate below that the uncertainty associated with the composition profiles built up by diffusion and other processes in competition is not presently a strong issue when it comes to infer structural parameters of sdB stars from their pulsation spectrum. 

\begin{figure}[!bht]
\begin{center}
\includegraphics[scale=0.45,angle=0]{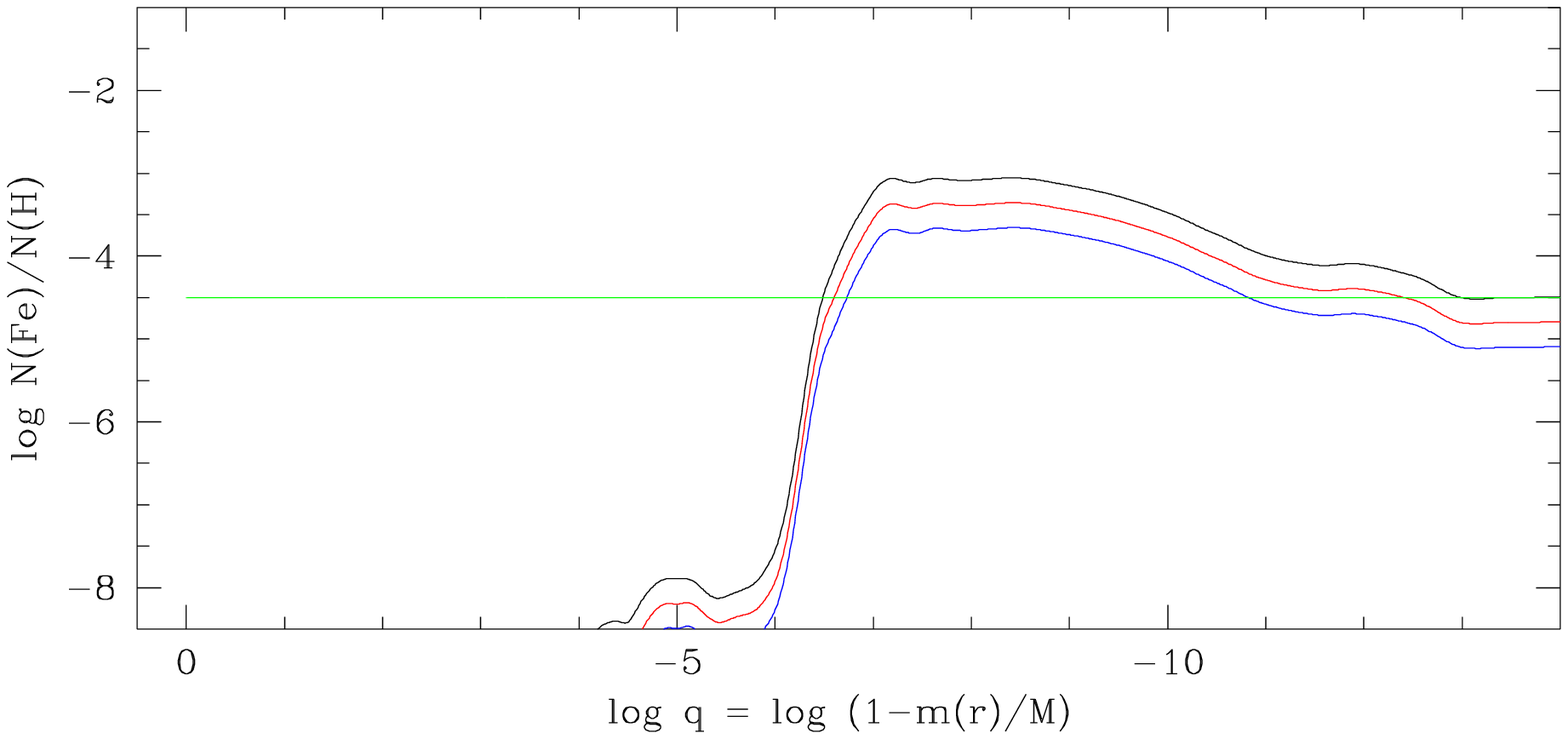}\\
\includegraphics[scale=0.45,angle=0]{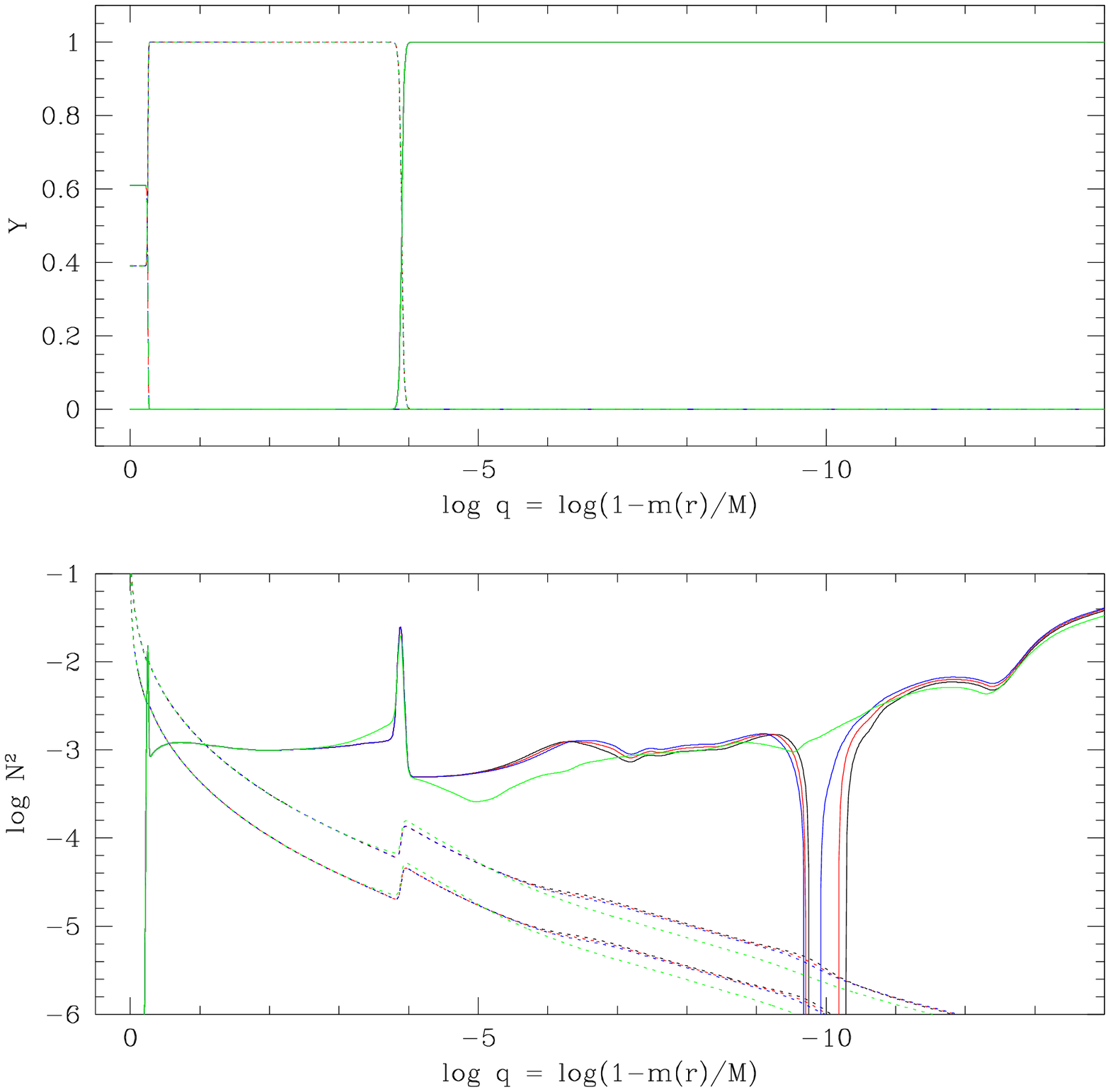}
\end{center}
\caption{\label{fB1}\textit{Upper panel:} Iron abundance profiles used in four, otherwise identical, test models. The black, red, blue, and green curves correspond to the profile at diffusive equilibrium between gravitational settling and radiative levitation, the profile when the amount of levitating iron is decreased by a factor of two, the profile after a decrease of factor of four, and the profile assuming a uniform iron distribution in solar proportion. \textit{Lower panel:} Corresponding profiles for the Brunt-V\"ais\"al\"a (solid curves) and Lamb (dotted curves for $l=1$ and $l=2$) frequencies.}  
\end{figure}

\begin{table*}[!bht]
\caption{\label{tabB1} Periods computed for a series of test models 
with various iron abundance or He/H transition profiles, and 
modified nuclear reaction rates.
}
\normalsize
\begin{center}\begin{tabular}{lcccccccccc}
\hline\hline
\multicolumn{2}{l}{Fe prof.$^a$ :} & $1\cdot \frac{N({\rm Fe})}{N(\rm H)}$$^{\dag}$ & & $\frac{1}{2}\cdot \frac{N({\rm Fe})}{N(\rm H)}$ & 
 $\frac{1}{4}\cdot \frac{N({\rm Fe})}{N(\rm H)}$ & Solar Fe & & $1\cdot \frac{N({\rm Fe})}{N(\rm H)}$ & & $1\cdot \frac{N({\rm Fe})}{N(\rm H)}$\tabularnewline
\multicolumn{2}{l}{He/H prof.$^b$ :} & sharp$^{\dag}$ & & sharp & sharp & sharp & & smooth & & sharp \tabularnewline
\multicolumn{2}{l}{Nucl. rates$^c$ :} & CF88$^{\dag}$ & & CF88 & CF88 & CF88 & & CF88 & & $2\times$CF88 \tabularnewline
 \hline
$\ell$ & $k$ & \multicolumn{9}{c}{Period (s)} \tabularnewline
\hline

&\\

0 &  5 &  95.361 & &   93.021  &  91.091  &   98.997 & &  95.558 & &  97.282\\
0 &  4 & 110.383 & &  107.854  & 105.720  &  114.996 & & 110.289 & & 112.633\\
0 &  3 & 121.878 & &  120.793  & 119.866  &  125.564 & & 122.073 & & 124.161\\
0 &  2 & 143.687 & &  139.815  & 136.578  &  152.200 & & 143.704 & & 146.740\\
0 &  1 & 167.983 & &  167.433  & 167.018  &  169.869 & & 167.967 & & 170.929\\
0 &  0 & 181.622 & &  180.359  & 179.550  &  215.586 & & 181.500 & & 185.170\\
&&\\
1 &  6 &  91.986 & &   89.748  &  87.667  &   95.255 & &  91.911 & &  93.815\\
1 &  5 & 101.879 & &  100.347  &  99.033  &  104.349 & & 102.002 & & 103.844\\
1 &  4 & 114.405 & &  111.923  & 110.009  &  120.653 & & 114.567 & & 116.759\\
1 &  3 & 138.376 & &  135.678  & 132.900  &  142.634 & & 138.265 & & 141.102\\
1 &  2 & 149.437 & &  147.657  & 146.628  &  155.379 & & 149.606 & & 152.381\\
1 &  1 & 180.242 & &  178.677  & 177.624  &  214.649 & & 180.120 & & 183.837\\
&&\\
2 &  5 &  96.122 & &   93.970  &  92.259  &   99.525 & &  96.370 & &  98.043\\
2 &  4 & 111.296 & &  108.611  & 106.408  &  116.728 & & 111.270 & & 113.597\\
2 &  3 & 125.484 & &  124.479  & 123.462  &  128.355 & & 125.595 & & 127.811\\
2 &  2 & 144.146 & &  140.426  & 137.428  &  152.362 & & 144.254 & & 147.202\\
2 &  1 & 176.711 & &  175.225  & 174.182  &  183.319 & & 176.626 & & 180.125\\
2 &  0 & 186.021 & &  185.753  & 185.606  &  213.268 & & 186.069 & & 189.270\\
&&\\
3 &  5 &  93.563 & &   91.148  &  89.037  &   97.147 & &  93.663 & &  95.449\\
3 &  4 & 106.995 & &  104.923  & 103.075  &  110.121 & & 106.922 & & 109.112\\
3 &  3 & 117.429 & &  115.788  & 114.560  &  122.408 & & 117.701 & & 119.732\\
3 &  2 & 141.849 & &  138.076  & 134.813  &  149.620 & & 141.908 & & 144.840\\
3 &  1 & 163.282 & &  162.570  & 162.053  &  165.622 & & 163.304 & & 166.224\\
3 &  0 & 180.063 & &  178.710  & 177.833  &  210.379 & & 180.184 & & 183.593\\
&&\\
4 &  5 &  91.650 & &   89.378  &  87.294  &   94.936 & &  91.627 & &  93.479\\
4 &  4 & 102.519 & &  100.850  &  99.374  &  105.011 & & 102.590 & & 104.508\\
4 &  3 & 114.444 & &  112.144  & 110.397  &  120.172 & & 114.697 & & 116.777\\
4 &  2 & 139.680 & &  136.190  & 133.039  &  146.380 & & 139.725 & & 142.574\\
4 &  1 & 156.689 & &  155.694  & 155.019  &  159.771 & & 156.786 & & 159.599\\
4 &  0 & 178.699 & &  177.224  & 176.246  &  190.230 & & 178.984 & & 182.218\\
&&\\
\hline
\multicolumn{9}{l}{$^\dag$ Standard model of reference}\\
\multicolumn{9}{l}{$^a$ Iron abundance profile (see Section \ref{exp1})}\\
\multicolumn{9}{l}{$^b$ Envelope He/H transition profile (see \ref{exp2})}\\
\multicolumn{9}{l}{$^c$ Nuclear reaction rates for the triple-$\alpha$ 
reaction (see \ref{exp3})}
\end{tabular}\end{center}
\normalsize
\end{table*}

The first experiment is based on a set of representative 3G models built with the same input parameters but assuming different iron abundance profiles (see Fig.~\ref{fB1}). The first model
represents our standard 3G structures that use iron profiles expected at diffusive equilibrium. Two more models are constructed by artificially reducing the amount of levitating iron by a factor of two
and four, respectively. A fourth model is built with a uniform iron abundance distribution in solar proportion. The purpose of this series of models is not to explore exhaustively all imaginable abundance profiles, which would of course be impossible to achieve. Instead, it allows us to sample a range of situations between two extreme cases where, on one side, no diffusion occurs at all (the uniform solar model) and, on the other side, full unperturbed diffusive equilibrium is reached (the standard model at diffusive equilibrium). We expect that the true profile is somewhere in between these two situations, the following results thus providing an accurate view of the magnitude of the induced effects in the context of asteroseismology.

\begin{table*}[!htbp]
\caption{\label{tabB2} Parameters derived for PG 1336$-$018 using various iron abundance profiles.}
\small
\begin{center}\begin{tabular}{lcccccc}
\hline\hline
Parameter & Uniform/Solar & $\frac{1}{4}\cdot \frac{N({\rm Fe})}{N({\rm H})}$
& $\frac{1}{2}\cdot \frac{N({\rm Fe})}{N({\rm H})}$ & 
$1\cdot \frac{N({\rm Fe})}{N({\rm H})}$ &
largest drift & likely drift\tabularnewline

\hline
&\\
$M/M_\odot$ & $0.471 \pm 0.009$ & $0.474 \pm 0.009$ & $0.474 \pm 0.009$ & $0.471 \pm 0.006$ & $+0.003\pm0.011$ & $+0.003\pm0.011$ \\
$\log g$    & $5.814 \pm 0.007$ & $5.795 \pm 0.008$ & $5.786 \pm 0.009$ & $5.775 \pm 0.007$ & $+0.039\pm0.010$ & $+0.020\pm0.011$\\
$R/R_\odot$ & $0.1406 \pm 0.0011$ & $0.1443 \pm 0.0012$ & $0.1459 \pm 0.0012$ & $0.1474 \pm 0.0009$ & $-0.0068\pm0.0014$ & $-0.0031\pm0.0015$\\
$\log q(H)$ & $-4.11 \pm 0.08$ & $-3.70 \pm 0.09$ & $-3.74 \pm 0.09$ & $-3.83 \pm 0.06$ & $-0.28\pm0.10$ & $+0.07\pm 0.11$ \\
$X(C+O)$    & $0.46 \pm 0.10$ & $0.48 \pm 0.10$ & $0.51 \pm 0.09$ & $0.58 \pm 0.06$ & $-0.12\pm0.12$ & $-0.10\pm0.12$\\ 
$L/L_\odot$ & $20.6 \pm 0.7$   & $21.8 \pm 0.7$  &  $22.3 \pm 0.8$ & $22.9 \pm 0.6$ & $-2.3\pm0.9$ & $-1.1\pm0.9$ \\ 
&&\\
$S^2$ (opt.) & 5.95 & 5.93 & 5.46 & 4.81 \\ 
$\overline{\Delta X/X}$ (\%) & 0.24 & 0.21 & 0.19 & 0.18 \\
$\overline{\Delta\nu}$ ($\mu$Hz) & 14.07 & 12.75 & 11.43 & 11.37 \\
&&\\
\hline

\end{tabular}\end{center}
\normalsize
\end{table*}

Clearly, from the noticeable differences observed in the profiles of the Brunt-V\"ais\"al\"a and Lamb frequencies in the corresponding models (lower panel of Fig.\ref{fB1}), the different iron profiles should have some impact on the oscillation modes. Table~\ref{tabB1} confirms this by providing the adiabatic periods of the low-degree ($\ell=0-4$) $p$-modes computed for each of the four models
illustrated in Fig.~\ref{fB1}. The changes in the iron profile generate noticeable differences in periods between the standard model at diffusive equilibrium (dubbed "$1\cdot N({\rm Fe})/N({\rm H})$" in
Table~\ref{tabB1}) and the other models. These however remain relatively limited for models were the amount of levitating iron has been decreased by a factor of two and four, suggesting that the impact of the exact chemical composition in the sampled range should remain small. Not unexpectedly, the model with a uniform Fe profile in solar proportion shows larger variations, possibly leading to more significant systematic drifts in the model parameters derived from asteroseismology. 

In order to quantify more precisely the real impact of these period changes, we reanalyzed the star PG 1336--018 following the same procedure as before (see Sect. \ref{analysis}), but using the various model assumptions about iron composition profiles discussed above. The results obtained with the standard  3G models (using an iron abundance profile at diffusive equilibrium), as discussed in Section \ref{analysis}, are summarized for convenience in the fifth column of Table~\ref{tabB2}. The probability density functions obtained for the main structural parameters of the star depending on the abundance profiles assumed for iron in the stellar models are shown in Fig.~\ref{fB3}, \ref{fB4}, and \ref{fB5} of Appendix \ref{annex2}. The corresponding statistical inferences for these parameters are given in Table~\ref{tabB2}. For each parameter, we also indicate in this table the largest systematic drift relative to the standard model (obtained by comparing with the model having a uniform and solar iron abundance) and the drift which is, in our view, more likely representative of the uncertainties associated with the metal abundance profiles in the envelope of sdB stars.

The main and obvious result of this exercise is that the overall impact of changing substantially the iron abundance profile is very small when it comes to infer the structural parameters of the star from 
asteroseismology. Systematic effects in the derived values, when they exist, remain quite limited in amplitude relative to the current precision of the measurements. The most affected parameters are: the radius with a drift of $-0.0068\pm0.0014$ $R_\odot$ ($4.9\sigma$) in the worst case, but more reasonably a systematic effect of $-0.0031\pm0.0015$ $R_\odot$ ($2.1\sigma$); the surface gravity $\log g$
with a drift of $+0.039\pm0.010$ dex ($3.9\sigma$) in the worst case, and more reasonably $+0.020\pm0.011$ dex ($1.8\sigma$); the H-rich envelope mass $\log q(H)$ with a drift of $-0.28\pm0.010$ dex
($2.8\sigma$) in the worst case, but more reasonably no significant effect; and the luminosity $L$ with a drift of $-2.3\pm0.9$ $L_\odot$ ($2.6\sigma$) in the worst case, and more reasonably $-1.1\pm0.9$
$L_\odot$ ($1.2\sigma$). Of particular interest, we find that the mass determination does not suffer of any systematics at all (the value $+0.003\pm 0.0011$ is consistent with no drift), thus strengthening the 
robustness of mass determinations of sdB stars based on asteroseismology. Consequently, the exact stratification of heavy metals that results from the competition of various diffusive and mixing
processes occurring in the star and which is subject to large uncertainties is clearly not a strong issue at the currently achieved precision for the stellar parameters derived from asteroseismology.

Nevertheless, we point out that the test case of PG 1336--018 is further enlightening here, because it provides a comparison with parameters independently derived from other techniques. Quite
interestingly, Table~\ref{tabB2} and Fig.~\ref{fB3}, \ref{fB4}, and \ref{fB5} show that our standard models assuming a nonuniform iron abundance profile at diffusive equilibrium lead to a seismic solution
that is the closest to the values derived for $\log g$ and $R$ from spectroscopy and orbital light curve analysis. Moreover, the best-fit solution to the observed periods is found to be noticeably better with these models, as illustrated in Table~\ref{tabB2} where the optimal $S^2$ values, mean relative period (or frequency) differences $\overline{\Delta X/X}$, and mean frequency differences $\overline{\Delta\nu}$ achieved for the optimal fits are compared. These facts suggest that the standard models are the most realistic structures among this batch of test models to represent the oscillation properties of PG 1336--018 and that the iron profile at diffusive equilibrium is currently the approximation that is the closest to the real iron distribution in the star. We however stress that improving the treatment of diffusion in the next generations of sdB models in order to obtain closer fits to the observed periods and increase the precision of the measured quantities is among our long term objectives.

\subsubsection{Impact of the envelope He/H transition profile}
\label{exp2}
As a sdB star evolves, diffusion affects the profile of the chemical transition between the H-rich envelope and the He mantle. From a relatively sharp interface at the beginning of core He-burning, the transition will slowly spread over time leading to smoother profiles (see, e.g., \citealt{2011MNRAS.418..195H}). 

In our static 2G and 3G models, the profile of the He/H transition zone is controlled through a single parameter that sets the maximum slope of the composition gradient. This parameter has been calibrated based on evolutionary models from \citet{1993ApJ...419..596D} that do not incorporate diffusion of helium. We therefore traditionally assume a sharp He/H transition in our standard stellar models used for asteroseismology and the impact of this assumption on the seismic solutions is a legitimate question. We demonstrate in the experiment below that the shape of this chemical transition indeed slightly affects the periods of the relevant oscillation modes, but at a level that has no practical consequence for deriving structural parameters by asteroseismology. 

\begin{figure}[!bht]
\begin{center}
\includegraphics[scale=0.45,angle=0]{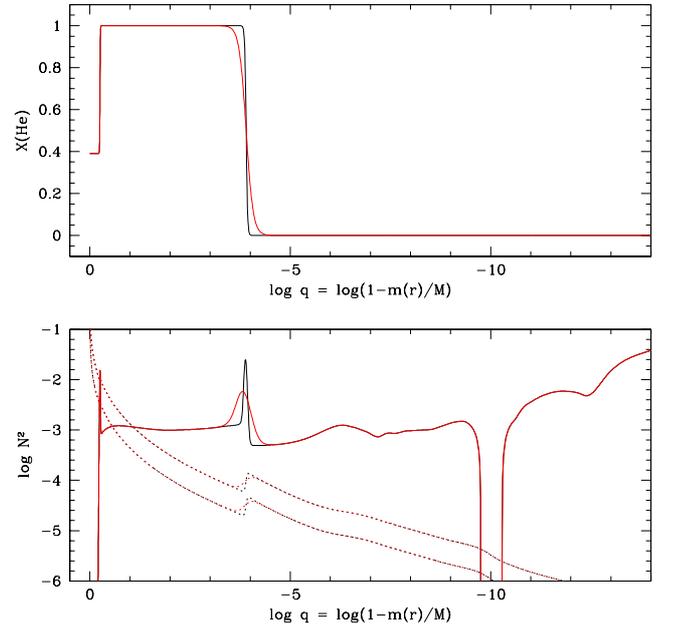}
\end{center}
\caption{\label{fB6}\textit{Upper panel:} Helium abundance profiles used in two, otherwise identical, test models. The curves in black and red respectively show a sharp and a smooth He/H transition profile at the envelope lower boundary. \textit{Lower panel:} Corresponding profiles for the Brunt-V\"ais\"al\"a (solid curves) and Lamb (dotted curves for $l=1$ and $l=2$) frequencies.}  
\end{figure}

We first considered two models with the same structural parameters (and our usual nonuniform iron abundance distribution from diffusive equilibrium) but with different He/H transition profiles (see the upper panel of Fig.~\ref{fB6}). The first model, which is the standard model of reference (see Section \ref{analysis}), has a sharp transition. The second model is built assuming a smooth transition. The
Brunt-V\"ais\"al\"a and Lamb frequency profiles shown in the lower panel of Fig.~\ref{fB6} clearly illustrate that the differences in model structure exclusively occur in the transition region, with a wider
Ledoux spike for the model having a smooth transition. The profiles remain nearly unchanged in the other regions of the star. It suggests that a rather limited impact on the pulsation periods occurs, which is confirmed in Table~\ref{tabB1} by comparing the third and seventh columns. Between the two models, the period differences for the low-order $p$-modes of interest remain very small. It is therefore
anticipated that the shape of He/H transition zone will have no significant impact on the structural parameters derived from asteroseismology.

\begin{table}[!htbp]
\caption{\label{tabB3} Parameters derived for PG 1336$-$018 using a sharp or a smooth He/H transition profile.}
\scriptsize
\begin{center}\begin{tabular}{lccc}
\hline\hline
Parameter & sharp transition & smooth transition & drift \tabularnewline

\hline
&\\
$M/M_\odot$ &  $0.471 \pm 0.006$   & $0.476 \pm 0.009$   & $+0.005\pm0.011$  \\
$\log g$    &  $5.775 \pm 0.007$   & $5.777 \pm 0.007$   & $+0.002\pm0.010$ \\
$R/R_\odot$ &  $0.1474 \pm 0.0009$ & $0.1476 \pm 0.0011$ & $+0.0002\pm0.0014$ \\
$\log q(H)$ &  $-3.83 \pm 0.06$    & $-3.79 \pm 0.08$    & $+0.04\pm0.10$ \\
$X(C+O)$    &  $0.58 \pm 0.06$     & $0.53 \pm 0.09$     & $-0.05\pm0.11$ \\ 
$L/L_\odot$ &  $22.9 \pm 0.6$      & $22.8 \pm 0.7$      & $-0.1\pm0.9$  \\ 
&&\\
$S^2$ (opt.) & 4.81 & 4.91 \\ 
$\overline{\Delta X/X}$ (\%) & 0.18 & 0.18 \\
$\overline{\Delta\nu}$ ($\mu$Hz) & 11.37 & 11.48 \\
&&\\
\hline

\end{tabular}\end{center}
\normalsize
\end{table}

We explicitly demonstrate the above assertion by reanalyzing PG 1336--018 with models featuring a smooth He/H transition. The probability density functions obtained for the main structural parameters of the star are provided in Fig.~\ref{fB7} (Appendix \ref{annex2}). The corresponding statistical inferences for these parameters are given in Table~\ref{tabB3} for comparisons with the values derived
with our standard models assuming a sharp transition. No significant drift in the derived stellar parameters is observed and the achieved quality of the optimal period fit is comparable. Consequently, the 
effects induced by diffusion on the mantle/envelope transition profile presently have no noticeable impact on deriving the structural parameters of sdB stars by asteroseismology, considering the actual precision of the fits. This can safely be regarded as a high order effect. 

\subsubsection{Impact of nuclear reaction rates}
\label{exp3}
In a last experiment, we investigated the effects of nuclear reaction rates on the inferred parameters of the star. Our 3G stellar models incorporate the network of nuclear reactions (in 
particular the triple-$\alpha$ and related reactions for He-burning) to compute the rate of energy production that will influence the overall thermal equilibrium of the star. For our purposes, the code still relies on the \cite{1988ADNDT..40..283C} nuclear reaction rates. However, newer rates from \cite{1999NuPhA.656....3A} have since been published which, at the temperatures relevant to He-burning in sdB stars, increase by nearly 10\% the triple-$\alpha$ rate and by a factor of 2 the rather uncertain $^{12}$C($\alpha$,$\gamma$)$^{16}$O rate.

In order to estimate the real impact of these changes on the seismic inferences of the stellar parameters, we constructed test models where the nuclear energy production rates are increased by 10\% for the triple-$\alpha$ reaction and doubled for the $^{12}$C($\alpha$,$\gamma$)$^{16}$O reaction. These are referred to as the "$2\times$CF88" models. The effect of this change on the pulsation periods is given in last column of Table~\ref{tabB1}. Compared to the model of reference, a slight nearly uniform increase of the periods is observed. This reflects the fact that for the same input parameters, the 3G model with the modified nuclear reaction rates leads to a star with a slightly larger radius, and therefore slightly less compact.

\begin{table}[!htbp]
\caption{\label{tabB4} Parameters derived for PG 1336$-$018 using models with modified reaction rates for the triple-$\alpha$ and $^{12}$C($\alpha$,$\gamma$)$^{16}$O reaction rates.}
\scriptsize
\begin{center}\begin{tabular}{lccc}
\hline\hline
Parameter & CF88 & $2\times$CF88 & drift \tabularnewline

\hline
&\\
$M/M_\odot$ &  $0.471 \pm 0.006$   & $0.474 \pm 0.010$   & $+0.003\pm0.012$  \\
$\log g$    &  $5.775 \pm 0.007$   & $5.775 \pm 0.008$   & $0.00\pm0.011$ \\
$R/R_\odot$ &  $0.1474 \pm 0.0009$ & $0.1479 \pm 0.0010$ & $+0.0005\pm0.0013$ \\
$\log q(H)$ &  $-3.83 \pm 0.06$    & $-3.84 \pm 0.09$    & $-0.01\pm0.11$ \\
$X(C+O)$    &  $0.58 \pm 0.06$     & $0.54 \pm 0.10$     & $-0.04\pm0.12$ \\ 
$L/L_\odot$ &  $22.9 \pm 0.6$      & $23.0 \pm 0.7$      & $+0.1\pm0.9$  \\ 
&&\\
$S^2$ (opt.) & 4.81 & 5.12 \\ 
$\overline{\Delta X/X}$ (\%) & 0.18 & 0.19 \\
$\overline{\Delta\nu}$ ($\mu$Hz) & 11.37 & 11.52 \\
&&\\
\hline

\end{tabular}\end{center}
\normalsize
\end{table}

We reanalyzed PG 1336--018 with the "$2\times$CF88" models, and the results are given in Fig.~\ref{fB8} (Appendix \ref{annex2}) showing the probability density functions obtained for the main structural parameters of the star. The corresponding statistical inferences for these parameters are given in Table~\ref{tabB4} for comparisons with the values derived with our standard model. No significant drift in the derived stellar parameters is observed and the quality of the optimal period fit is comparable. Consequently, the impact of updating the nuclear reaction rates from \cite{1988ADNDT..40..283C}  to \cite{1999NuPhA.656....3A} on the values derived from asteroseismology is not expected to be important. This will be done in future versions of our static models aimed at detailed seismic analyses of sdB stars, but clearly this is currently not a major source of uncertainty for our current purposes.

\section{Summary and conclusion}

In this paper we presented new stellar models of so-called third generation (3G), aimed at performing detailed asteroseismic studies of hot B subdwarf pulsators. These models are static structures realistically representing the star from the surface to its center, contrary to the former generation of envelope models of second generation (2G) that were missing the innermost regions. This improvement was a strong necessity in order to exploit with asteroseismology the $g$-mode sdB pulsators that have been and are currently observed from space with CoRoT and \textit{Kepler}. The new models are constructed assuming hydrostatic and thermal equilibrium for a given set of model parameters, including a chemical composition of the core. They allow for the computation of accurate periods/frequencies for the $p$-modes and the $g$-modes (including possible mixed modes), while the former generation of envelope models was limited to the $p$-modes in terms of accuracy.  

In order to test the reliability of these new 3G models for asteroseismology, we reanalyzed the pulsating sdB star PG1336--018 with them. This object is a close eclipsing sdB+dM system where the sdB
component is a $p$-mode pulsator. This very rare configuration allows us to test some of the stellar parameters inferred from asteroseismology with the values obtained independently from the orbital light curve analysis of \citet{2007A&A...471..605V}. Our new seismic results show that a very strong consistency persists at that level, as it was shown to be the case with our former 2G models used in the seismic study of this star \citep{2008A&A...489..377C}. Our new analysis provides updated estimates of the structural parameters of the sdB star in PG1336--018 based on more realistic models (see
Table~\ref{tabcomp}). The derived parameters for the star do not differ much from those already inferred in \citet{2008A&A...489..377C}, but more importantly they demonstrate that the asteroseismic inferences based on both 2G and 3G models are both \textit{precise and accurate}, at least for the stellar parameters that could be tested, which includes the mass and radius of the star. This is an important result in view, for instance, of our current attempts to construct an empirical mass distribution for comparison with predictions of various formation channels of sdB stars \citep{2012A&A...539A..12F}. We also demonstrated that known model uncertainties do not affect, at the current level of accuracy, the structural parameters derived from asteroseismology. However, this is one of our long term goal to include improved physics in stellar models, in particular the treatment of diffusion, in order  to obtain closer fits to the observed pulsations and increase the precision of the derived quantities by asteroseismology.

Our approach based on the minimization of a merit function has sometimes been criticized in the past, arguing that the best-fit seismic model is not necessary the most representative model of the star. We show in this analysis, through an improved treatment and estimation of the statistical significance of the inferred parameters, that the optimal model that best match the observed periods of PG1336--018 is not an outlier of the statistical distribution of potential solutions. It can therefore safely be considered as the model that best represents the star in consideration, as we used to consider it in our previous
analyses.

Overall, we can conclude that the main goal of this paper, which was to present and demonstrate the basic validity of the new third generation models used for asteroseismology of sdB stars, is fully achieved. Such models have passed an important test showing that they are up to this task.

\begin{acknowledgements}
This work was supported in part by the Natural Sciences and Engineering Research Council of Canada. G.F. also acknowledges the contribution of the Canada Research Chair Program. S.C. thanks the Programme National de Physique Stellaire (PNPS, CNRS/INSU, France). This work was granted access to the HPC resources of CALMIP under the allocation 2012-p0205.
\end{acknowledgements}

\bibliographystyle{aa}
\bibliography{Biblio/references}

\appendix

\section{Period fit and mode identification from the optimal model of PG~1336--018}

The period match and mode identification obtained for the best-fit model of PG 1336-018 are given in Table \ref{modeid}. The relative and absolute differences for each pair of the 25 observed/theoretical modes are provided, in period and frequency, $\Delta X/X$ (in \%), $\Delta P$ (in sec), and $\Delta \nu$ (in $\mu$Hz). On average, the relative dispersion between the observed and theoretical
periods is 0.18\%, which is similar to the best-fit model of second generation uncovered by \citet{2008A&A...489..377C} that led to 0.17\%. 

Independent constraints on the mode identification in PG 1336$-$018 exist from time-resolved spectroscopy \citep{2009A&A...505..239V}. Their detailed line-profile analysis excluded the $(\ell,|m|) =$ (3,3), (1,0), (2,1) or (2,0) modes for the 5435 $\mu$Hz pulsation ($f_{5}$ in Table \ref{modeid}), although data were too noisy to uniquely identify this pulsation with the radial or sectoral dipole/quadrupole modes. The 5435 $\mu$Hz $f_{5}$ pulsation is identified as the fundamental radial mode $(k,\ell,m)=(0,0,0)$ in Table \ref{modeid}, which is consistent with the results of \citealt{2009A&A...505..239V} (see in particular their Fig.~13).

It is also interesting to look back at the three uncertain periods of \citet{2003MNRAS.345..834K} that were not included in the seismic search: 184.04 s ($f_{10}$), 178.96 s ($f_{13}$), and 173.59 s ($f_8$). $f_{10}$ and $f_8$ all have an acceptable counterpart in the best-fit model theoretical spectrum (see Table \ref{modeid}). $f_{10}$ can be associated with the ($l,k,m$)$=$(2,0,$-$1) mode for a relative
dispersion $\Delta X/X = -0.33\%$, and $f_8$ finds a counterpart at $\Delta X/X = -0.26\%$ with the ($l,k,m$)$=$(4,0,$-$2) mode. These matches do not significantly affect the overall quality of the fit. A
$l=3$ mode (not shown here) is needed to account for the last uncertain period, 178.96 s ($f_{13}$). We recall however that the existence of these periods need to be confirmed by additional observations.

The mode identification of the best-fit model respects quite well, qualitatively, the expected amplitude hierarchy. On average, one may expect that modes of higher $l$ would have lower apparent amplitudes even though this argument can hardly be applied to individual identifications, since intrinsic amplitudes are not known and may vary significantly from one mode to another. According to the mode identification in Table \ref{modeid}, the observed average amplitudes $\overline{A_l}$ are: $\overline{A_0} = $ 0.145\% (for $l=$ 0 modes), $\overline{A_1} = $ 0.230\%, $\overline{A_2} = $ 0.183\%, and $\overline{A_4} = $ 0.081\%. This significant drop for $l=4$ modes corresponds well to the theoretical expectation that modes with $l=4$ have a much lower visibility than the $l \leq 2$ modes \citep{2005ApJS..161..456R,2007A&A...476.1317R}.

\scriptsize
\onecolumn
\begin{center}
\begin{longtable}{ccrccccccccl}
\caption{\label{modeid}Mode identification and details of the frequency fit obtained for the optimal solution. The mean relative dispersion of the fit is $\overline{\Delta X/X} = 0.18 \%$ (or $\overline{\Delta P} = 0.30 $ s and $\overline{\Delta \nu} = 11.366 $ $\mu$Hz; $S^2=4.812$).}
\\ 
\hline\hline
 & & & $\nu_{\rm obs}$ & $\nu_{\rm th}$ & $P_{\rm obs}$ & $P_{\rm th}$ &
$\Delta X/X$ & $\Delta P$  & $\Delta \nu$ & Amplitude &
Id.\tabularnewline
 $l$ & $k$ & $m$ & ($\mu$Hz) & ($\mu$Hz) & (s) & (s) & ($\%$) & (s) &
($\mu$Hz) & (\%) & \tabularnewline
\hline

 & \tabularnewline
\endfirsthead

\caption{continued.}\\
\hline\hline
 & & & $\nu_{\rm obs}$ & $\nu_{\rm th}$ & $P_{\rm obs}$ & $P_{\rm th}$ &
$\Delta X/X$ & $\Delta P$  & $\Delta \nu$ & Amplitude &
Id.\tabularnewline
 $l$ & $k$ & $m$ & ($\mu$Hz) & ($\mu$Hz) & (s) & (s) & ($\%$) & (s) &
($\mu$Hz) & (\%) & \tabularnewline
\hline

 & \tabularnewline
\endhead

 & \tabularnewline
\hline

\endfoot

 & \tabularnewline
\hline

\endlastfoot

0 & $2$ & $ 0$ & ... & 6849.952 & ... & 145.99 & ... & ... & ... & ...
&\tabularnewline
0 & $1$ & $ 0$ & 5891.363 & 5896.601 & 169.74 & 169.59 & $+0.09$ &
$+0.15$ & $-5.238$ & $0.0900$ & $f_{15}$\tabularnewline
0 & $0$ & $ 0$ & 5435.373 & 5431.524 & 183.98 & 184.11 & $-0.07$ &
$-0.13$ & $+3.849$ & $0.2000$ & $f_{5}$\tabularnewline
 & \tabularnewline
 \hline
 & \tabularnewline
1 & $3$ & $-1$ & ... & 7249.631 & ... & 137.94 & ... & ... & ...
&\tabularnewline
1 & $3$ & $ 0$ & 7108.836 & 7138.460 & 140.67 & 140.09 & $+0.41$ &
$+0.58$ & $-29.623$ & $0.0900$ & $f_{14}$\tabularnewline
1 & $3$ & $+1$ & ... & 7027.288 & ... & 142.30 & ... & ... & ...
&\tabularnewline
 & \tabularnewline
1 & $2$ & $-1$ & ... & 6721.396 & ... & 148.78 & ... & ... & ... & ...
&\tabularnewline
1 & $2$ & $ 0$ & ... & 6609.834 & ... & 151.29 & ... & ... & ... & ...
&\tabularnewline
1 & $2$ & $+1$ & ... & 6498.272 & ... & 153.89 & ... & ... & ... & ...
&\tabularnewline
 & \tabularnewline
1 & $1$ & $-1$ & 5585.656 & 5581.761 & 179.03 & 179.15 & $-0.07$ &
$-0.12$ & $+3.895$ & $0.4000$ & $f_{2}$\tabularnewline
1 & $1$ & $ 0$ & 5470.759 & 5469.172 & 182.79 & 182.84 & $-0.03$ &
$-0.05$ & $+1.587$ & $0.0600$ & $f_{22}$\tabularnewline
1 & $1$ & $+1$ & 5369.416 & 5356.584 & 186.24 & 186.69 & $-0.24$ &
$-0.45$ & $+12.832$ & $0.3700$ & $f_{3}$\tabularnewline
 & \tabularnewline
 \hline
 & \tabularnewline
2 & $3$ & $-2$ & ... & 8079.349 & ... & 123.77 & ... & ... & ... & ...
&\tabularnewline
2 & $3$ & $-1$ & 7948.494 & 7976.309 & 125.81 & 125.37 & $+0.35$ &
$+0.44$ & $-27.815$ & $0.0600$ & $f_{24}$\tabularnewline
2 & $3$ & $ 0$ & 7880.842 & 7873.268 & 126.89 & 127.01 & $-0.10$ &
$-0.12$ & $+7.573$ & $0.0700$ & $f_{19}$\tabularnewline
2 & $3$ & $+1$ & ... & 7770.228 & ... & 128.70 & ... & ... & ... & ...
&\tabularnewline
2 & $3$ & $+2$ & ... & 7667.188 & ... & 130.43 & ... & ... & ... & ...
&\tabularnewline
  & \tabularnewline
2 & $2$ & $-2$ & ... & 7051.422 & ... & 141.82 & ... & ... & ... & ...
&\tabularnewline
2 & $2$ & $-1$ & ... & 6940.368 & ... & 144.08 & ... & ... & ... & ...
&\tabularnewline
2 & $2$ & $ 0$ & ... & 6829.314 & ... & 146.43 & ... & ... & ... & ...
&\tabularnewline
2 & $2$ & $+1$ & ... & 6718.259 & ... & 148.85 & ... & ... & ... & ...
&\tabularnewline
2 & $2$ & $+2$ & ... & 6607.205 & ... & 151.35 & ... & ... & ... & ...
&\tabularnewline
 & \tabularnewline
2 & $1$ & $-2$ & 5757.384 & 5785.532 & 173.69 & 172.84 & $+0.49$ &
$+0.84$ & $-28.148$ & $0.4700$ & $f_{1}$\tabularnewline
2 & $1$ & $-1$ & ... & 5686.870 & ... & 175.84 & ... & ... & ... & ...
&\tabularnewline
2 & $1$ & $ 0$ & 5598.477 & 5588.209 & 178.62 & 178.95 & $-0.18$ &
$-0.33$ & $+10.268$ & $0.1700$ & $f_{6}$\tabularnewline
2 & $1$ & $+1$ & ... & 5489.548 & ... & 182.16 & ... & ... & ... & ...
&\tabularnewline
2 & $1$ & $+2$ & 5392.289 & 5390.887 & 185.45 & 185.50 & $-0.03$ &
$-0.05$ & $+1.402$ & $0.2500$ & $f_{4}$\tabularnewline
 & \tabularnewline
2 & $0$ & $-2$ & 5505.698 & 5504.174 & 181.63 & 181.68 & $-0.03$ &
$-0.05$ & $+1.525$ & $0.0800$ & $f_{17}$\tabularnewline
2 & $0$ & $-1$ & ... & 5416.040 & [184.04] & 184.64 & ... & ... & ... &
... & [$f_{10}$]\tabularnewline
2 & $0$ & $ 0$ & ... & 5327.907 & ... & 187.69 & ... & ... & ... & ...
&\tabularnewline
2 & $0$ & $+1$ & ... & 5239.773 & ... & 190.85 & ... & ... & ... & ...
&\tabularnewline
2 & $0$ & $+2$ & ... & 5151.640 & ... & 194.11 & ... & ... & ... & ...
&\tabularnewline
 & \tabularnewline
 \hline
 & \tabularnewline
4 & $5$ & $-4$ & ... & 11203.734 & ... & 89.26 & ... & ... & ... & ...
&\tabularnewline
4 & $5$ & $-3$ & ... & 11092.302 & ... & 90.15 & ... & ... & ... & ...
&\tabularnewline
4 & $5$ & $-2$ & ... & 10980.871 & ... & 91.07 & ... & ... & ... & ...
&\tabularnewline
4 & $5$ & $-1$ & ... & 10869.439 & ... & 92.00 & ... & ... & ... & ...
&\tabularnewline
4 & $5$ & $ 0$ & ... & 10758.007 & ... & 92.95 & ... & ... & ... & ...
&\tabularnewline
4 & $5$ & $+1$ & ... & 10646.576 & ... & 93.93 & ... & ... & ... & ...
&\tabularnewline
4 & $5$ & $+2$ & ... & 10535.144 & ... & 94.92 & ... & ... & ... & ...
&\tabularnewline
4 & $5$ & $+3$ & ... & 10423.713 & ... & 95.94 & ... & ... & ... & ...
&\tabularnewline
4 & $5$ & $+4$ & 10314.595 & 10312.281 & 96.95 & 96.97 & $-0.02$ &
$-0.02$ & $+2.314$ & $0.0500$ & $f_{26}$\tabularnewline
 & \tabularnewline
 ... & ... & ... & ... & ... & ... & ... & ... & ... & ... & ...
\tabularnewline
 & \tabularnewline
4 & $2$ & $-4$ & ... & 7495.823 & ... & 133.41 & ... & ... & ... & ...
&\tabularnewline
4 & $2$ & $-3$ & 7412.898 & 7386.019 & 134.90 & 135.39 & $-0.36$ &
$-0.49$ & $+26.880$ & $0.0600$ & $f_{23}$\tabularnewline
4 & $2$ & $-2$ & ... & 7276.214 & ... & 137.43 & ... & ... & ... & ...
&\tabularnewline
4 & $2$ & $-1$ & ... & 7166.409 & ... & 139.54 & ... & ... & ... & ...
&\tabularnewline
4 & $2$ & $ 0$ & 7071.136 & 7056.605 & 141.42 & 141.71 & $-0.21$ &
$-0.29$ & $+14.531$ & $0.1300$ & $f_{9}$\tabularnewline
4 & $2$ & $+1$ & ... & 6946.800 & ... & 143.95 & ... & ... & ... & ...
&\tabularnewline
4 & $2$ & $+2$ & ... & 6836.995 & ... & 146.26 & ... & ... & ... & ...
&\tabularnewline
4 & $2$ & $+3$ & ... & 6727.191 & ... & 148.65 & ... & ... & ... & ...
&\tabularnewline
4 & $2$ & $+4$ & ... & 6617.386 & ... & 151.12 & ... & ... & ... & ...
&\tabularnewline
 & \tabularnewline
4 & $1$ & $-4$ & ... & 6726.790 & ... & 148.66 & ... & ... & ... & ...
&\tabularnewline
4 & $1$ & $-3$ & ... & 6626.935 & ... & 150.90 & ... & ... & ... & ...
&\tabularnewline
4 & $1$ & $-2$ & ... & 6527.080 & ... & 153.21 & ... & ... & ... & ...
&\tabularnewline
4 & $1$ & $-1$ & ... & 6427.225 & ... & 155.59 & ... & ... & ... & ...
&\tabularnewline
4 & $1$ & $ 0$ & ... & 6327.371 & ... & 158.04 & ... & ... & ... & ...
&\tabularnewline
4 & $1$ & $+1$ & ... & 6227.516 & ... & 160.58 & ... & ... & ... & ...
&\tabularnewline
4 & $1$ & $+2$ & 6163.328 & 6127.661 & 162.25 & 163.19 & $-0.58$ &
$-0.94$ & $+35.667$ & $0.0500$ & $f_{28}$\tabularnewline
4 & $1$ & $+3$ & ... & 6027.806 & ... & 165.90 & ... & ... & ... & ...
&\tabularnewline
4 & $1$ & $+4$ & 5916.110 & 5927.951 & 169.03 & 168.69 & $+0.20$ &
$+0.34$ & $-11.842$ & $0.1000$ & $f_{11}$\tabularnewline
 & \tabularnewline
4 & $0$ & $-4$ & ... & 5969.956 & ... & 167.51 & ... & ... & ... & ...
&\tabularnewline
4 & $0$ & $-3$ & ... & 5857.737 & ... & 170.71 & ... & ... & ... & ...
&\tabularnewline
4 & $0$ & $-2$ & ... & 5745.518 & [173.59] & 174.05 & ... & ... & ... &
... & [$f_8$]\tabularnewline
4 & $0$ & $-1$ & 5621.135 & 5633.299 & 177.90 & 177.52 & $+0.22$ &
$+0.38$ & $-12.164$ & $0.0700$ & $f_{20}$\tabularnewline
4 & $0$ & $ 0$ & 5516.633 & 5521.080 & 181.27 & 181.12 & $+0.08$ &
$+0.15$ & $-4.447$ & $0.1000$ & $f_{12}$\tabularnewline
4 & $0$ & $+1$ & 5401.026 & 5408.861 & 185.15 & 184.88 & $+0.14$ &
$+0.27$ & $-7.835$ & $0.0600$ & $f_{25}$\tabularnewline
4 & $0$ & $+2$ & ... & 5296.642 & ... & 188.80 & ... & ... & ... & ...
&\tabularnewline
4 & $0$ & $+3$ & ... & 5184.423 & ... & 192.89 & ... & ... & ... & ...
&\tabularnewline
4 & $0$ & $+4$ & ... & 5072.205 & ... & 197.15 & ... & ... & ... & ...
&\tabularnewline
 & \tabularnewline
4 & $-1$ & $-4$ & 5444.251 & 5450.925 & 183.68 & 183.46 & $+0.12$ &
$+0.22$ & $-6.674$ & $0.1700$ & $f_{7}$\tabularnewline
4 & $-1$ & $-3$ & 5356.473 & 5338.149 & 186.69 & 187.33 & $-0.34$ &
$-0.64$ & $+18.324$ & $0.0800$ & $f_{16}$\tabularnewline
4 & $-1$ & $-4$ & 5218.934 & 5225.373 & 191.61 & 191.37 & $+0.12$ &
$+0.24$ & $-6.439$ & $0.0600$ & $f_{21}$\tabularnewline
4 & $-1$ & $-1$ & 5111.168 & 5112.598 & 195.65 & 195.60 & $+0.03$ &
$+0.05$ & $-1.430$ & $0.0500$ & $f_{27}$\tabularnewline
4 & $-1$ & $ 0$ & ... & 4999.822 & ... & 200.01 & ... & ... & ... & ...
&\tabularnewline
4 & $-1$ & $+1$ & 4885.198 & 4887.046 & 204.70 & 204.62 & $+0.04$ &
$+0.08$ & $-1.848$ & $0.0700$ & $f_{18}$\tabularnewline
4 & $-1$ & $+2$ & ... & 4774.270 & ... & 209.46 & ... & ... & ... & ...
&\tabularnewline
4 & $-1$ & $+3$ & ... & 4661.494 & ... & 214.52 & ... & ... & ... & ...
&\tabularnewline
4 & $-1$ & $+4$ & ... & 4548.718 & ... & 219.84 & ... & ... & ... & ...
&\tabularnewline

\end{longtable}
\end{center}
\twocolumn
\normalsize

\section{Probability density functions of stellar parameters with known model uncertainties}
\label{annex2}
\begin{figure*}[!ht]
\begin{center}
\includegraphics[scale=0.37,angle=0]{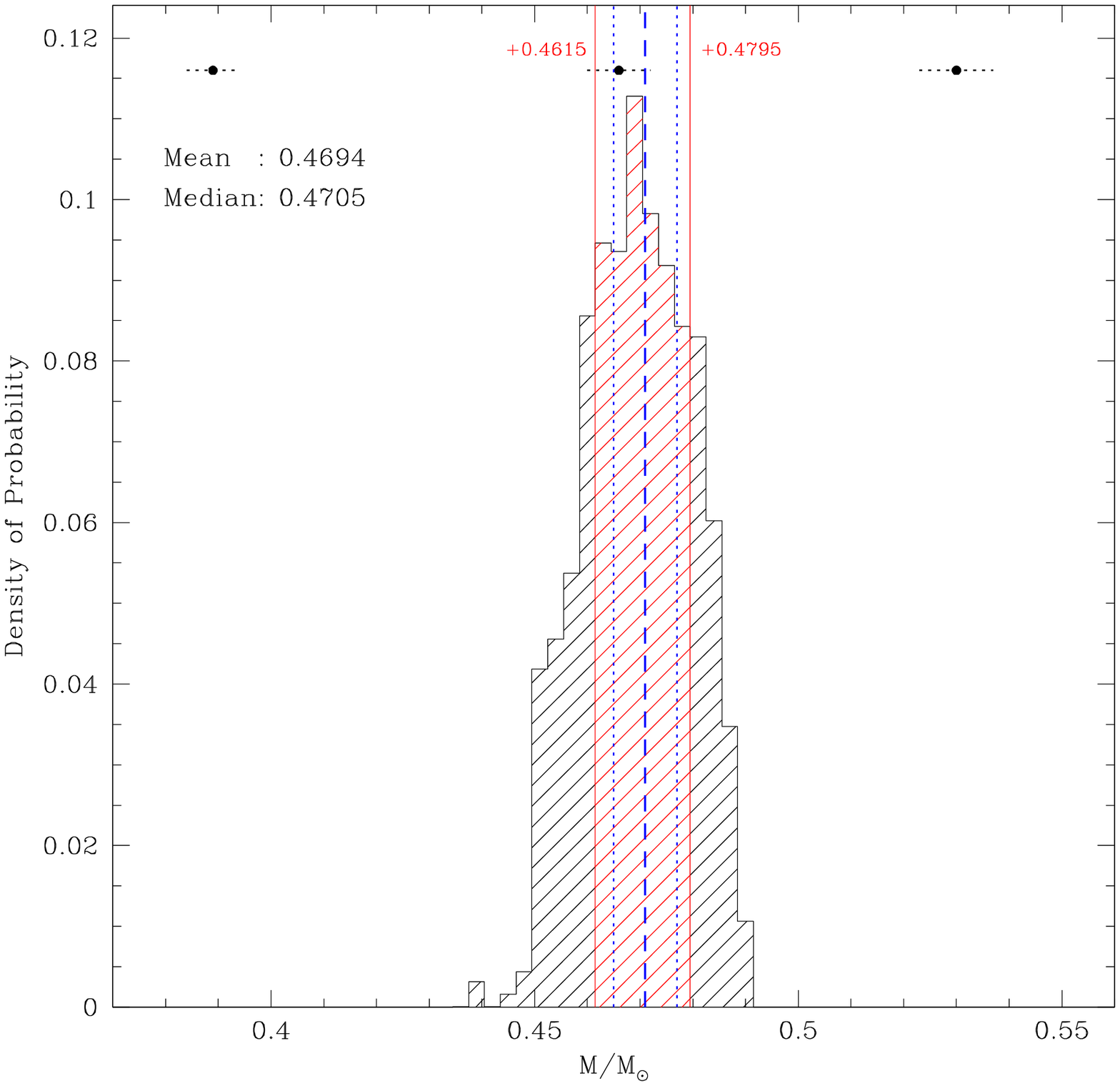}
\includegraphics[scale=0.37,angle=0]{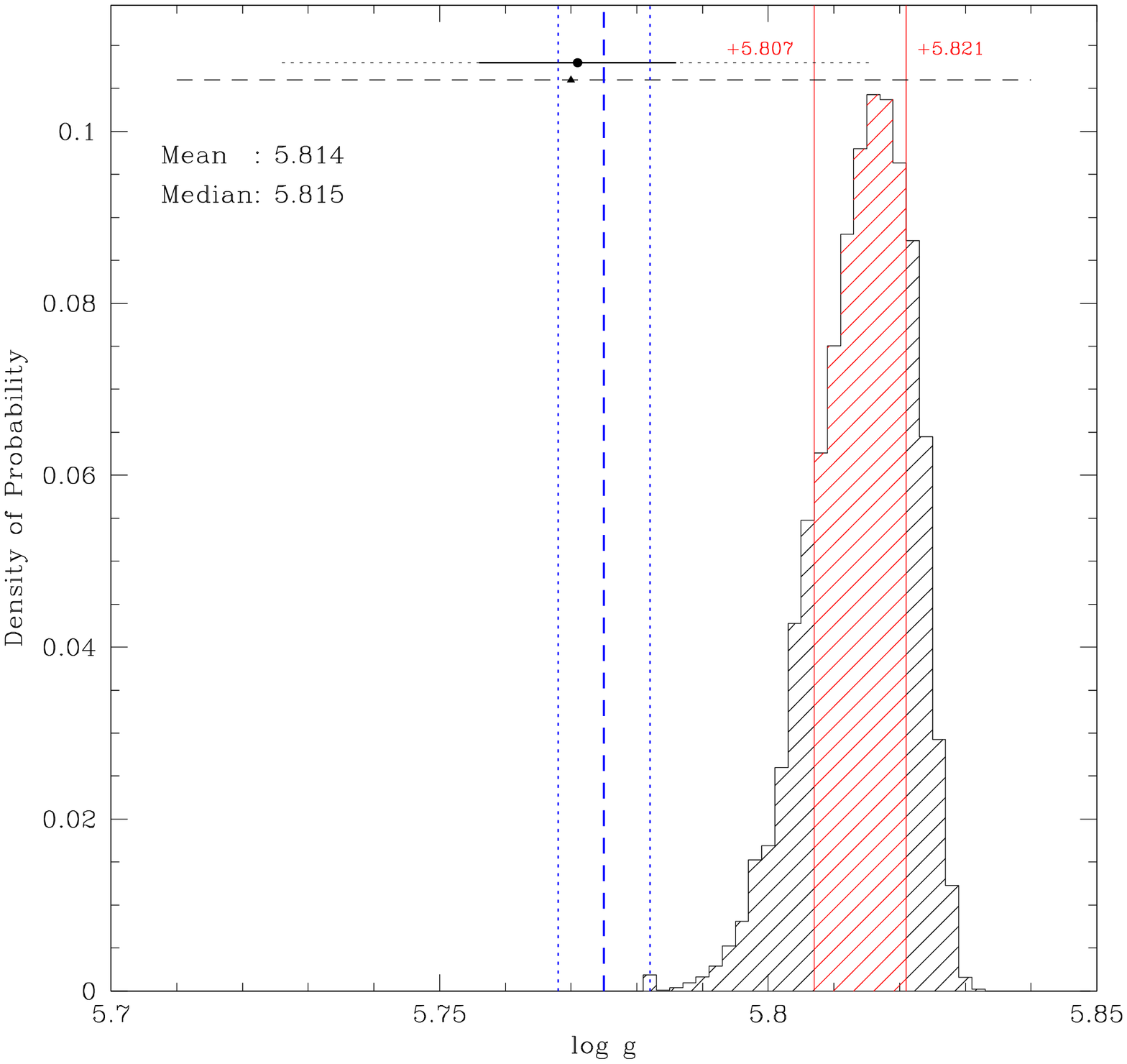}\\
\includegraphics[scale=0.37,angle=0]{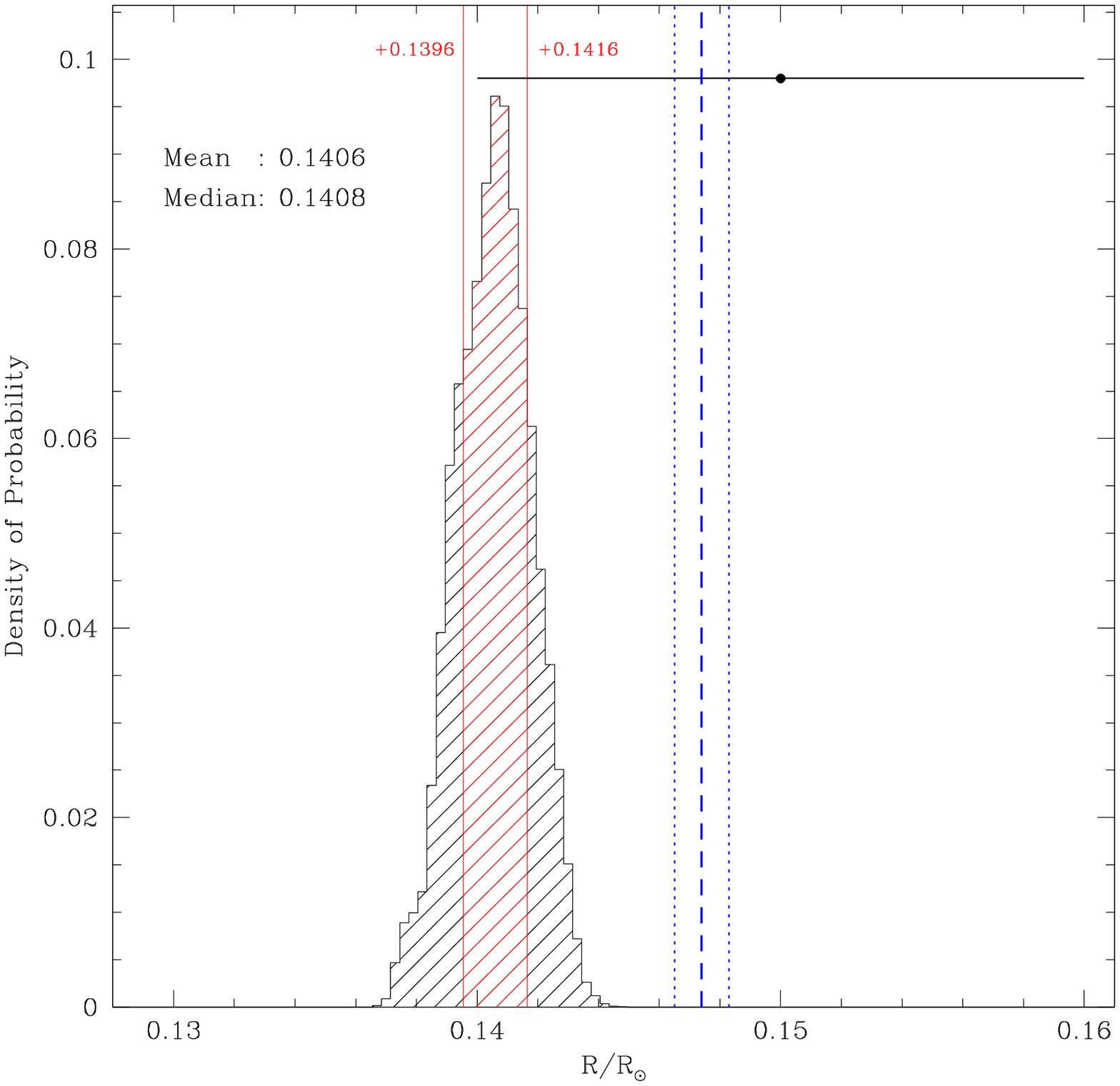}
\includegraphics[scale=0.37,angle=0]{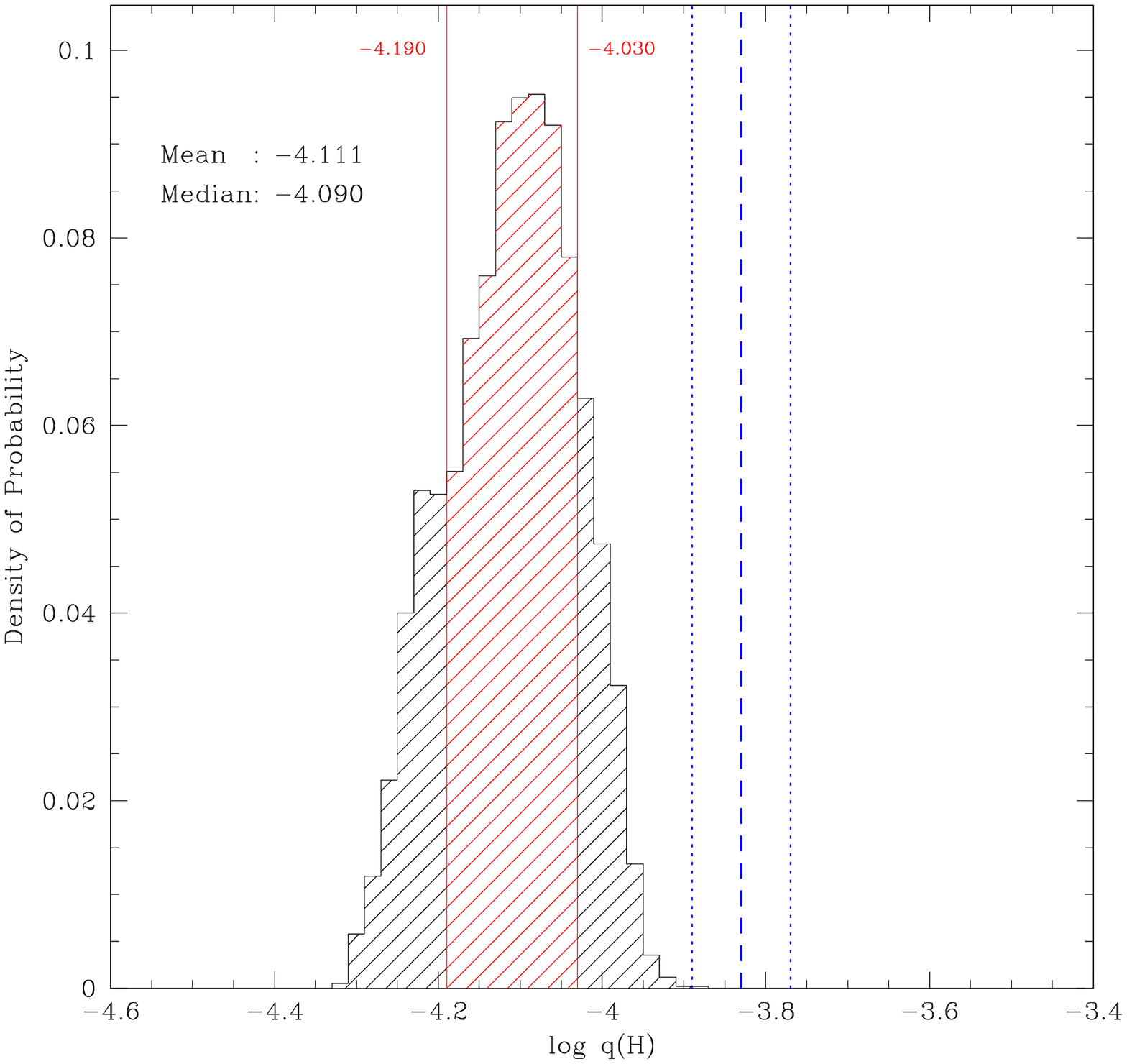}\\
\includegraphics[scale=0.37,angle=0]{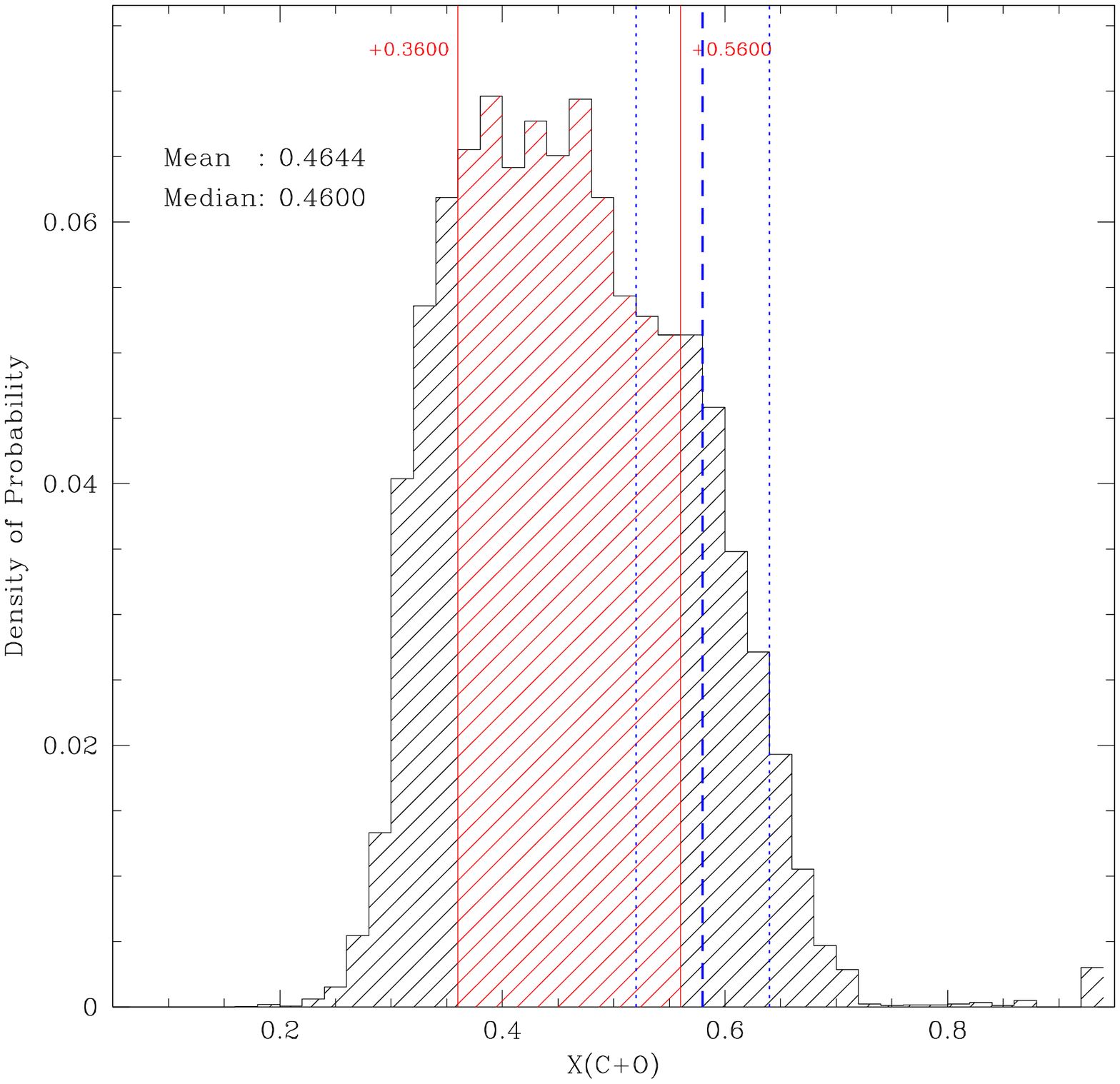}
\includegraphics[scale=0.37,angle=0]{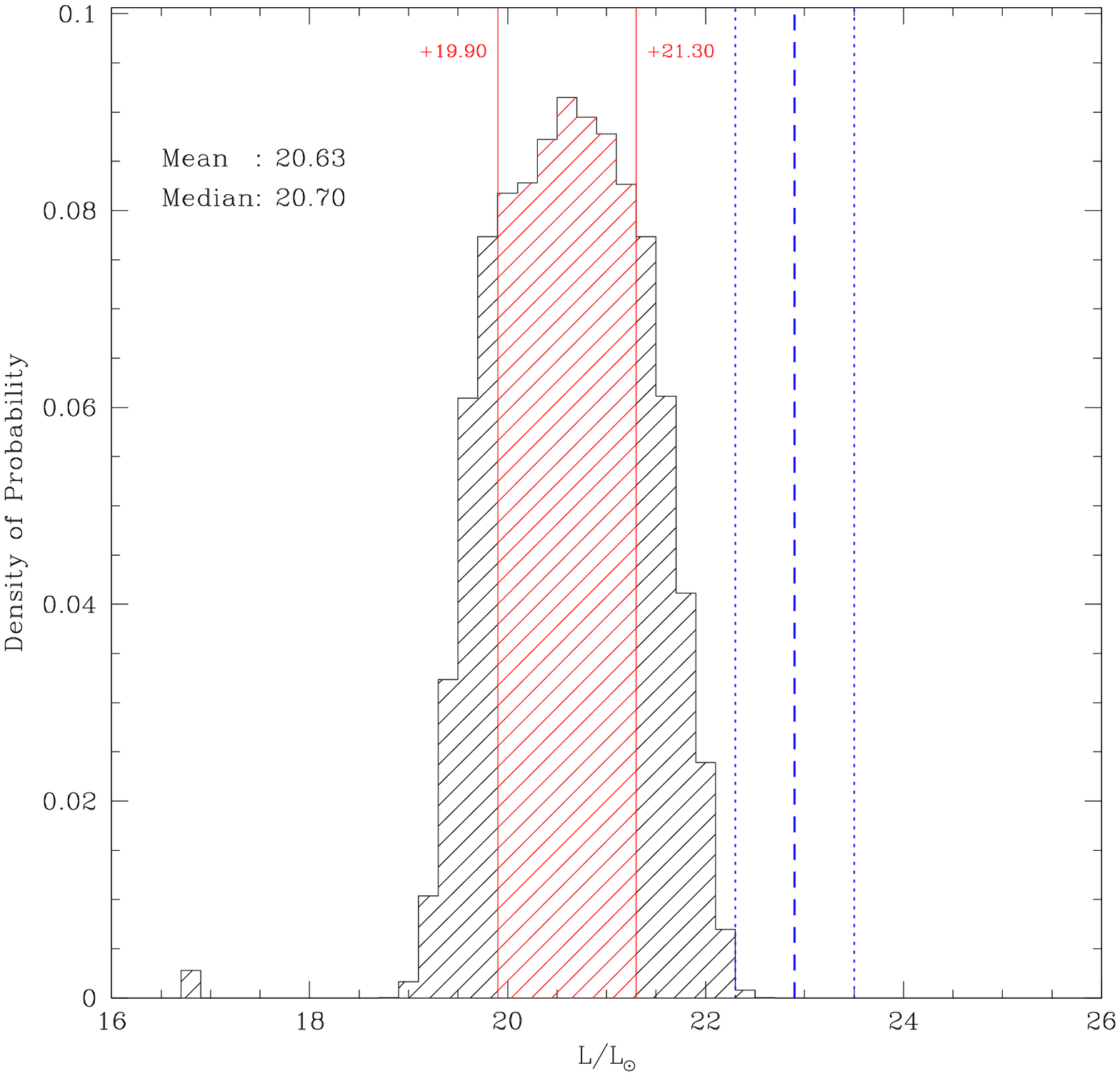}
\end{center}
\caption{\label{fB3}Probability density functions derived from asteroseismology using models with a uniform distribution if iron in solar proportions. The parameters inferred for the sdB component of PG 1336--018 are, from the upper-left panel to the lower-right panel, the mass $M$, $\log g$, the radius $R$, loq $q$(H), log $q$(core), $X_{\rm core}$(C+O), and $L_*/L_{\odot}$. The red hatched regions
between the two vertical solid red lines shows the 1$\sigma$ range containing 68.3\% of the distribution. In the panel showing the mass, the filled circles with the dotted lines are the 3 orbital solutions 
for the mass of the sdB component proposed by Vu{\v c}kovi\'c et al. (2007) with their $1\sigma$ uncertainties. In the panel showing $\log g$, The filled circle with the solid line and dotted line indicate the value derived from spectroscopy with its $1\sigma$ and $3\sigma$ uncertainties, respectively. Finally, in the panel showing the radius, The filled circle with the solid horizontal line indicates the $1\sigma$ error associated to the relevant orbital solution from Vu{\v c}kovi\'c et al. (2007). In each panel, the blue vertical dashed and dotted lines show the seismic solution and its $1\sigma$ uncertainty obtained with our standard 3G models.  
}  
\end{figure*}

\begin{figure*}[!ht]
\begin{center}
\includegraphics[scale=0.37,angle=0]{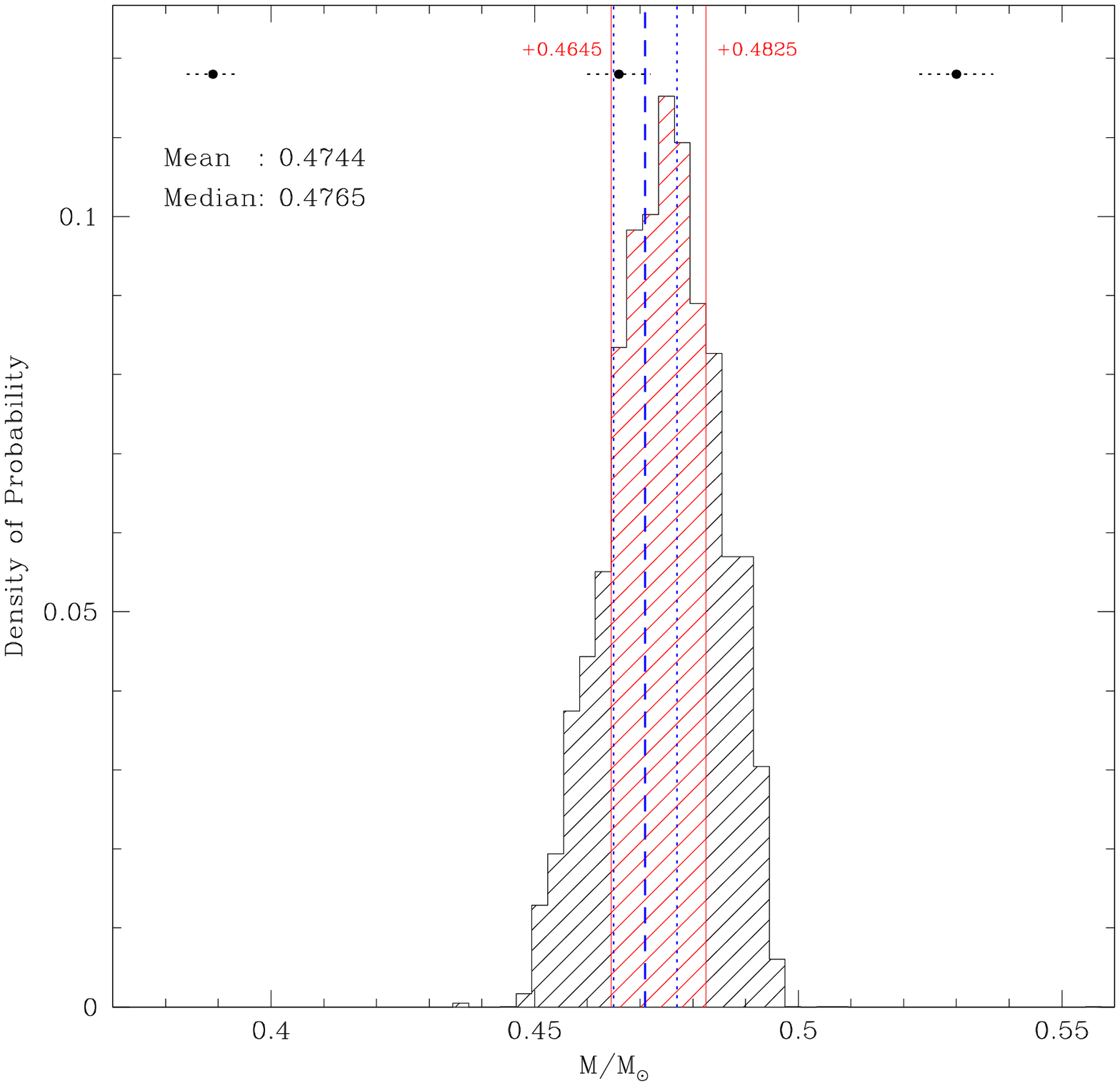}
\includegraphics[scale=0.37,angle=0]{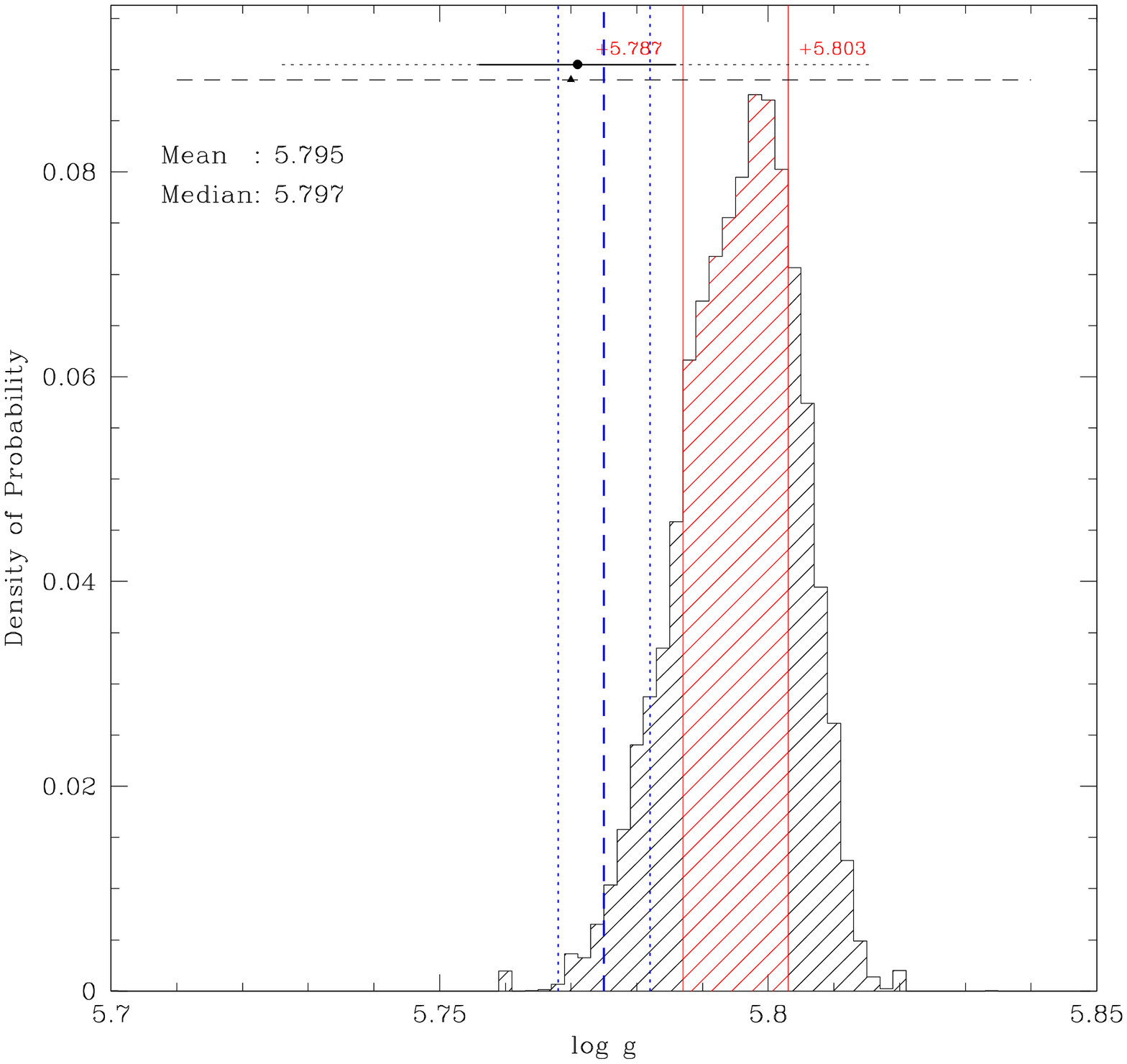}\\
\includegraphics[scale=0.37,angle=0]{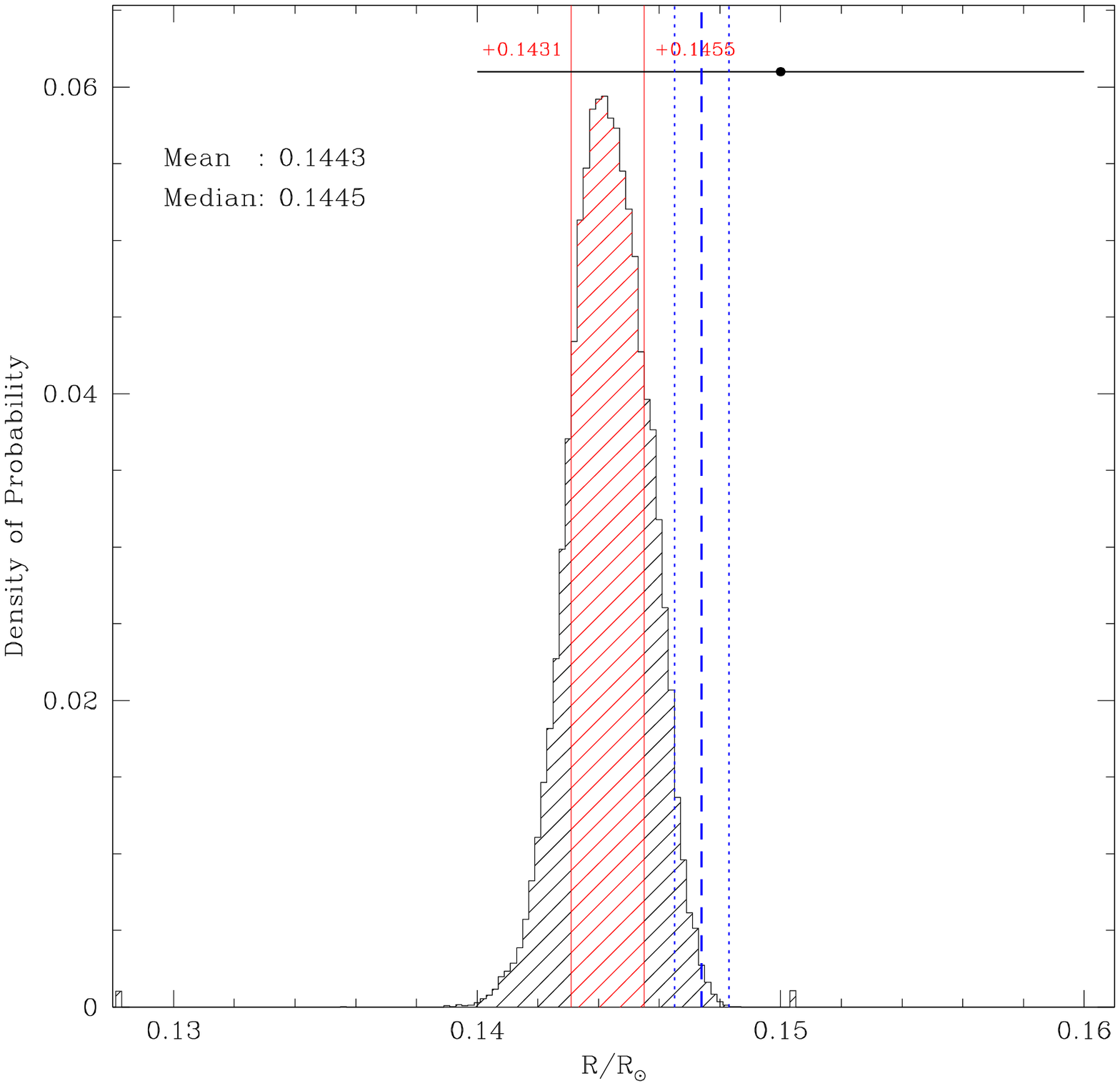}
\includegraphics[scale=0.37,angle=0]{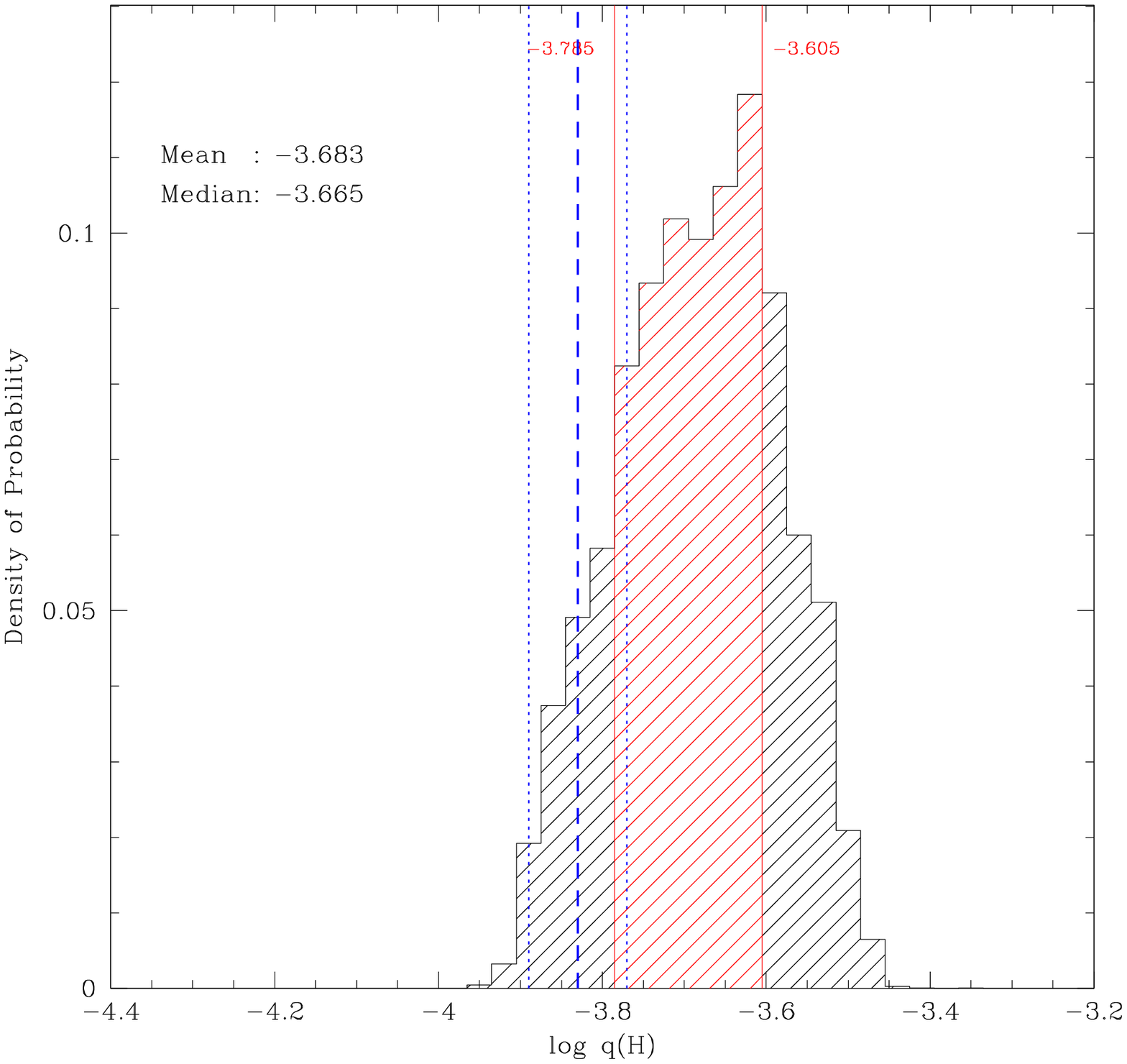}\\
\includegraphics[scale=0.37,angle=0]{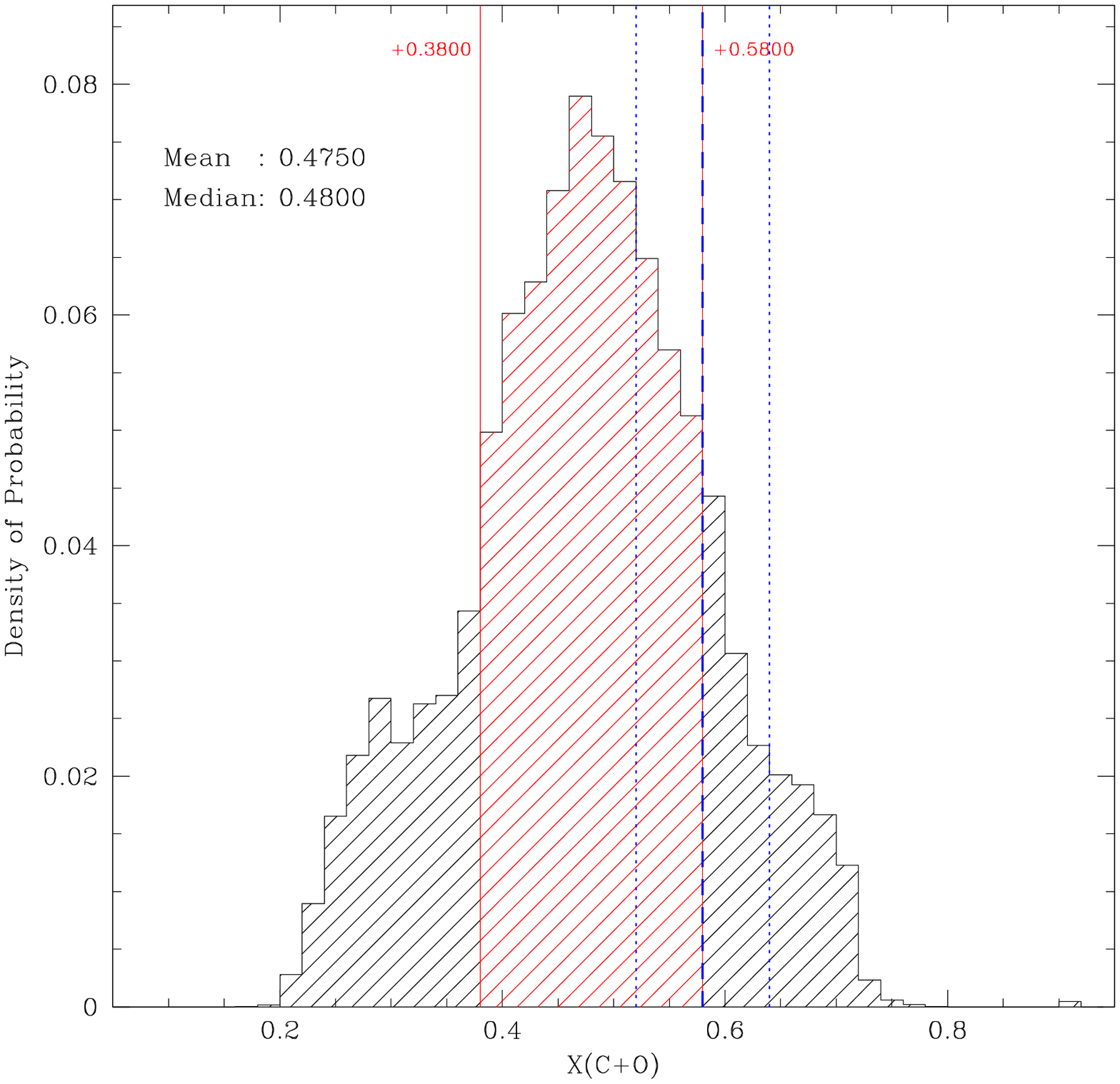}
\includegraphics[scale=0.37,angle=0]{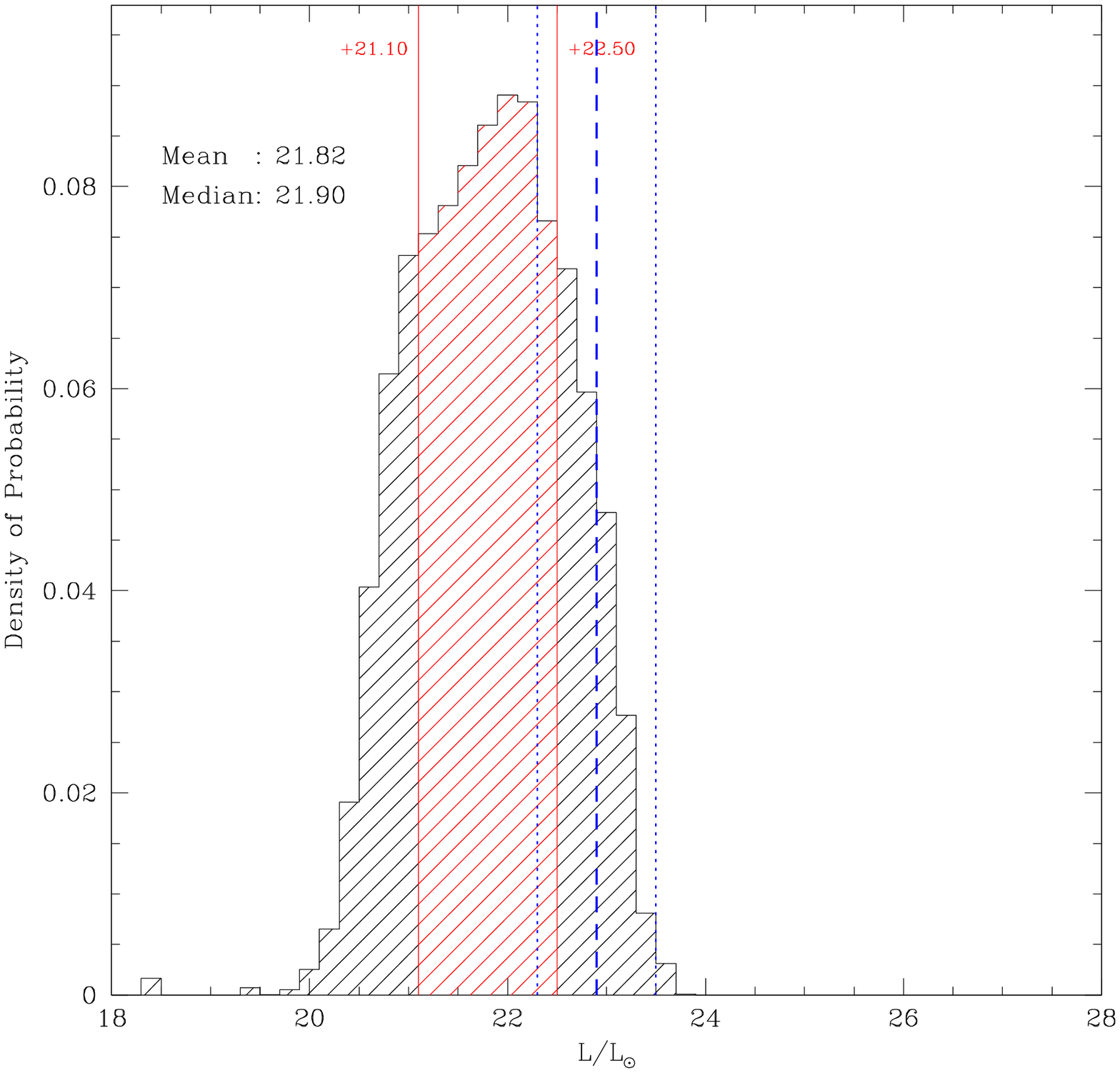}
\end{center}
\caption{\label{fB4} Same as Fig. \ref{fB3}, but using models with a nonuniform iron abundance profile reduced by a factor of four relative to the value expected at diffusive equilibrium.}  
\end{figure*}

\begin{figure*}[!ht]
\begin{center}
\includegraphics[scale=0.37,angle=0]{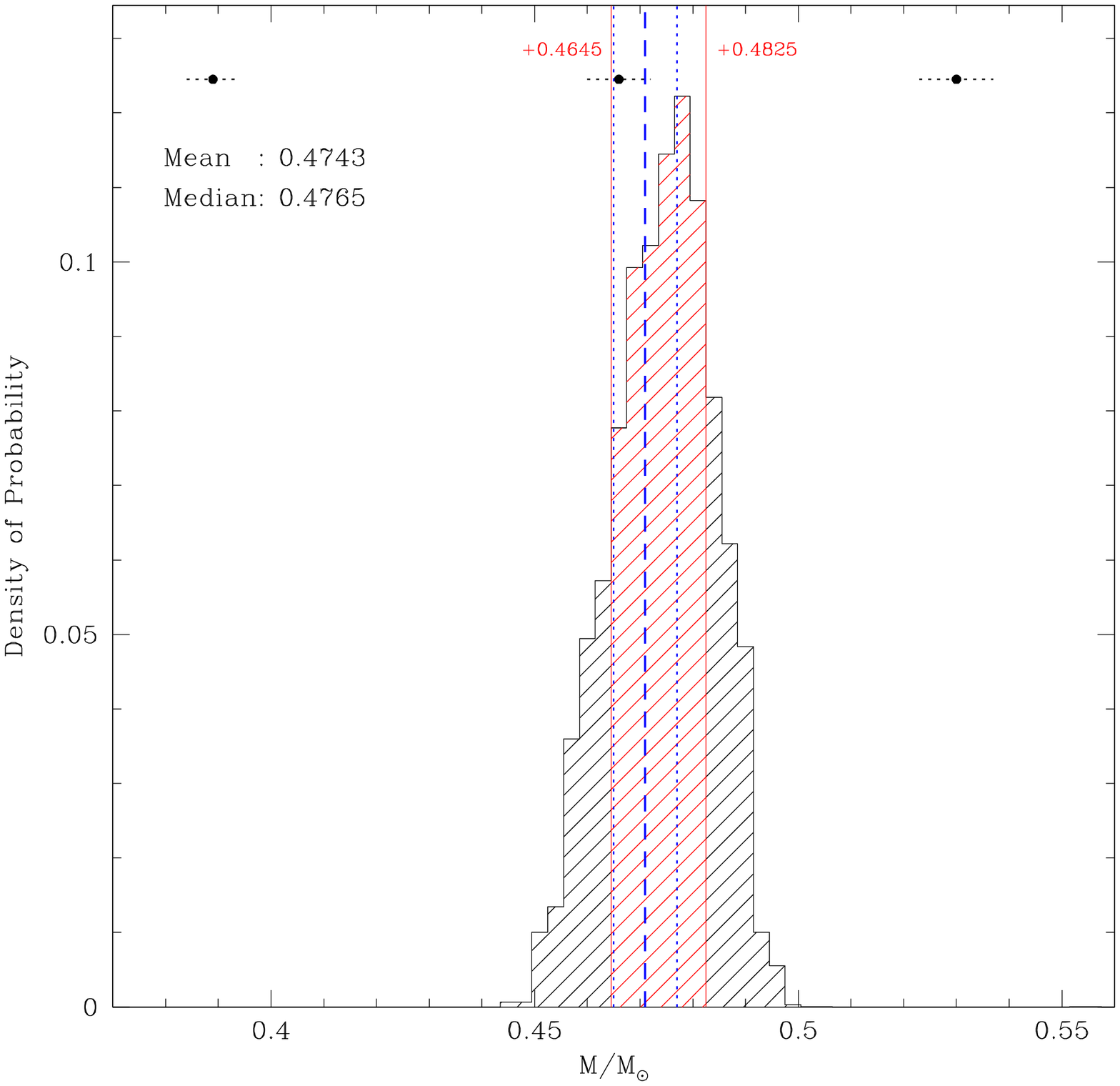}
\includegraphics[scale=0.37,angle=0]{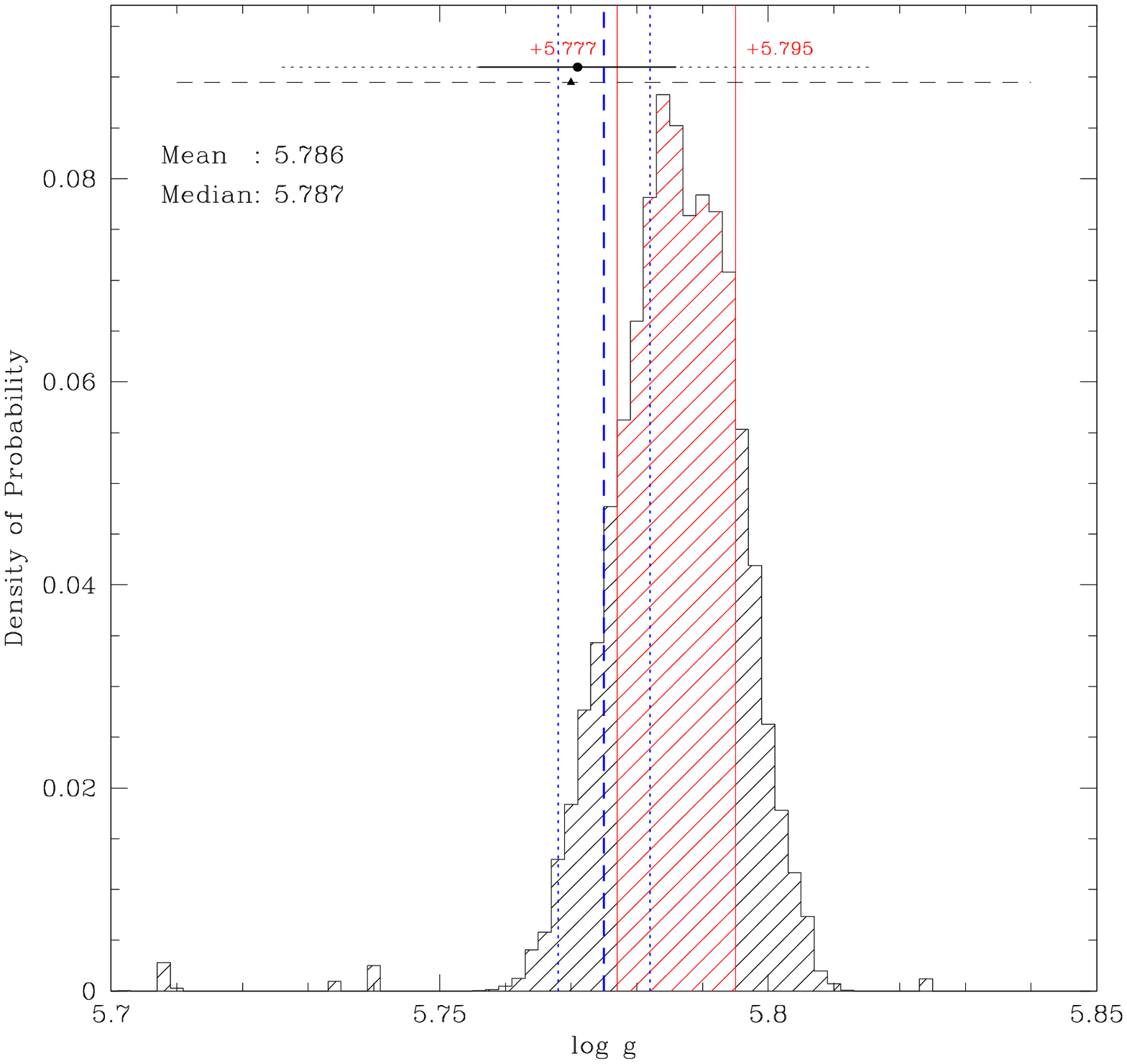}\\
\includegraphics[scale=0.37,angle=0]{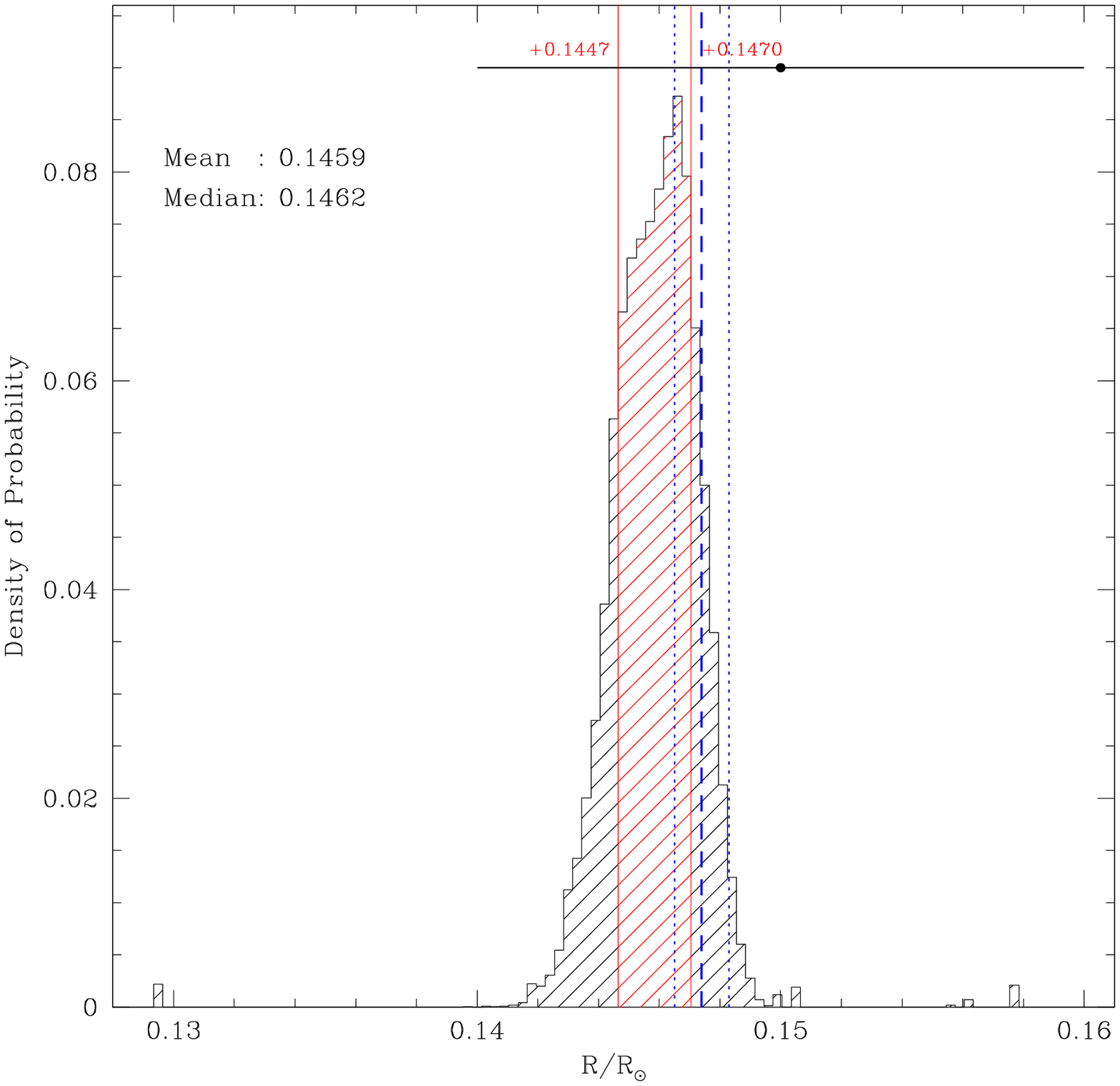}
\includegraphics[scale=0.37,angle=0]{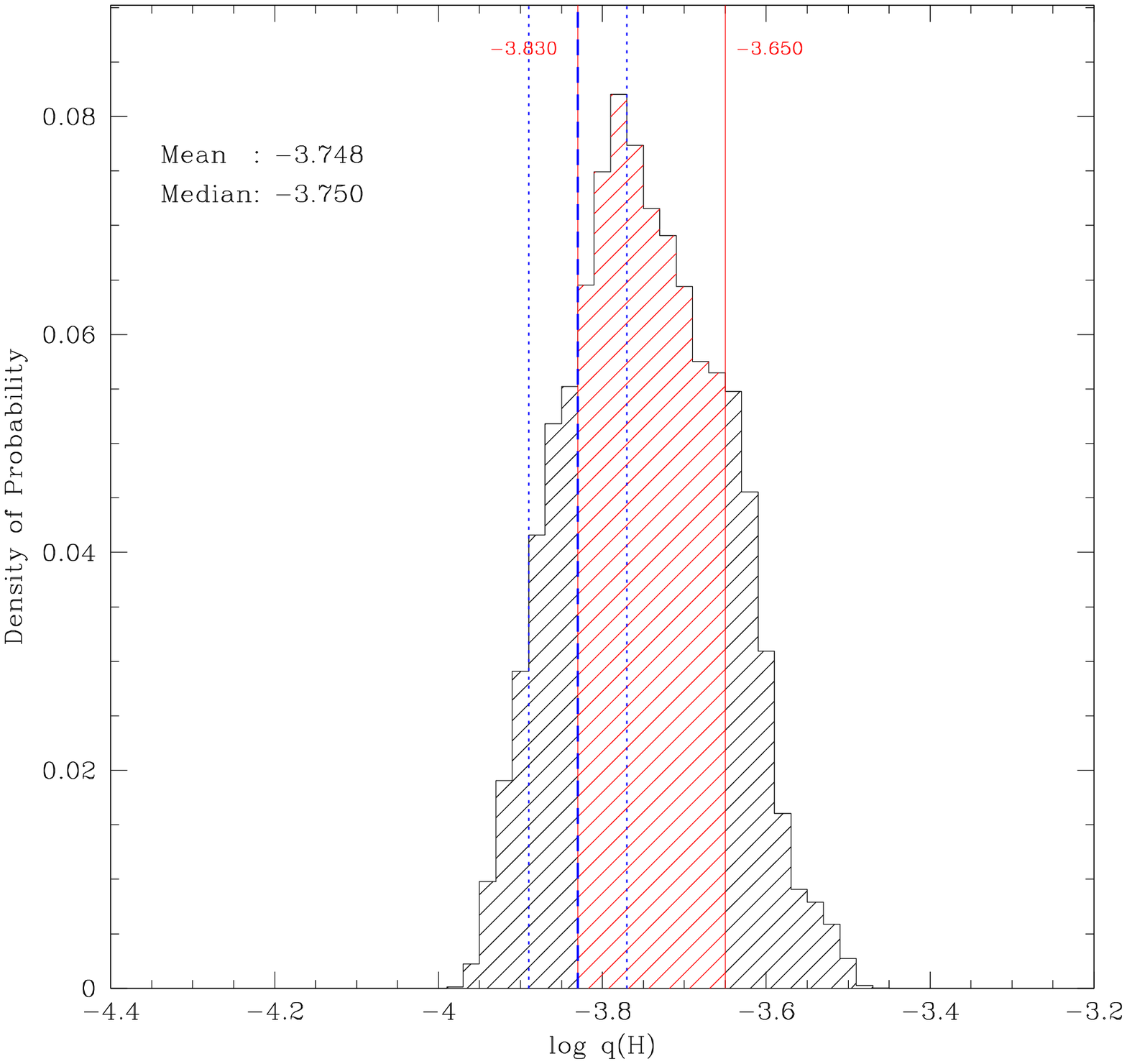}\\
\includegraphics[scale=0.37,angle=0]{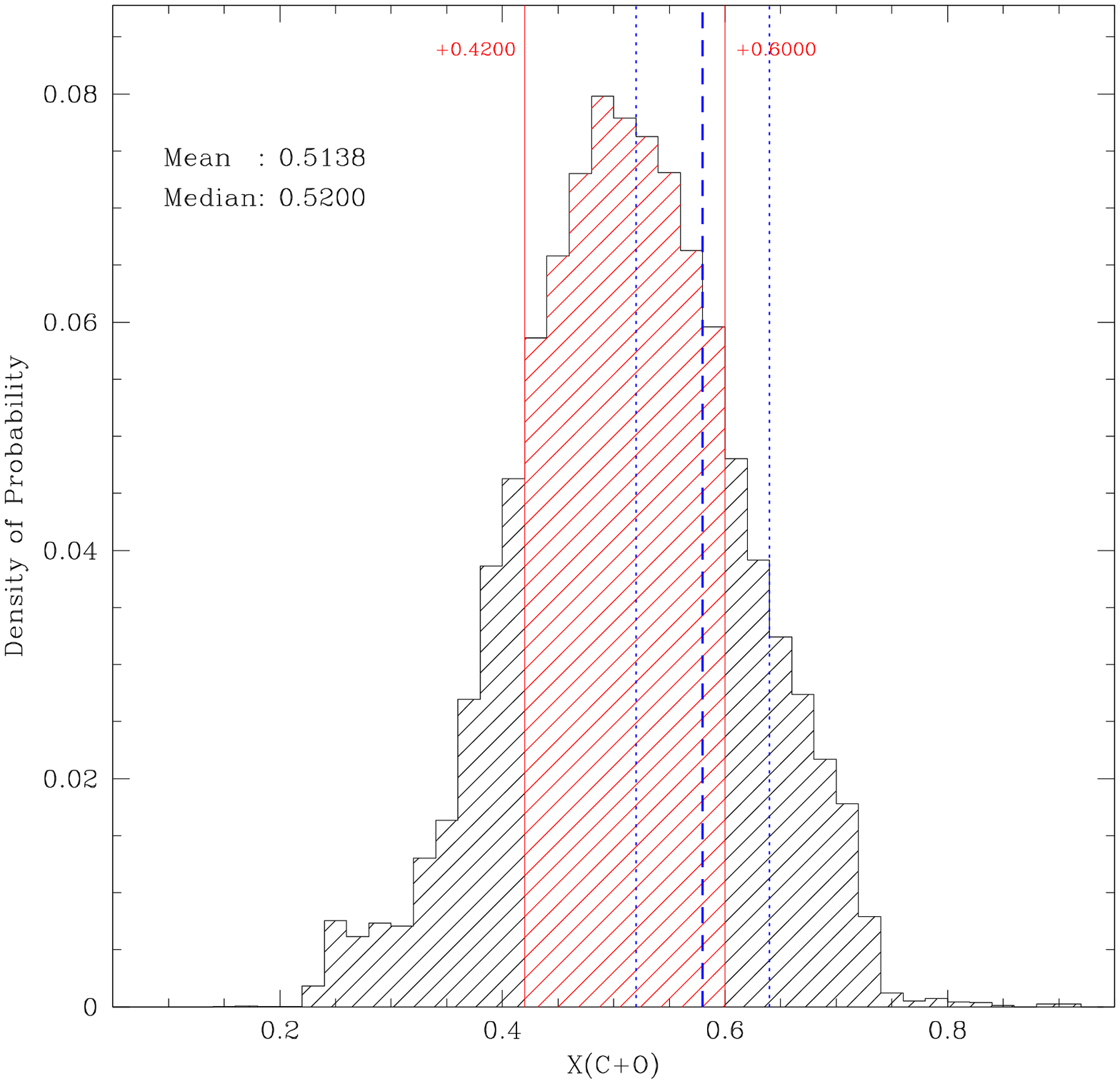}
\includegraphics[scale=0.37,angle=0]{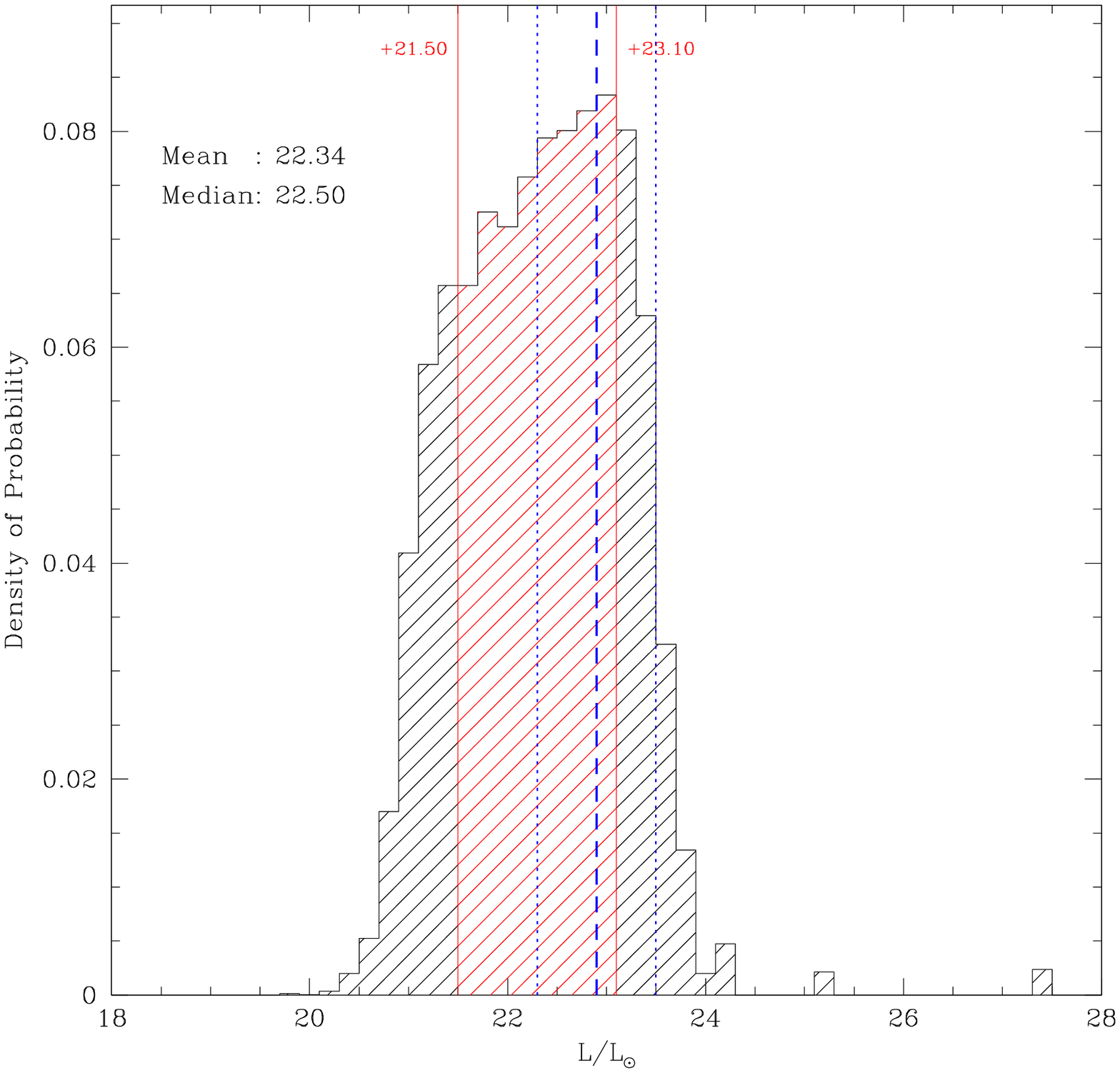}
\end{center}
\caption{\label{fB5} Same as Fig. \ref{fB3}, but using models with a nonuniform iron abundance profile reduced by a factor of two relative to the value expected at diffusive equilibrium.}  
\end{figure*}

\begin{figure*}[!ht]
\begin{center}
\includegraphics[scale=0.37,angle=0]{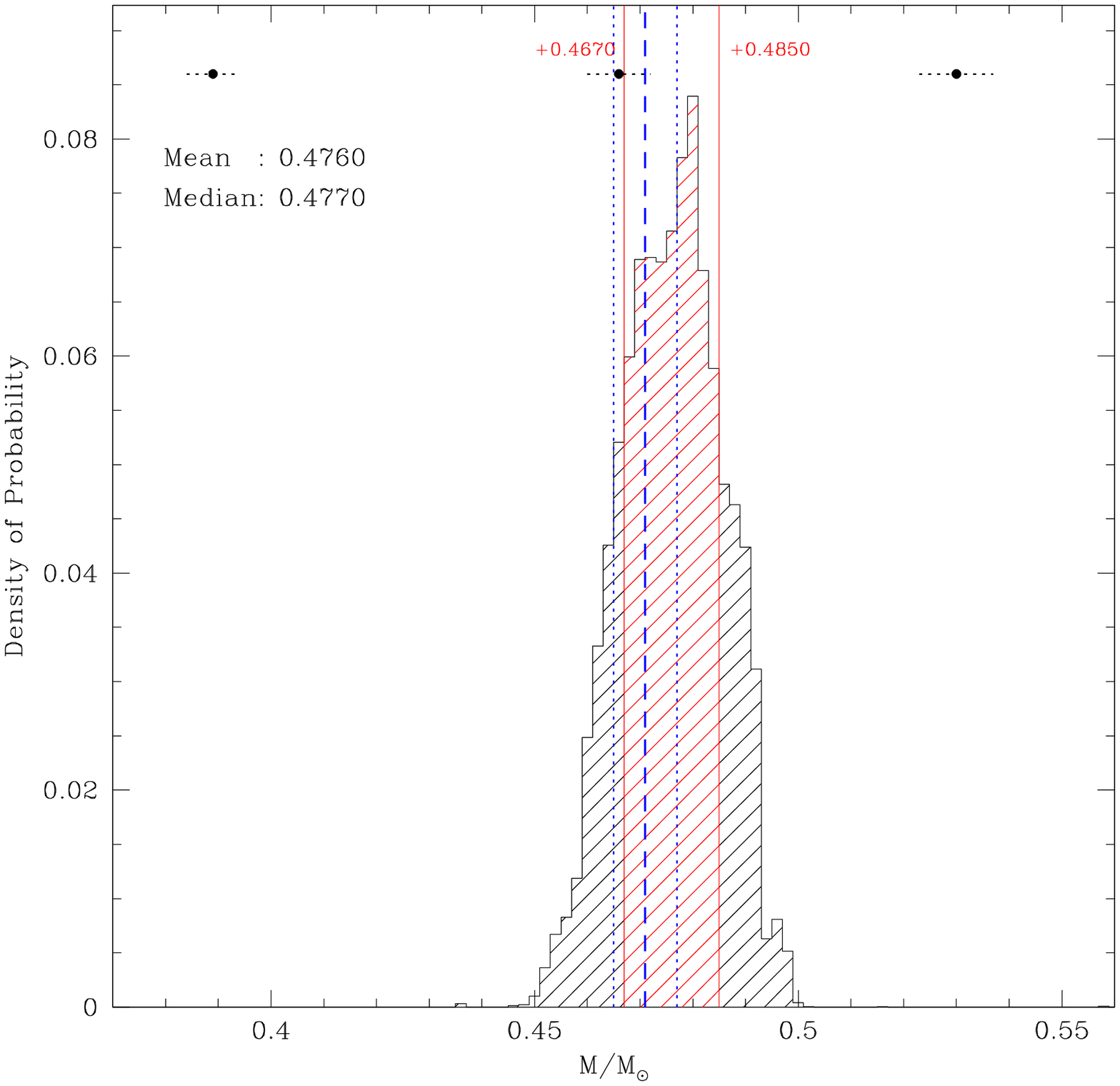}
\includegraphics[scale=0.37,angle=0]{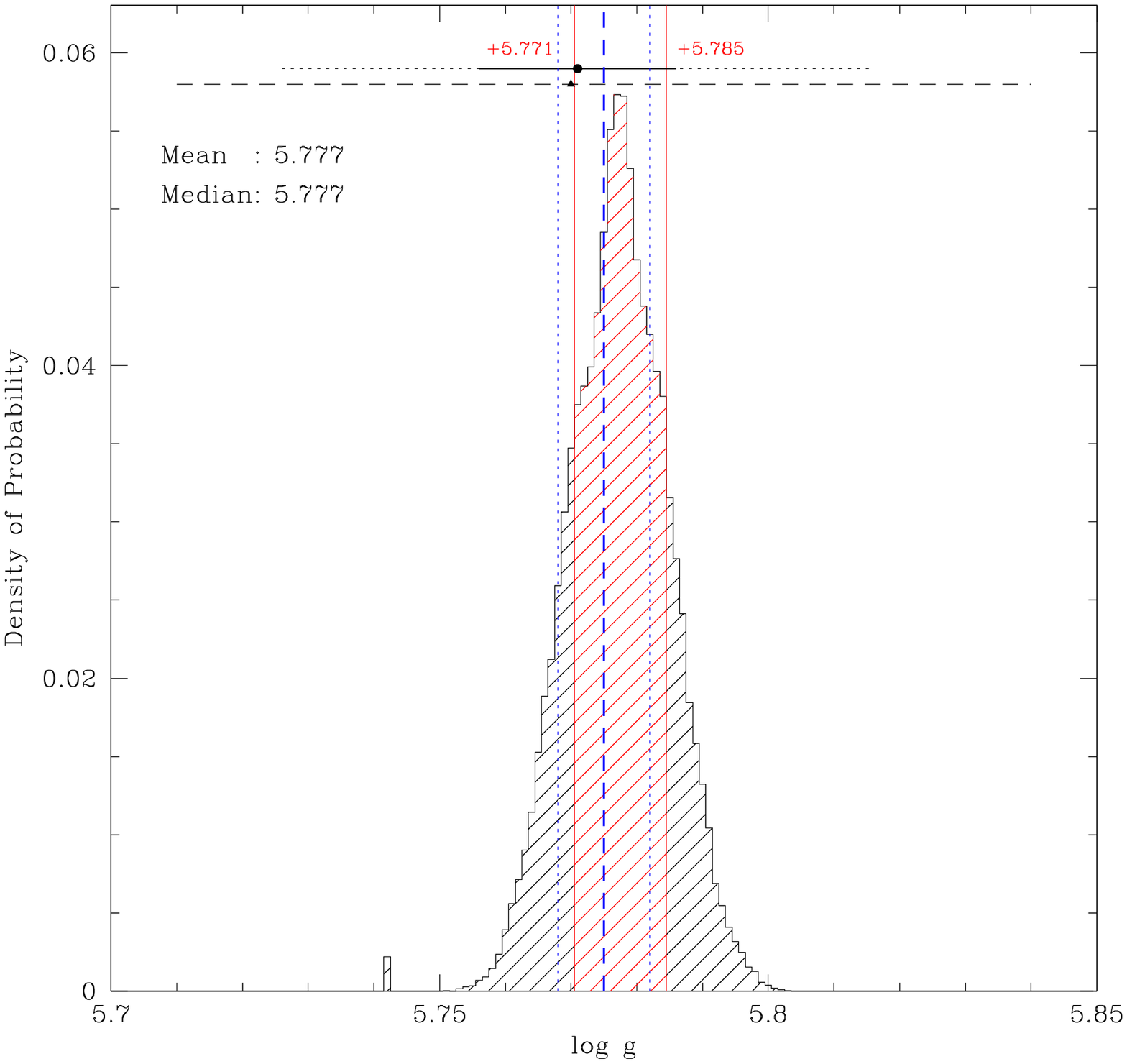}\\
\includegraphics[scale=0.37,angle=0]{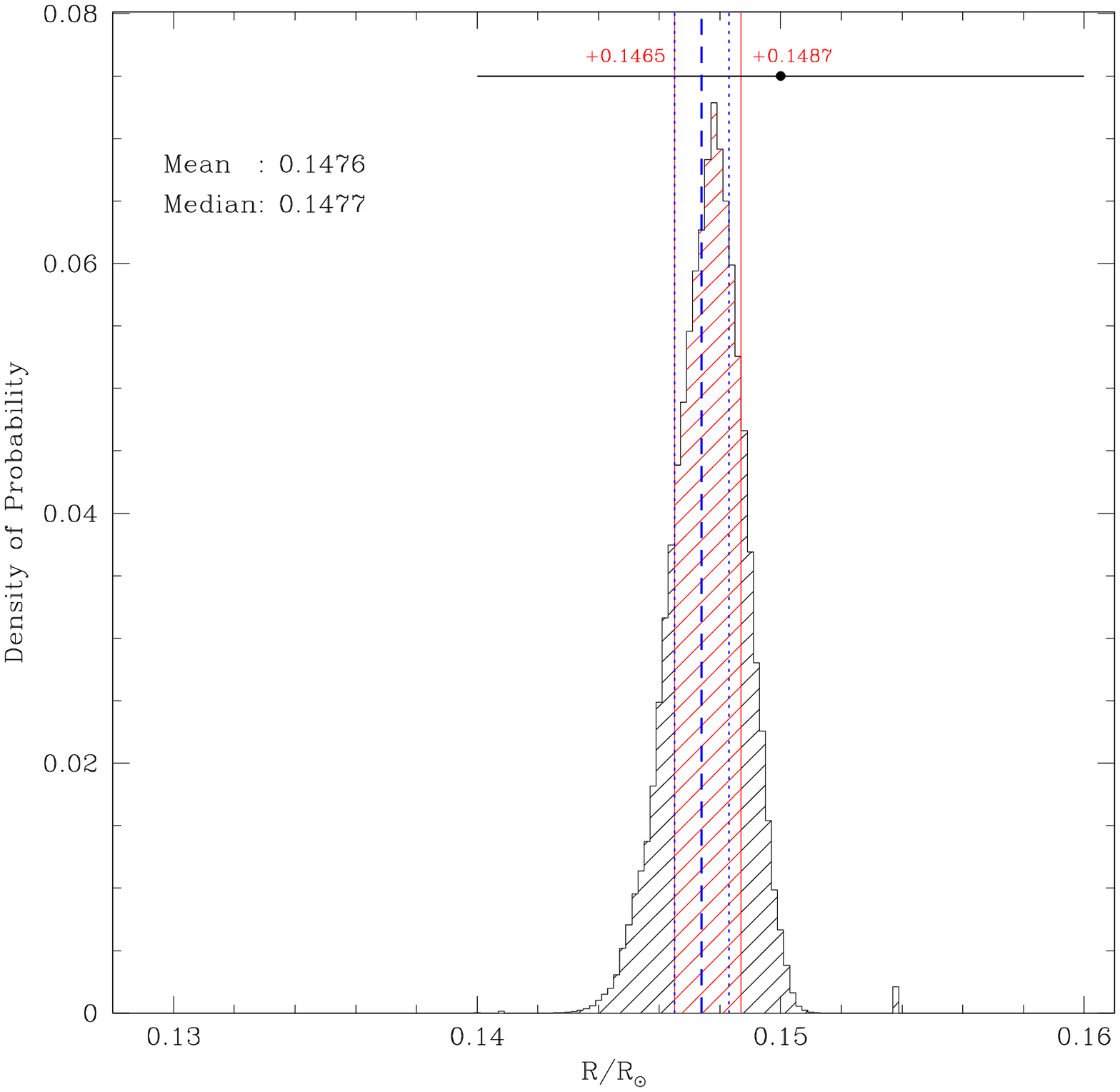}
\includegraphics[scale=0.37,angle=0]{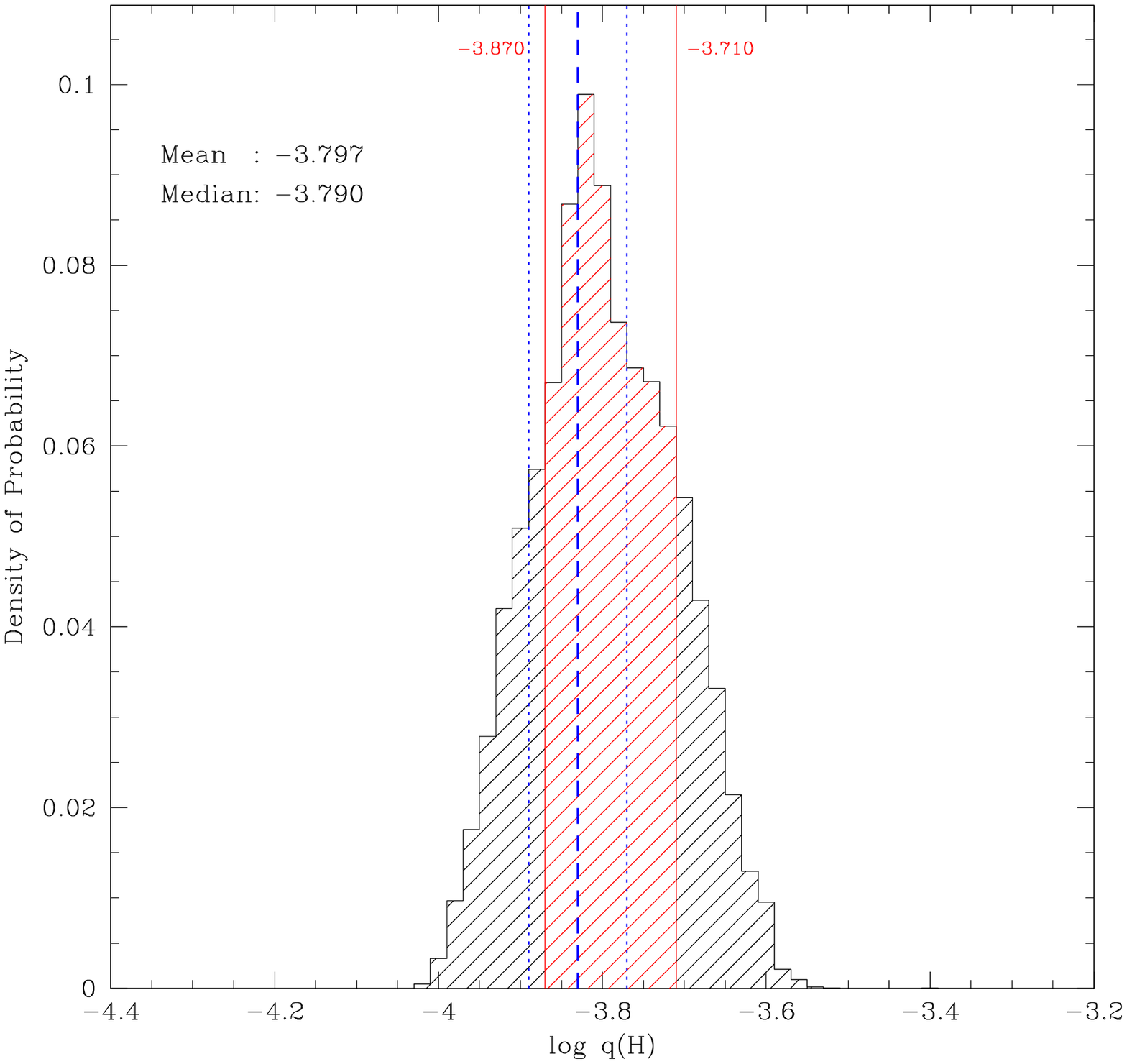}\\
\includegraphics[scale=0.37,angle=0]{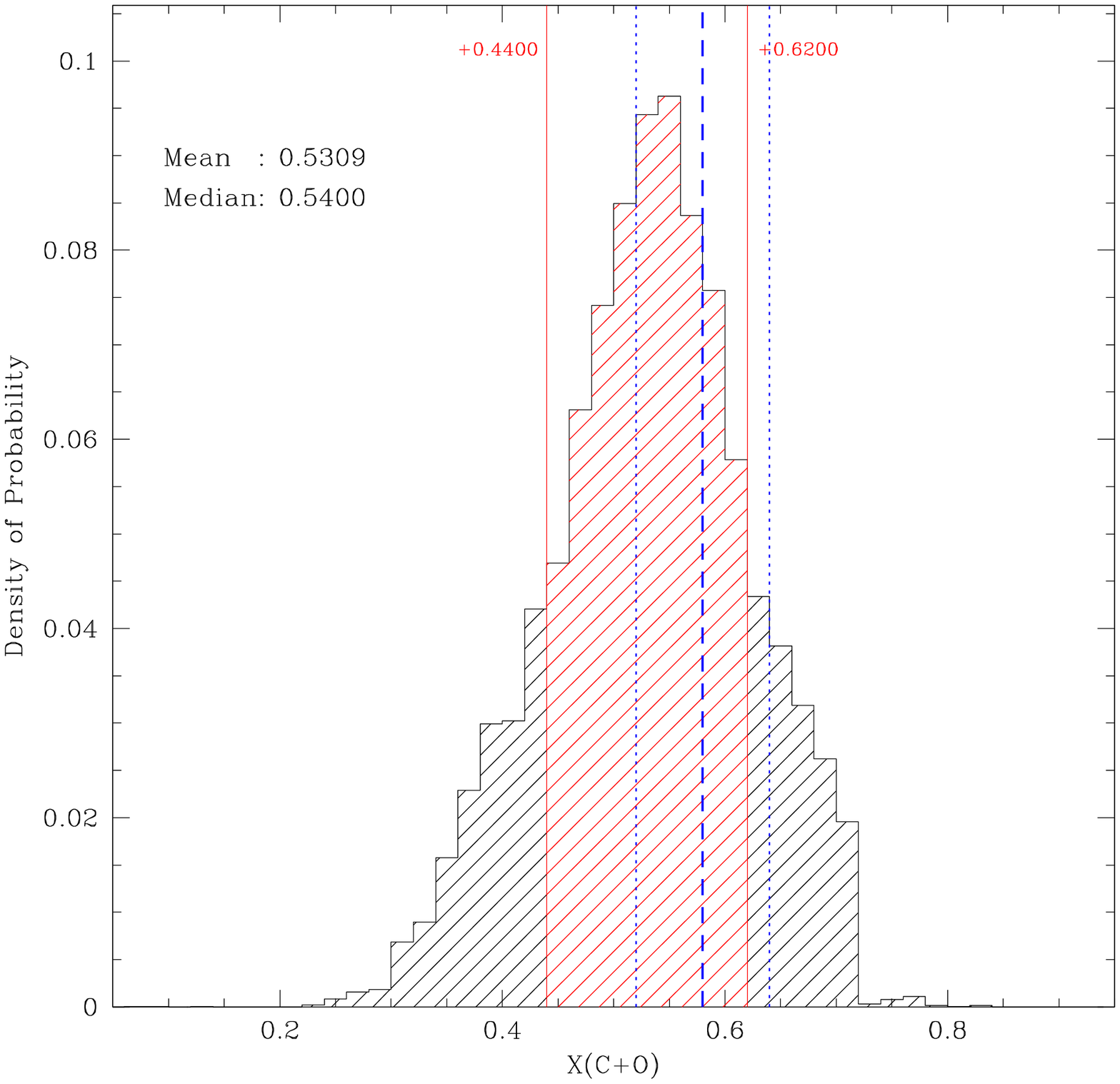}
\includegraphics[scale=0.37,angle=0]{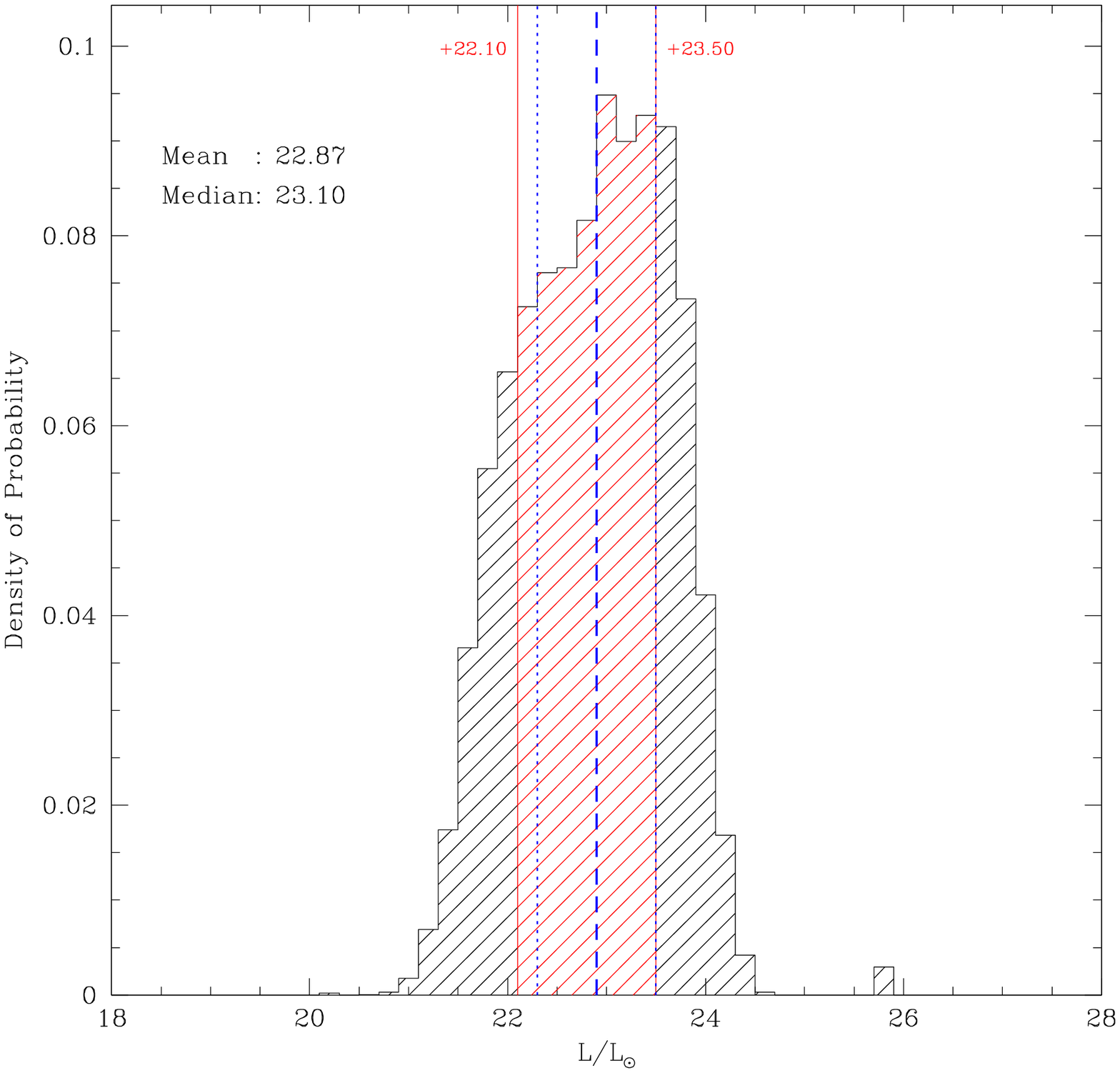}
\end{center}
\caption{\label{fB7} Same as Fig. \ref{fB3}, but using models with a smooth He/H transition.}  
\end{figure*}

\begin{figure*}[!ht]
\begin{center}
\includegraphics[scale=0.37,angle=0]{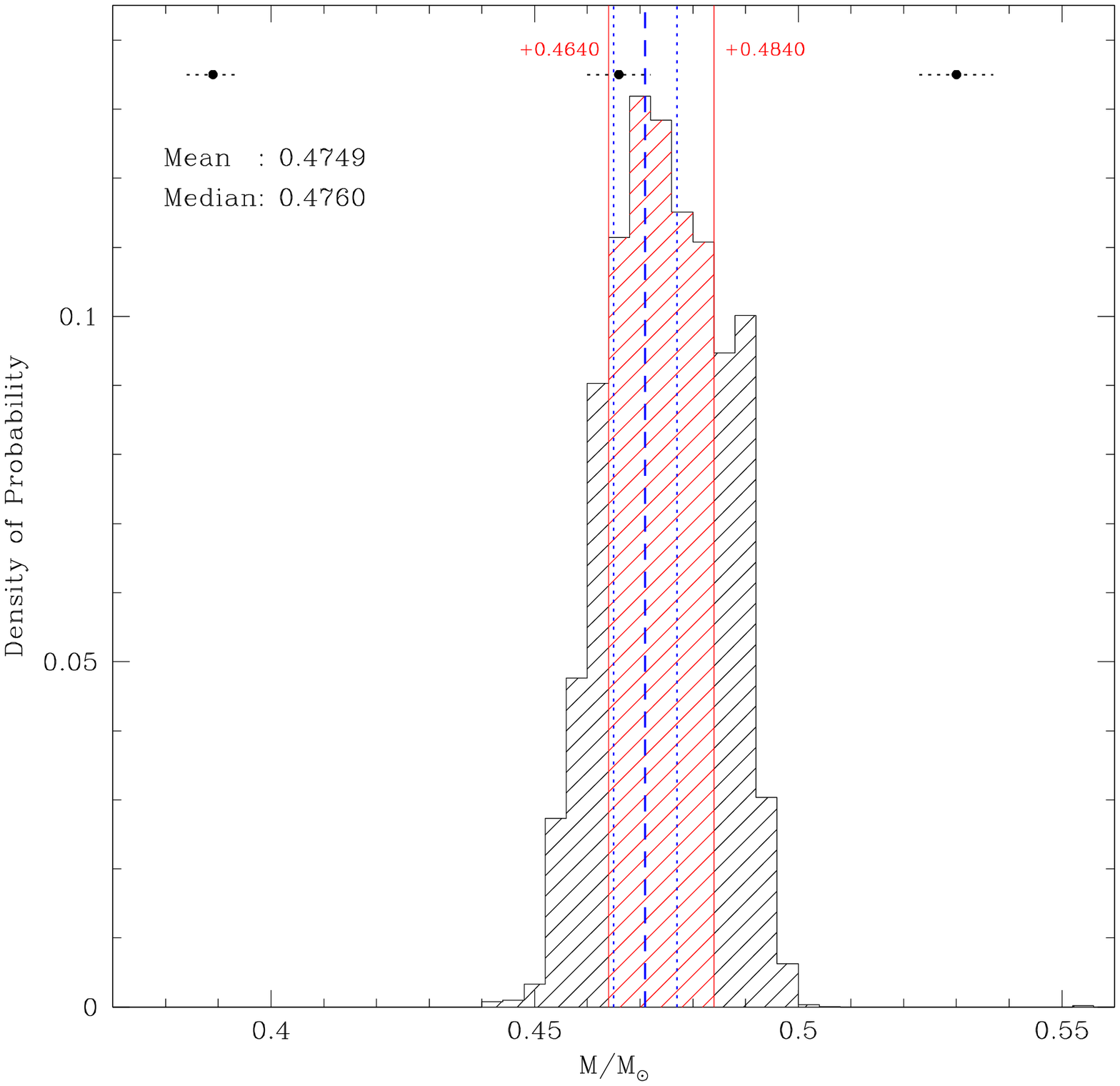}
\includegraphics[scale=0.37,angle=0]{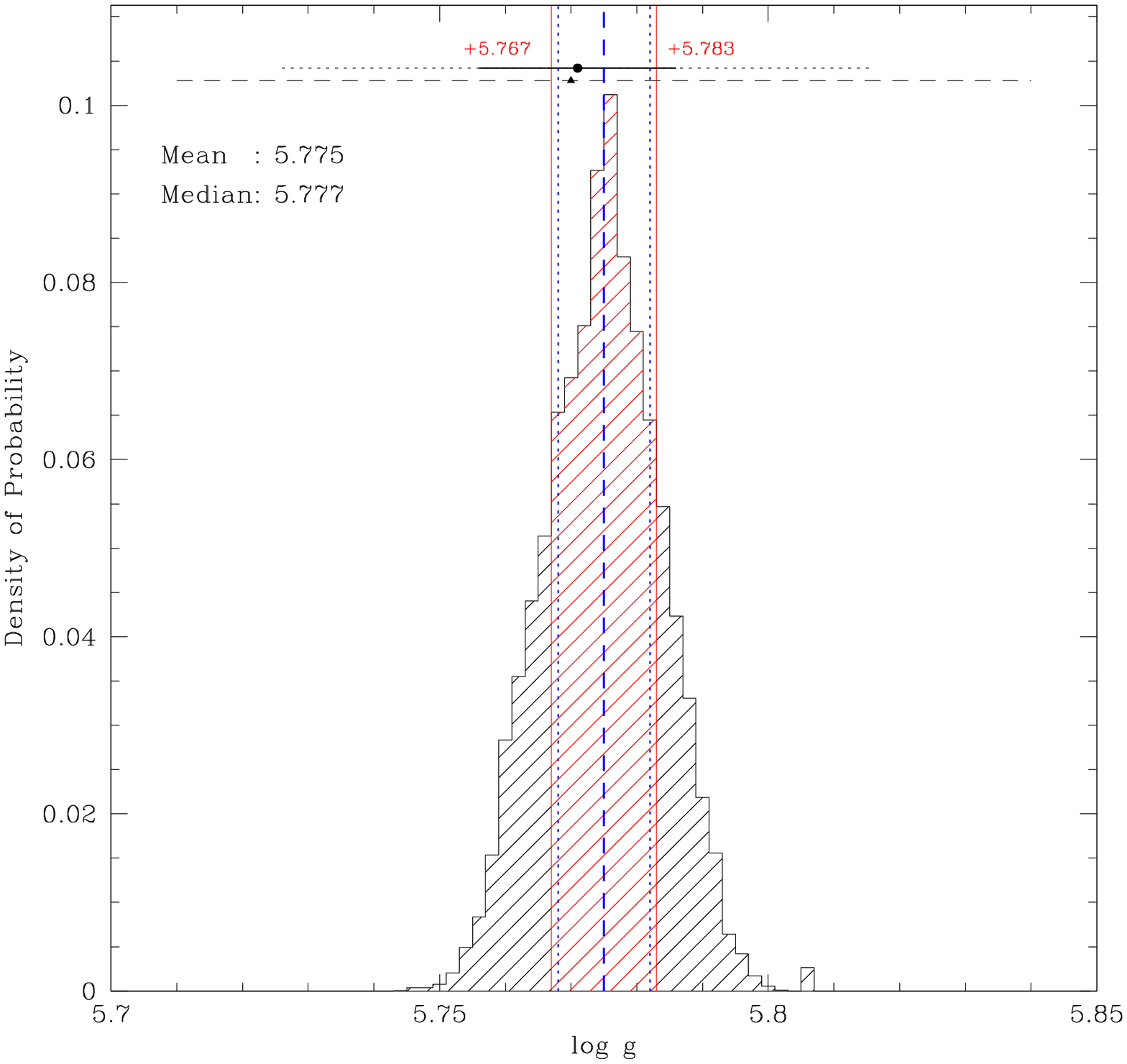}\\
\includegraphics[scale=0.37,angle=0]{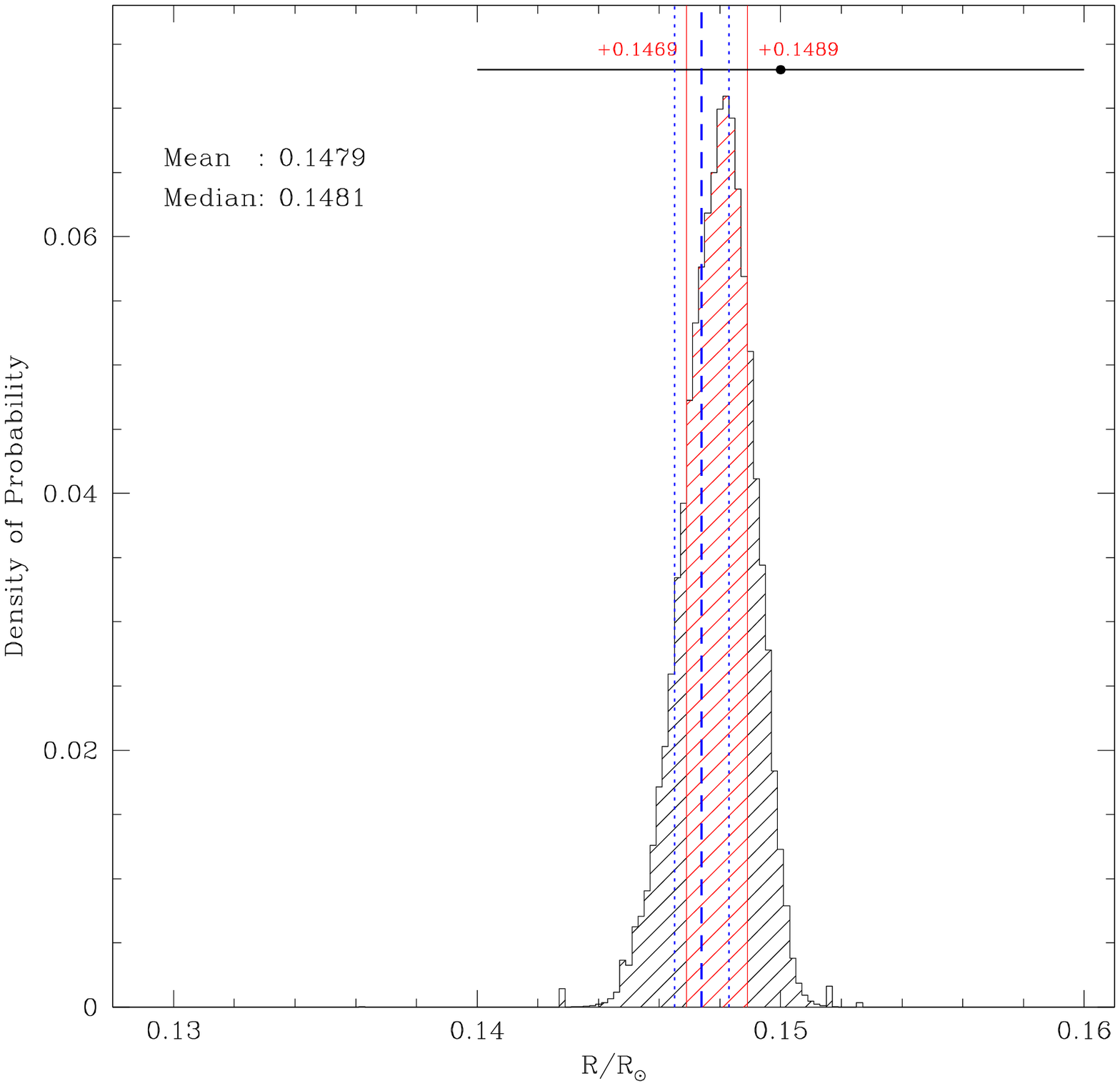}
\includegraphics[scale=0.37,angle=0]{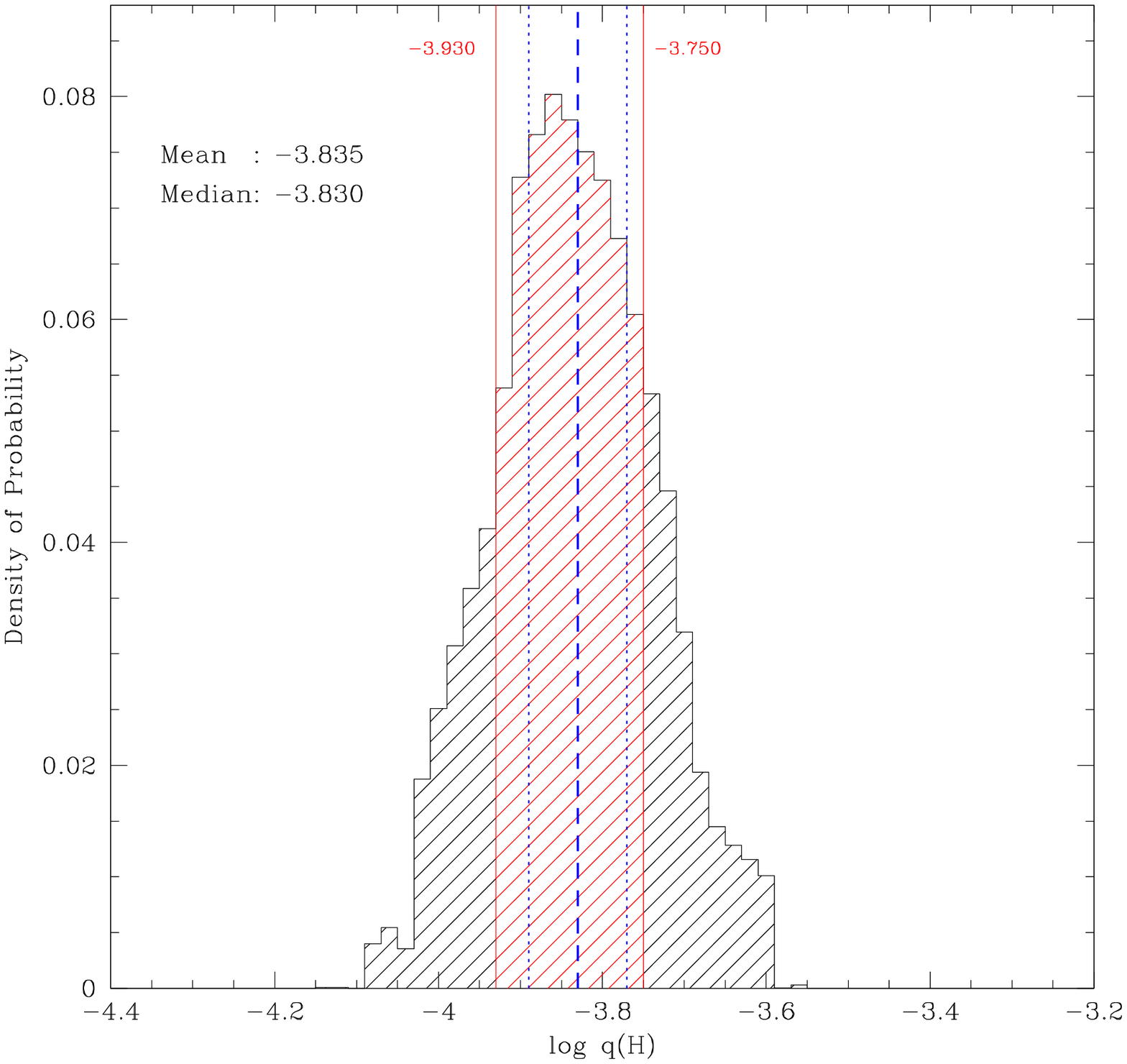}\\
\includegraphics[scale=0.37,angle=0]{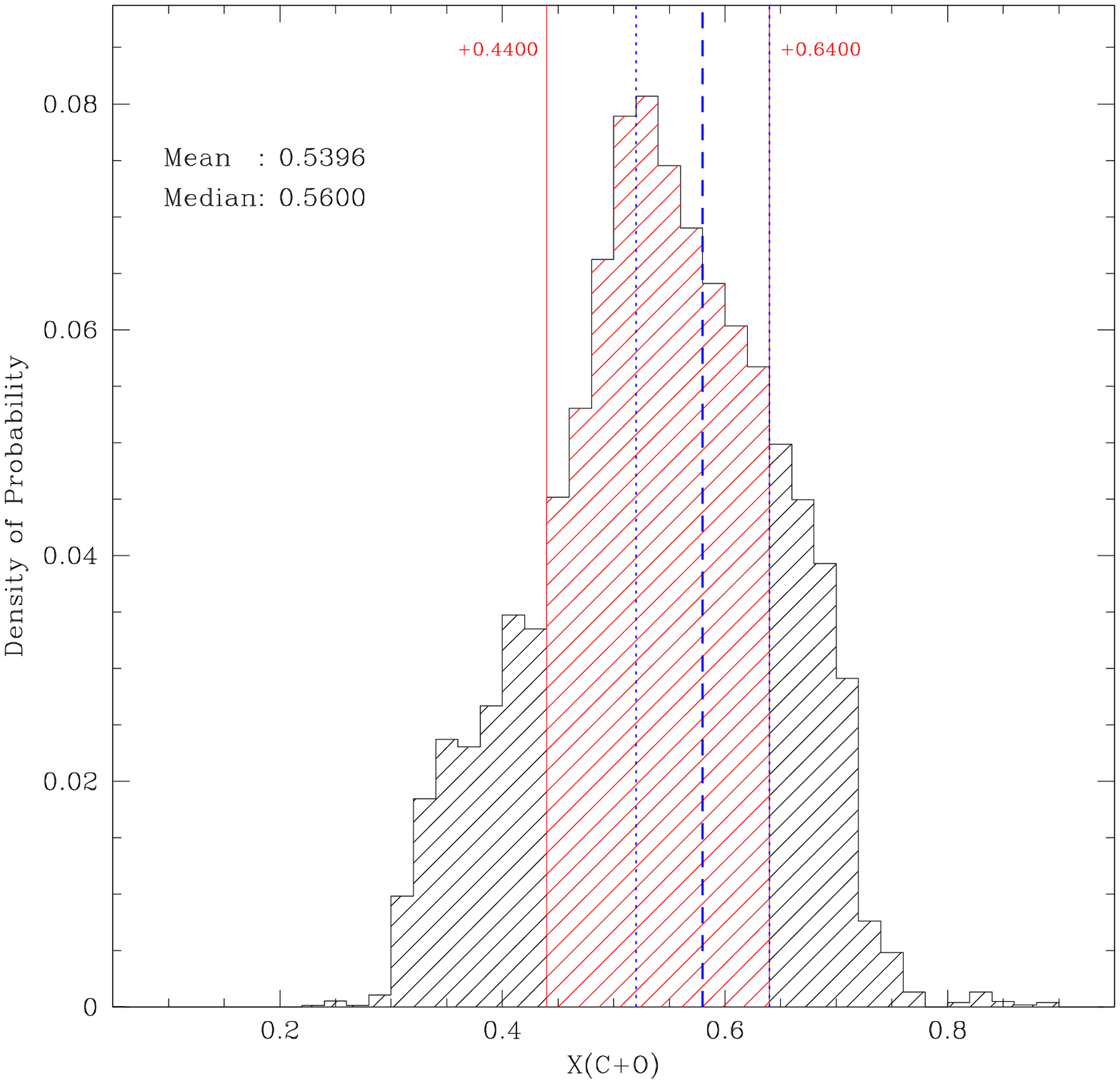}
\includegraphics[scale=0.37,angle=0]{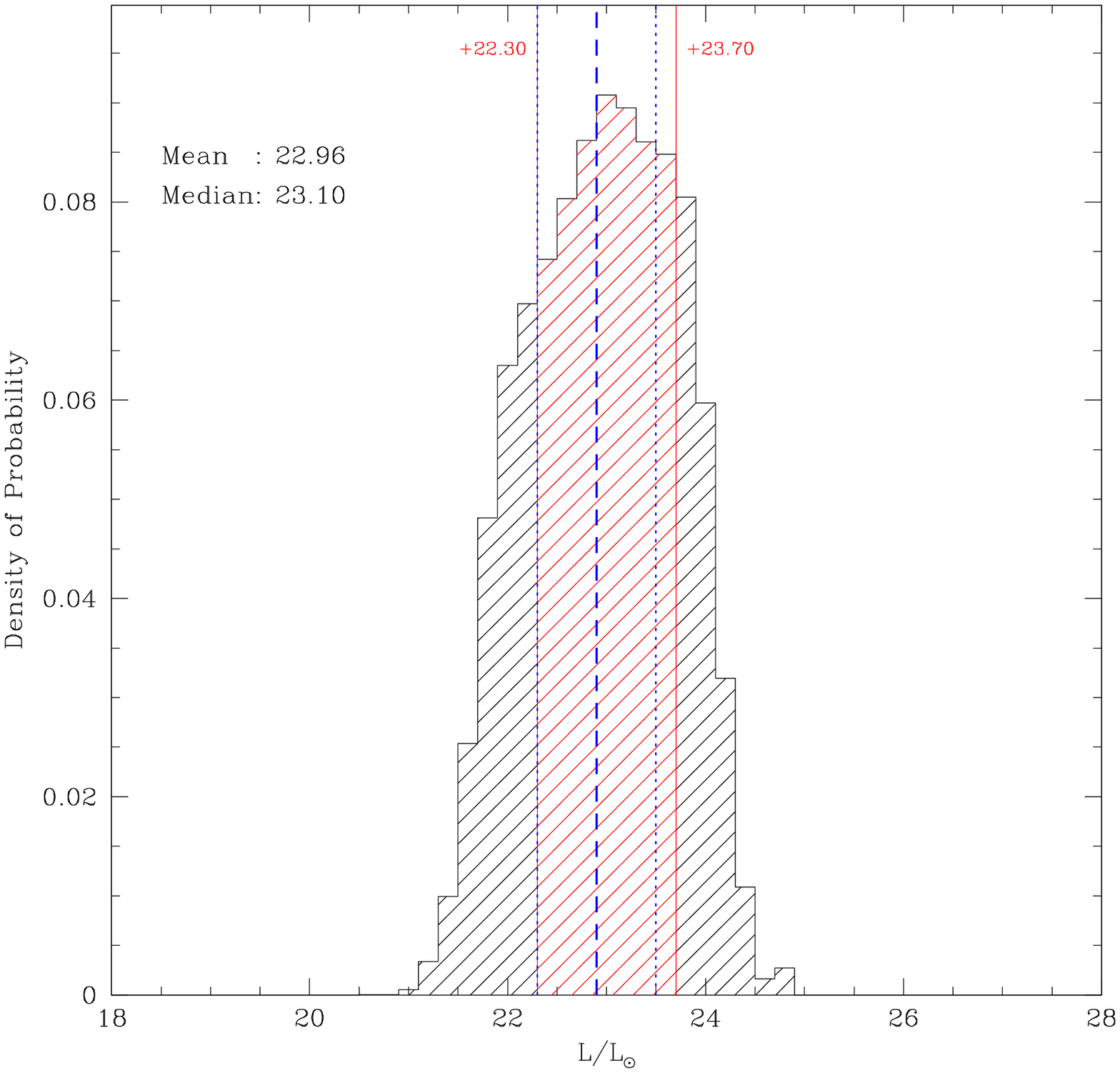}
\end{center}
\caption{\label{fB8} Same as Fig. \ref{fB3}, but using models with modified rates for the triple-$\alpha$ and $^{12}$C($\alpha$,$\gamma$)$^{16}$O nuclear reactions.}  
\end{figure*}

\end{document}